\documentclass[aps,prd,amsmath,11pt,superscriptaddress,floatfix,nofootinbib]{revtex4-1}
\usepackage{epsfig,amssymb,amsfonts,amsmath,mathtools,bm,color,xcolor,graphicx,braket,adjustbox,esint,upgreek}
\usepackage{multirow}
\usepackage[utf8]{inputenc}
\usepackage{setspace}
\usepackage{dcolumn,kantlipsum,soul}
\usepackage{rotating}
\usepackage[colorlinks=true,linkcolor=blue]{hyperref}
\allowdisplaybreaks

\newcommand{\lag}{\mathcal{L}}
\newcommand{\nno}{\nonumber\\}
\usepackage{natbib}
\usepackage{graphicx}

\newcommand{\be}{\begin{equation}}
\newcommand{\ee}{\end{equation}}
\newcommand{\bea}{\begin{eqnarray}}
\newcommand{\eea}{\end{eqnarray}}
\newcommand{\beas}{\begin{eqnarray*}}
\newcommand{\eeas}{\end{eqnarray*}}

\def\vec#1{\boldsymbol{#1}}

\newcommand{\pc}{P_c}
\newcommand{\jp}{J/\psi p}
\newcommand{\sigh}{\Sigma_c^{(*)}\bar{D}^{(*)}}
\newcommand{\sigd}{\Sigma_c\bar{D}}

\newcommand{\lamh}{\Lambda_c\bar{D}^{(*)}}
\newcommand{\etacp}{\eta_c p}
\newcommand{\etacn}{\eta_c p}

\newcommand{\blue}[1]{{\color[rgb]{0,0,1}{#1}}}

\newcommand{\bonn}{\affiliation{Helmholtz-Institut f\"ur Strahlen- und Kernphysik and\\ Bethe Center for Theoretical Physics, Universit\"at Bonn, D-53115 Bonn, Germany}}

\newcommand{\ific}{\affiliation{Instituto de F\'isica Corpuscular (centro mixto CSIC-UV),\\
Institutos de Investigaci\'on de Paterna, Apartado 22085, 46071, Valencia, Spain}
}

\newcommand{\rub}{\affiliation{Institut f\"ur Theoretische Physik II, Ruhr-Universit\"at Bochum, D-44780 Bochum, Germany}}

\newcommand{\fzj}{\affiliation{Institute for Advanced Simulation, Institut f\"ur Kernphysik and\\ J\"ulich Center for Hadron Physics, Forschungszentrum J\"ulich, D-52425 J\"ulich, Germany}}

\newcommand{\itp}{\affiliation{CAS Key Laboratory of Theoretical Physics, Institute of Theoretical Physics,\\ Chinese Academy of Sciences,  Zhong Guan Cun East Street 55, Beijing 100190, China}}

\newcommand{\ucas}{\affiliation{School of Physical Sciences, University of Chinese Academy of Sciences, Beijing 100049, China}}

\newcommand{\itep}{\affiliation{Institute for Theoretical and Experimental Physics  NRC  “Kurchatov Institute”, Moscow 117218, Russia}}

\newcommand{\lebedev}{\affiliation{P.N. Lebedev Physical Institute of the Russian Academy of Sciences, 119991, Leninskiy Prospect 53, Moscow, Russia}}

\newcommand{\tsu}{\affiliation{Tbilisi State University, 0186 Tbilisi, Georgia}}

\newcommand{\murcia}{\affiliation{Departamento de F{\'i}sica, Universidad de Murcia, E-30071 Murcia, Spain}}

\newcommand{\scnu}{\affiliation{{Guangdong Provincial Key Laboratory of Nuclear Science,}\\ Institute of Quantum Matter, South China Normal University, Guangzhou 510006, China}}

\newcommand{\ihep}{\affiliation{Institute of High Energy Physics, Chinese Academy of Sciences, Beijing 100049, China}}

\newcommand{\snsc}{\affiliation{Guangdong-Hong Kong Joint Laboratory of Quantum Matter,\\Southern Nuclear Science Computing Center, South China Normal University, Guangzhou 510006, China}}

\begin{document}

\title{Revisiting the nature of the $P_c$ pentaquarks}

\author{Meng-Lin Du}\email{du@hiskp.uni-bonn.de}
\bonn\ific

\author{Vadim Baru}\email{vadim.baru@tp2.rub.de}
\rub \itep \lebedev

\author{Feng-Kun Guo}\email{fkguo@itp.ac.cn}
\itp \ucas

\author{Christoph~Hanhart}\email{c.hanhart@fz-juelich.de}
\fzj

\author{Ulf-G.~Mei{\ss}ner}\email{meissner@hiskp.uni-bonn.de}
\bonn \fzj \tsu

\author{Jos\'e A. Oller}\email{oller@um.es}
\murcia

\author{Qian Wang}\email{qianwang@m.scnu.edu.cn}
\scnu\snsc
\ihep

\begin{abstract}

The nature of the three narrow hidden-charm pentaquark $P_c$ states, 
i.e., $\pc(4312)$, $\pc(4440)$ and $\pc(4457)$, 
is under intense discussion since their discovery from the updated analysis of  the process $\Lambda_b^0\to J/\psi p K^-$ by LHCb. 
In this work we extend our previous coupled-channel approach [Phys. Rev. Lett. \textbf{124}, 072001 (2020)],  in which the  $P_c$ states are treated as $\sigh$ molecules,
 by including the $\Lambda_c\bar{D}^{(*)}$ and $\etacn$ as explicit inelastic channels  in addition to the $J/\psi p$, as required by unitarity and heavy quark spin symmetry (HQSS), respectively.    Since  inelastic parameters are very badly constrained by the current data, three calculation schemes are considered:  (a) scheme~I with pure contact interactions between the elastic, i.e., $\sigh$, and inelastic channels and without the
$\Lambda_c \bar D^{(*)}$ interactions, (b) scheme~II, where the one-pion exchange (OPE) is added to scheme~I, and (c) scheme~III, where the $\Lambda_c \bar D^{(*)}$ interactions are included in addition. 
It is shown that to obtain cutoff independent results, OPE in the multichannel system is to be supplemented  with $S$-wave-to-$D$-wave mixing contact terms. 
As a result, in line with our previous analysis, we demonstrate that the experimental data for the  $J/\psi p$ invariant mass distribution are consistent with the   interpretation 
of the  $\pc(4312)$ and $\pc(4440)/\pc(4457)$ as   $\Sigma_c\bar{D}$ and $\Sigma_c \bar{D}^{*}$ hadronic molecules, respectively,  and  that the data show
clear evidence for a new narrow state, $\pc(4380)$, 
identified as a $\Sigma_c^* \bar D$ molecule, which
should exist as a consequence of HQSS.  
While two statistically  equally good  solutions are found in scheme I, only one of these solutions with the  quantum numbers of the $\pc(4440)$ and $\pc(4457)$  being $J^P=3/2^-$ and $1/2^-$, respectively, survives  the requirement of  regulator independence once the OPE is included.
Moreover, we predict the line shapes in the elastic and inelastic channels  and demonstrate that those related to the $\pc(4440)$ and the $\pc(4457)$ in the $\Sigma_c^{(*)}\bar{D}$ and $\etacn$ mass distributions from $\Lambda_b^0\to \Sigma_c^{(*)}\bar{D} K^-$ and  $\Lambda_b^0\to \etacn K^-$ will allow one to confirm the quantum numbers given above, once the data are available. 
We also investigate possible pentaquark signals in the
$\Lambda_c \bar{D}^{(*)}$ final states. 

\end{abstract}

\maketitle

\section{Introduction}

A quantitative understanding of hadron spectra plays a key role
 in our understanding of the strong interaction as well as its fundamental underlying theory, 
 i.e., quantum chromodynamics (QCD). For a long time, hadrons were believed to be composed of
 either a pair of quark and antiquark ($q\bar{q}$) or three quarks ($qqq$) 
 according to the conventional quark model. However, the confinement property of QCD in principle
allows for the existence of any color neutral object,
such as multiquarks, hybrids, glueballs, hadronic molecules and so on,
 which are usually called exotic states. Up to now, a tremendous number of candidates for 
 having an exotic structure has been observed~\cite{Chen:2016qju,Chen:2016spr,Esposito:2016noz,Hosaka:2016pey,Dong:2017gaw,Olsen:2017bmm,Guo:2017jvc,Cerri:2018ypt,Kou:2018nap,Brambilla:2019esw,Guo:2019twa,Liu:2019zoy,Yuan:2019zfo}.
 Among them the observation of the $\pc(4380)$ and $\pc(4450)$~\cite{Aaij:2015tga},
 as the first observation of pentaquarks,
 in 2015 builds a landmark. They decay strongly into $\jp$ and thus contain 
at least five quarks, $c\bar{c}uud$. The existence of these two pentaquark states is consistent with a model-independent
reanalysis of the same data sample~\cite{Aaij:2016phn} and a full 
amplitude analysis of the $\Lambda_b^0\to \jp \pi^-$ decay~\cite{Aaij:2016ymb}. 
A later analysis based on an order-of-magnitude larger data sample, 
i.e., the combined data set collected in Run 1 and Run 2 by the LHCb Collaboration,
shows that the $\pc(4450)$ structure consists of two narrow overlapping peaks, i.e., $\pc(4440)$ and $\pc(4457)$, and a third narrow peak $\pc(4312)$ emerges~\cite{Aaij:2019vzc}. 
However, the verification of the broad $\pc(4380)$ reported in 2015  awaits a complete amplitude analysis. Numerous theoretical interpretations of the nature of the pentaquarks followed these discoveries,
including hadronic molecules~\cite{Chen:2019bip,Chen:2019asm,Guo:2019fdo,Liu:2019tjn,He:2019ify,Guo:2019kdc,Shimizu:2019ptd,Xiao:2019mst,Xiao:2019aya,Wang:2019nwt,Meng:2019ilv,Wu:2019adv,Xiao:2019gjd,Voloshin:2019aut,Sakai:2019qph,Wang:2019hyc,Yamaguchi:2019seo,Liu:2019zvb,Lin:2019qiv,Wang:2019ato,Gutsche:2019mkg,Burns:2019iih,Du:2019pij,Wang:2019spc,Xu:2020gjl,Kuang:2020bnk,Peng:2020xrf,Peng:2020gwk,Xiao:2020frg,Dong:2021juy,Peng:2021hkr}, 
compact pentaquarks states~\cite{Ali:2019npk,Zhu:2019iwm,Wang:2019got,Giron:2019bcs,Cheng:2019obk,Stancu:2019qga,Kuang:2020bnk}, hadro-charmonia~\cite{Eides:2015dtr,Eides:2019tgv,Anwar:2018bpu}, and cusp effects~\cite{Kuang:2020bnk}.

The proximity of the $\Sigma_c\bar{D}^{(*)}$ thresholds
to these narrow pentaquark structures suggests that
the corresponding two-hadron states play an important role in the dynamics of the pentaquarks, hinting at an  interpretation of their structure as hadronic molecules.
In the most common hadronic molecular picture, the $\pc(4312)$ is an $S$-wave $\Sigma_c\bar{D}$ bound state, while the $\pc(4440)$ and $\pc(4457)$ are bound states of $\Sigma_c\bar{D}^*$ with different spin structures, see e.g., Refs.~\cite{Liu:2019tjn,Xiao:2019aya,Sakai:2019qph,Du:2019pij,Xiao:2020frg}.  The origin of the peak from the $\pc(4312)$ is attributed to a virtual state of $\Sigma_c\bar{D}$ in Ref.~\cite{Fernandez-Ramirez:2019koa} based on an amplitude analysis,\footnote{As a consequence of being a near-threshold virtual state, the peak for the $\pc(4312)$ should behave as a sharp threshold cusp. It is due to the convolution with the energy resolution that the peak shown in Ref.~\cite{Fernandez-Ramirez:2019koa} is smooth.} which only fits to data around the $\Sigma_c\bar{D}$ threshold. In Ref.~\cite{Kuang:2020bnk}, final state interactions are constructed based on a $K$-matrix including the channels $\jp$-$\Sigma_c\bar{D}$-$\Sigma_c\bar{D}^*$.
The analysis  suggests that the $\pc(4312)$ is a $\Sigma_c\bar{D}$ molecule, while
the $\pc(4440)$ could be a compact pentaquark state, 
and the $\pc(4457)$ could be caused by the cusp effect.\footnote{A strong threshold cusp effect normally requires the existence of a near-threshold pole in an unphysical Riemann sheet (RS)~\cite{Guo:2014iya,Guo:2019twa,Dong:2020hxe}.}

While the proximity of the narrow $P_c$ peaks to the $\Sigma_c\bar{D}^{(*)}$ make the 
molecular interpretation very compelling, at least some peaking structures in the 
$\jp$ mass distributions can be also generated by the triangle singularities~\cite{Guo:2015umn,Liu:2015fea,Mikhasenko:2015vca,Bayar:2016ftu,Aaij:2019vzc,Guo:2019twa}. 
A triangle singularity arises when all intermediate particles in a triangle loop are (nearly) on-mass-shell. Therefore its location is quite sensitive to the masses and widths of the involved particles~\cite{Guo:2019twa}. 
The potential triangle singularities were discussed in Ref.~\cite{Aaij:2019vzc} for the 
three $P_c$ structures. Considering the realistic widths of the exchanged resonances, the $P_c(4312)$ 
and $\pc(4440)$ structures are unlikely to be caused by triangle singularities. However, the 
$\pc(4457)$ structure could in principle be produced by a triangle diagram with  $D_{s1}^*(2860), \Lambda_c(2595)$ and $\bar{D}^{*0}$ in the intermediate state. 
It is worth noticing that even if 
$P_c(4312)$ and $P_c(4440)$ are not generated by triangle singularities, this does not necessarily mean that triangle diagrams play no role in producing these states. 
In the present work, however, we assume a pointlike production mechanism of  the $P_c$ states and focus on a coupled-channel approach for the final-state interactions of the $\Sigma^{(*)} \bar D^{(*)}$ as well as the inelastic channels. 
This means that in this work the possible effects of the triangle singularities on the pentaquark production are absorbed into the parameters of the production vertex. 
A study of the dynamical role of  triangle diagrams will be the focus of future studies, once more data become available.   

To make further progress it is necessary to systematically 
investigate the implications of the symmetries of QCD, here, most importantly, heavy quark spin symmetry (HQSS).
In simple terms, under the assumption that the observed
pentaquarks are of molecular nature,
the symmetry allows one to predict additional states, the so-called spin partners. To be concrete,
HQSS predicts seven $\pc$ states which are classified into two heavy quark spin multiplets.  Three of these seven  correspond to the ones reported by LHCb~\cite{Xiao:2013yca,Liu:2018zzu,Liu:2019tjn,Du:2019pij}:
While the $\pc(4312)$ is unambiguously assigned to the $J^P=\frac12^-$ $\Sigma_c\bar{D}$ bound state, 
there are two possible spin structures for the $\pc(4440)$ and $\pc(4457)$ identified as the $\Sigma_c\bar{D}^*$ bound states, namely $J^P=\frac12^-$ and $J^P=\frac32^-$~\cite{Guo:2019kdc,Liu:2019tjn,Du:2019pij}, 
and their spin assignment is not uniquely fixed by HQSS alone~\cite{Chen:2019asm,Valderrama:2019chc,He:2019ify,Liu:2019zvb,Du:2019pij}.
In Ref.~\cite{Du:2019pij},
the $\jp$ invariant mass distribution of the $\Lambda_b^0\to \jp K^-$ process 
was described in the molecular scenario. In particular,
the $\sigh$ channels (in what follows called elastic channels, since their thresholds 
are close to the $\pc$ states) and the $\jp$ channel (inelastic channel) were included dynamically. 
The interactions between elastic channels were constrained 
by HQSS, which guarantees that there are only two momentum-independent contact potentials, reflecting
that the number of independent multiplets is two.
The effect from additional inelastic channels, such as the $\lamh$ and $\etacn$ channels and so on, was absorbed into additional imaginary parts of the two contact potentials in the spirit of an optical potential.
For the case of only contact potentials, two different solutions, 
corresponding to scenarios $A$ and $B$ in Ref.~\cite{Liu:2019tjn}, were
found describing the data almost equally well. 
Each of them gives seven poles in the $\sigh$ scattering amplitudes,
however, with different pole locations. Meanwhile, as
soon as the one-pion-exchange (OPE) potentials was included, 
only one solution was found which suggests that $\pc(4440)$ and $\pc(4457)$
couple dominantly to the $\Sigma_c\bar{D}^*$ with quantum numbers $J^P=\frac32^-$ and $\frac12^-$,
respectively~\cite{Du:2019pij}. In both fits, the $\pc(4312)$ couples dominantly to the $\Sigma_c\bar{D}$
channel with $J^P=\frac12^-$. In addition, evidence for an additional narrow state, also called for by HQSS,
around $4.38$ GeV was found in the data with $J^P=\frac32^-$, which couples dominantly to the $\Sigma_c^*\bar{D}$ (see also Ref.~\cite{Xiao:2019mst,Xiao:2020frg}). 

The mentioned imaginary parts of the contact terms violate unitarity. In this work we overcome this shortcoming by including explicitly the supposedly most prominent additional inelastic channels, namely, $\lamh$ and $\etacn$
making the mentioned imaginary parts of the contact terms obsolete.
The inclusion of the $\etacp$ channel explicitly is also necessary from HQSS, since the $\eta_c$  and $J/\psi$  are in the same HQSS multiplet. 
While the interactions of $\etacn$ and $\jp$ with the elastic channels are of a short-range character, $\lamh$ can also interact with $\Sigma^{(*)}\bar D^{(*)}$ via one-pion
exchange. Thus, when the $\Lambda_c\bar{D}$ and $\Lambda_c\bar{D}^*$ channels are included explicitly,  three-body effects from the $\Lambda_c\bar{D}\pi$ intermediate states have to be taken
into account in both the $\Sigma^*\bar D^{(*)}\to \Lambda_c \bar D^{(*)}$ transition potentials and the $\Sigma_c^{(*)}$ self-energy~\cite{Aaron:1969my} (for recent discussions on the subject see Refs.~\cite{Mai:2017vot,Zhang:2021hcl}). 
The $\Sigma_c^{(*)}$ width may have a sizable effect on the widths of the $\sigh$ molecular states. This is
especially important for those states that have a prominent $\Sigma_c^*\bar{D}^{(*)}$
component,  since the $\Sigma_c^*$ width, which is around 15 MeV~\cite{Zyla:2020zbs}, is of the same order as the widths of the pentaquarks. Among them, the narrow $\pc(4380)$ attracts special interest, 
since its confirmation would provide strong support for the molecular picture. 
Equipped with the mentioned extensions of the approach, in this work we are also in a position to investigate  
the invariant mass distributions in all the elastic and inelastic channels to understand what can be learned from the corresponding data, once they exist. 

Before we proceed, a remark on the treatment of the OPE in effective field theories (EFTs) is in order. 
As discussed in Ref.~\cite{Baru:2015nea}, the OPE potential  is well defined in the sense of an EFT only in connection with contact operators, the dependence of which on the regulator is dictated by the renormalisation group. Thus, the
question if the OPE alone provides sufficient binding to produce  shallow bound states cannot be addressed
from an EFT perspective.   In addition, to preserve HQSS in a coupled-channel problem with OPE, 
all relevant channels need to be coupled with each other: The omission of the off-diagonal transitions between  some particle channels or partial waves  leads to a strong violation of HQSS~\cite{Baru:2016iwj}. 
To the best of our knowledge, the aforementioned conditions are not satisfied in the previous studies available in the literature, where the OPE was included, see, e.g., Refs.~\cite{Nieves:2012tt,Chen:2015loa,Yamaguchi:2019seo,Burns:2019iih,Liu:2019zvb,Valderrama:2019chc,Meng:2019ilv,Wang:2019ato,Yamaguchi:2016ote}. 
However, these conditions are crucial to unearth the true impact of the OPE on the formation of the molecular states. Thus, in what follows, we aim at the construction of a field theoretically consistent formalism,  where  the renormalisation of the OPE is carried out in a way consistent with  the requirements of
renormalisation group invariance and HQSS. 

It was recently noticed in studies of the $X(3872)$ and $Z_b$ 
systems~\cite{Baru:2017gwo,Baru:2016iwj} that, in line with what is well-known for the nucleon-nucleon interaction~\cite{Ericson:1988gk}, the most prominent contribution from the OPE 
originates from the tensor force inducing significant $S$-$D$ transitions.
However, the situation in the heavy quark sector is more complicated than in nuclear EFT
because  transitions between the different channels connected
by HQSS, can be separated by more than 100 MeV due to the $D^*$-$D$ and $\Sigma^*_c$-$\Sigma_c$ mass differences.
The resulting sizable momentum scales
enhance further the role of the $S$-$D$ transitions. 
The analysis of the line shapes relevant for the $Z_b(10610)/Z_b(10650)$ and their spin partners~\cite{Wang:2018jlv,Baru:2019xnh} showed that, to remove the strong regulator dependence caused by the high-momentum contribution from the
$S$-$D$ OPE transitions, the formally next-to-leading-order (NLO) contact term for the $S$-$D$ transitions is
required to be promoted to leading order (LO). Meanwhile, the NLO $S$-$S$ contact terms are numerically marginal, consistent with the expectations. 
In line with the observations of Refs.~\cite{Wang:2018jlv,Baru:2019xnh},  if we omit the $S$-$D$ counter terms in the pentaquark system, we also observe a strong dependence on the regulator, as soon as the OPE is included in the $\Sigma_c^{(*)}\bar{D}^{(*)}\to \Sigma_c^{(*)}\bar{D}^{(*)}$  and $\Sigma_c^{(*)}\bar{D}^{(*)}\to\Lambda_c\bar{D}^{(*)}$ transition potentials. 
Following Refs.~\cite{Wang:2018jlv,Baru:2019xnh}, one expects that introducing the $S$-$D$ contact terms allows us to obtain regulator-independent results and, at the same time, to arrive at a satisfying description of the data. 
 However, only one of the two statistically almost equivalent solutions present in the pionless formulation survives  the strict requirements of renormalizability upon the inclusion of the OPE together with the $S$-$D$ term.  
The second solution still shows a strong regulator dependence, which can not be cured by the $S$-$D$ counter terms only, and, accordingly, is treated here as unreliable.
 The cutoff independent solution suggests that the quantum numbers of the $P_c(4440)$ and $P_c(4457)$ 
should be $J^P=\frac32^-$ and $\frac12^-$, respectively,  in line with Ref.~\cite{Du:2019pij}. Moreover, we predict the possible line shapes of the $\sigh$ 
and $\eta_cp$ mass distributions from $\Lambda_b^0\to \sigh K^-$ and $\Lambda_b^0\to \etacp K^-$, 
which can be used to test the hadronic molecular nature of the $P_c$ states, determine their quantum numbers 
experimentally and verify the existence of the missing $P_c$ states.

The paper is organized as follows. In Sec.~\ref{sec:potentials} we derive the effective interaction potentials for the dynamical channels and parametrize the relevant weak production vertices. In particular, in Subsection~\ref{subsec:heavylight}, we discuss the purely contact potentials in the basis of heavy-light spin degrees of freedom. In Subsections~\ref{subsec:Lagrangian} and \ref{subsec:counterterm}, we derive the OPE and the $S$-$D$ contact terms from the effective Lagrangian. The relativistic Lippmann-Schwinger equations (LSEs) with the dynamical width of $\Sigma_c^{(*)}$ are given in Sec.~\ref{sec:lse}. The fitting schemes to the $\jp$ mass distributions are presented in Sec.~\ref{sec:fit}. 
Section~\ref{subsec:contact} is devoted to a study, in which   only contact potentials are considered.
The OPE potentials are included in Sec.~\ref{subsec:ope} and the explicit inclusion of the $\lamh$ is investigated in Sec.~\ref{subsec:lamh}.
The pole positions and their effective couplings are also explicitly presented for schemes~I and II. 
In addition,  we also predict the line shapes for the $\sigh$ and $\eta_c p$ (and $\lamh$ for scheme~III) mass distributions in the
$\Lambda_b^0\to \sigh K^-$ and $\Lambda_b^0\to \eta_c p K^-$ (and $\Lambda_b^0\to \lamh K^-$) processes.  We summarize in Sec.~\ref{sec:summary}.

\section{Effective potentials}\label{sec:potentials}

In order to fully exploit the implications of HQSS
and to study the role of the $\lamh$ channels and the effect of the $\Sigma_c^{(*)}$ width 
stemming from the decay $\Sigma_c^{(*)}\to \Lambda_c\pi$, we extend the framework of Ref.~\cite{Du:2019pij} to dynamically include the $\lamh$ and $\eta_c p$ channels in addition to the $\sigh$ and $J/\psi p$ channels.  The effective transition potentials for those channels entering the LSEs contain short-range contact terms as well as OPE.
In this section, both the contact and OPE potentials are presented employing the constraints from HQSS. 

\subsection{Contact potentials}\label{subsec:heavylight}

The short-ranged contact potentials can be derived 
either from the effective Lagrangians or from the decomposition of the spin 
structures of the charm-anticharm and light degrees of freedom in the 
heavy-quark limit following, e.g., Ref.~\cite{Voloshin:2011qa}. As the later one is more
intuitive and easier to be applied to the bare production amplitudes,
we extract the LO contact potentials and the bare production amplitudes
based on the heavy-light spin structures.
Afterwards, we also show the construction based on the Lagrangian method --- while it provides identical results 
it is easier extendable to higher orders.

Along this line, we expand the two-particle basis 
in terms of heavy-light spin structure $|s_Q\otimes j_\ell\rangle$, 
with $s_Q$ and $j_\ell$ representing the total spin of the heavy 
quarks and the total angular momentum of light degrees of freedom, respectively. 
In the heavy-light spin structure basis, the $\Sigma_c^{(*)}$ and $\bar{D}^{(*)}$ are
 $\left|\frac12 \otimes 1\right\rangle$ and $\left|\frac12\otimes\frac12\right\rangle$ spin multiplets, 
 respectively, while the $\Lambda_c^+$ corresponds to a $\left|\frac12\otimes 0\right\rangle$ singlet. 
The higher partial-wave amplitudes for the orbital angular momentum $\ell$ are 
suppressed by a factor $(p/\Lambda)^{2\ell}$, where $p$ is the center-of-mass (c.m.) frame three-momentum of the system and $\Lambda$ denotes some typical hard scale commonly assumed to be of the order of 1~GeV. 
Accordingly, the short-ranged $\sigh$ and $\lamh$ potentials should be dominated by $S$ waves, although, 
given the large range of momenta involved in the study,  effects of the $S$-$D$ mixing may also become important, as discussed below.  The $S$-wave $\sigh$ and $\lamh$ systems can be cast in terms of $|s_Q\otimes j_\ell\rangle$ as~\cite{Sakai:2019qph,Xiao:2013yca,Du:2019pij}
\begin{eqnarray}
\left(\begin{array}{c}
|\Sigma_{c}\bar{D}\rangle\\
|\Sigma_{c}\bar{D}^{*}\rangle\\
|\Sigma_{c}^{*}\bar{D}^{*}\rangle\\
|\Lambda_c\bar{D}\rangle\\
|\Lambda_c\bar{D}^{*}\rangle
\end{array}\right)_{\frac{1}{2}} = \left(\begin{array}{ccccc}
\frac{1}{2} & \frac{1}{2\sqrt{3}} & \sqrt{\frac{2}{3}} & 0 &0 \\
\frac{1}{2\sqrt{3}} & \frac{5}{6} & -\frac{\sqrt{2}}{3}& 0 &0 \\
\sqrt{\frac{2}{3}} & -\frac{\sqrt{2}}{3} & -\frac{1}{3}& 0 &0 \\
0 & 0 & 0 & -\frac12 & \frac{\sqrt{3}}{2} \\
0 & 0 & 0 & \frac{\sqrt{3}}{2} & \frac12 
\end{array}\right)\left(\begin{array}{c}
|0\otimes\frac{1}{2}\rangle\\
|1\otimes\frac{1}{2}\rangle\\
|1\otimes\frac{3}{2}\rangle\\
|0\otimes \frac12\rangle^\prime\\
|1\otimes \frac12\rangle^\prime 
\end{array}\right),~~\label{eq:HL1} 
\end{eqnarray}
\begin{eqnarray}
\left(\begin{array}{c}
|\Sigma_{c}\bar{D}^{*}\rangle\\
|\Sigma_{c}^{*}\bar{D}\rangle\\
|\Sigma_{c}^{*}\bar{D}^{*}\rangle\\
|\Lambda_c\bar{D}^*\rangle 
\end{array}\right)_{\frac{3}{2}} = \left(\begin{array}{cccc}
\frac{1}{\sqrt{3}} & -\frac{1}{3} & \frac{\sqrt{5}}{3}&0\\
-\frac{1}{2} & \frac{1}{\sqrt{3}} & \frac{1}{2}\sqrt{\frac{5}{3}} &0 \\
\frac{1}{2}\sqrt{\frac{5}{3}} & \frac{\sqrt{5}}{3} & -\frac{1}{6} & 0\\
0 & 0 & 0 & 1 
\end{array}\right)\left(\begin{array}{c}
|0\otimes\frac{3}{2}\rangle\\
|1\otimes\frac{1}{2}\rangle\\
|1\otimes\frac{3}{2}\rangle\\
|1\otimes\frac12\rangle^\prime 
\end{array}\right),~~\label{eq:HL2}
\end{eqnarray}
\begin{eqnarray}
|\Sigma_{c}^{*}\bar{D}^{*}\rangle_{\frac{5}{2}}=\bigg|1\otimes\frac{3}{2}\bigg\rangle,\label{eq:HL3}
\end{eqnarray}
where the subscripts on the left-hand side represent the total angular momentum $J=\frac12$, $\frac32$, $\frac52$.
The superscript $^\prime$ on the right-hand side denotes the heavy-light structures of the $\lamh$ systems. 
The rotation matrices in Eqs.~\eqref{eq:HL1}-\eqref{eq:HL3} will be denoted as $R^J$ in the following. Because the $\Sigma_c^{(*)}$ and $\Lambda_c$ are in different light spin multiplets, one needs four independent parameters to describe all transitions between various channels. Therefore, we introduce 
\bea
C_\frac12 \equiv \bigg\langle s_Q \otimes \frac12\bigg| \hat{\mathcal{H}}_I \bigg| s_Q\otimes \frac12 \bigg\rangle, \quad C_\frac32 \equiv \bigg\langle s_Q \otimes \frac32\bigg| \hat{\mathcal{H}}_I \bigg| s_Q\otimes \frac32 \bigg\rangle,
\eea
for the transitions between the $\sigh$ channels, and 
\bea
C_\frac12^{\prime\prime} \equiv \prescript{\prime}{}{\bigg\langle} s_Q\otimes \frac12 \bigg| \hat{\mathcal{H}}_I \bigg| s_Q\otimes \frac12\bigg\rangle^\prime,
\eea
for those between the $\lamh$ channels with $\hat{\mathcal{H}}_I$ the effective Hamiltonian respecting HQSS. For the transition between $\sigh$ and $\lamh$, we define
\bea
C_\frac12^{\prime} \equiv \prescript{\prime}{}{\bigg\langle} s_Q\otimes \frac12 \bigg| \hat{\mathcal{H}}_I \bigg| s_Q\otimes \frac12\bigg\rangle = \bigg\langle s_Q\otimes \frac12 \bigg| \hat{\mathcal{H}}_I \bigg| s_Q\otimes \frac12\bigg\rangle^\prime.
\eea
In the heavy quark limit, the contact interactions defined above are independent of $s_Q=0$ or 1. By virtue of the decomposition of Eqs.~\eqref{eq:HL1}-\eqref{eq:HL3}, the leading order contact potentials read
\bea\label{eq:contactpotential1}
V^{\frac{1}{2}^{-}}_{C}=\left(\begin{array}{ccccc}
\frac{1}{3}C_{\frac{1}{2}}+\frac{2}{3}C_{\frac{3}{2}} & \frac{2}{3\sqrt{3}}C_{\frac{1}{2}}-\frac{2}{3\sqrt{3}}C_{\frac{3}{2}} & \frac{1}{3}\sqrt{\frac{2}{3}}C_{\frac{1}{2}}-\frac{1}{3}\sqrt{\frac{2}{3}}C_{\frac{3}{2}} & 0 & \frac{1}{\sqrt{3}}C^\prime_{\frac12}\\
\frac{2}{3\sqrt{3}}C_{\frac{1}{2}}-\frac{2}{3\sqrt{3}}C_{\frac{3}{2}} & \frac{7}{9}C_{\frac{1}{2}}+\frac{2}{9}C_{\frac{3}{2}} & -\frac{\sqrt{2}}{9}C_{\frac{1}{2}}+\frac{\sqrt{2}}{9}C_{\frac{3}{2}} & \frac{1}{\sqrt{3}}C^\prime_{\frac12} & \frac23 C^\prime_{\frac12} \\
\frac{1}{3}\sqrt{\frac{2}{3}}C_{\frac{1}{2}}-\frac{1}{3}\sqrt{\frac{2}{3}}C_{\frac{3}{2}} & -\frac{\sqrt{2}}{9}C_{\frac{1}{2}}+\frac{\sqrt{2}}{9}C_{\frac{3}{2}} & \frac{8}{9}C_{\frac{1}{2}}+\frac{1}{9}C_{\frac{3}{2}} & -\sqrt{\frac23}C^\prime_{\frac12} & \frac{\sqrt{2}}{3}C^\prime_{\frac12} \\
0 & \frac{1}{\sqrt{3}}C^\prime_{\frac12} & -\sqrt{\frac23}C^\prime_{\frac12} & C_\frac12^{\prime\prime} & 0 \\
\frac{1}{\sqrt{3}}C^\prime_{\frac12} &  \frac23 C^\prime_{\frac12} & \frac{\sqrt{2}}{3}C^\prime_{\frac12} & 0 & C_\frac12^{\prime\prime}
\end{array}\right),
\eea
\bea\label{eq:contactpotential2}
V^{\frac{3}{2}^{-}}_{C}=\left(\begin{array}{cccc}
\frac{1}{9}C_{\frac{1}{2}}+\frac{8}{9}C_{\frac{3}{2}} & -\frac{1}{3\sqrt{3}}C_{\frac{1}{2}}+\frac{1}{3\sqrt{3}}C_{\frac{3}{2}} & -\frac{\sqrt{5}}{9}C_{\frac{1}{2}}+\frac{\sqrt{5}}{9}C_{\frac{3}{2}} & -\frac13 C^\prime_{\frac12} \\
-\frac{1}{3\sqrt{3}}C_{\frac{1}{2}}+\frac{1}{3\sqrt{3}}C_{\frac{3}{2}} & \frac{1}{3}C_{\frac{1}{2}}+\frac{2}{3}C_{\frac{3}{2}} & +\frac{1}{3}\sqrt{\frac{5}{3}}C_{\frac{1}{2}}-\frac{1}{3}\sqrt{\frac{5}{3}}C_{\frac{3}{2}} & \frac{1}{\sqrt{3}}C^\prime_{\frac12} \\
-\frac{\sqrt{5}}{9}C_{\frac{1}{2}}+\frac{\sqrt{5}}{9}C_{\frac{3}{2}} & \frac{1}{3}\sqrt{\frac{5}{3}}C_{\frac{1}{2}}-\frac{1}{3}\sqrt{\frac{5}{3}}C_{\frac{3}{2}} & \frac{5}{9}C_{\frac{1}{2}}+\frac{4}{9}C_{\frac{3}{2}} & \frac{\sqrt{5}}{3}C^\prime_{\frac12}  \\
-\frac13 C^\prime_{\frac12} & \frac{1}{\sqrt{3}}C^\prime_{\frac12} & \frac{\sqrt{5}}{3}C^\prime_{\frac12} & C_{\frac12}^{\prime\prime}
\end{array}\right),
\eea
\bea\label{eq:contactpotential3}
V^{\frac{5}{2}^{-}}_{C}  =C_{\frac{3}{2}}.
\eea

The transition between the elastic channels and the inelastic channels $\jp$ and $\eta_c p$
can be obtained in a similar way. In the present work, we include both $S$ and $D$ waves for 
the $\jp$ and $\etacn$ systems, since the three-momenta of the proton could be as large as 0.9~GeV and thus the $D$ wave can be as important as the $S$ wave in the energy region of interest. 
While the $\left|1\otimes\frac12\right\rangle$ component only couples
to the $\jp$ in $S$ wave in the heavy quark limit, the $\left|1\otimes\frac32\right\rangle$
only couples to the $\jp$ in $D$ wave~\cite{Du:2019pij}. The $\left|0\otimes\frac12\right\rangle$ and the $\left|0\otimes\frac32\right\rangle$ components couple to the $\eta_c p$ channel in $S$ wave and $D$ wave, respectively. 
The coupling strengths of the $S$-wave $\sigh$ to the $\jp$ and $\etacn$ 
channels are related to each other via HQSS. We therefore need to introduce only two coupling constants,
\bea\label{eq:inelasticcoupling}
g_S &\equiv & \left.\left.\left\langle 1\otimes \frac12 \right|\hat{\mathcal{H}}_I \right| \jp\right\rangle_S = \left.\left.\left\langle 0\otimes\frac12\right|\hat{\mathcal{H}}_I \right|\etacn \right\rangle_S,\nno
 g_D k^2& \equiv & \left.\left.\left\langle 1\otimes \frac32 \right|\hat{\mathcal{H}}_I \right| \jp\right\rangle_D = \left.\left.\left\langle 0\otimes \frac32 \right|\hat{\mathcal{H}}_I \right| \etacn\right\rangle_D,
\eea
where $k$ is the magnitude of the $J/\psi$ ($\eta_c$) momentum in the $\jp$ ($\etacn$) rest-frame.
Employing Eqs.~\eqref{eq:HL1}-\eqref{eq:HL3},
the constants $g_S$
and $g_D$ allow one to write the transition potentials $\mathcal{V}_{\alpha i}$ and $\mathcal{V}_{\alpha i}^{\prime}$ between the $\alpha$th elastic channel and 
the  inelastic channels $\jp$ and $\etacn$, respectively, 
as
\begin{align}\label{eq:transitioneine}
&\mathcal{V}^J_{\alpha 1} = g_SR_{\alpha 2}^J, \quad &\mathcal{V}^J_{\alpha 2}(k)&=g_D k^2R_{\alpha 3}^J,
  \quad &J &=\frac12,\,\frac32, \nonumber\\
&\mathcal{V}^J_{\alpha 1} = 0, \quad &\mathcal{V}^J_{\alpha 2}(k)&=g_D k^2, \quad &J &=\frac52 , \nonumber\\
&\mathcal{V}^{\prime J}_{\alpha 1} = g_SR_{\alpha 1}^J, \quad 
&\mathcal{V}^{\prime J}_{\alpha 2}(k)& =0, 
\quad  &  J & = \frac12, \nonumber\\
& \mathcal{V}^{\prime J}_{\alpha 1} =0 &\mathcal{V}^{\prime J}_{\alpha 2}(k)&=g_D k^2R_{\alpha 1}^J,
  \quad &J & = \frac32, \nonumber\\
  &\mathcal{V}^{\prime J}_{\alpha 1} = 0, \quad &\mathcal{V}^{\prime J}_{\alpha 2}(k)&=0, \quad &J &=\frac52 ,
\end{align}
where $\alpha = 1,2,3$ for $J=\frac 12, \frac 32$ and $\alpha=1$ for $J=\frac 52$, and  $i=1$, 2 denote the $S$ and $D$ wave in the inelastic channels, respectively.
The direct $\jp$ and $\etacn$ interactions can be neglected since  they are Okubo-Zweig-Iizuka suppressed and were found very weak in a recent lattice
QCD calculation~\cite{Skerbis:2018lew} (see also the estimate based on a coupled-channel model in Ref.~\cite{Du:2020bqj} leading to a $J/\psi p$ scattering length of the order of milli-fm). A similar assumption about  neglecting the direct interactions between the inelastic $\Upsilon \pi$ and $h_b \pi$ channels was shown to be consistent with data for the $Z_b(10610)$ and $Z_b(10650)$ in a combined analysis of the $\Upsilon(10860)$ decays~\cite{Wang:2018jlv}, and with the Dalitz plot analysis of the $\Upsilon \pi\pi$ final states from the $\Upsilon(10860)$~\cite{Baru:2020ywb}.
In addition, we assume that the effect from the interaction of the   $\Lambda_c \bar{D}^{(*)}$ with the $J/\psi p$ and $\eta_c p$ channels is suppressed and can also be neglected. 
As long as the direct $\jp$ and $\etacn$ interactions are neglected, the effect of these
channels on the elastic channels can be included through an additional contribution~\cite{Hanhart:2015cua,Albaladejo:2015lob,Guo:2016bjq,Wang:2018jlv,Baru:2019xnh,Du:2019pij} to the potentials among the elastic channels. 
While the real parts of this contribution can be absorbed by redefining the contact terms $C_\frac12$ and $C_\frac32$~\cite{Baru:2019xnh}, 
the imaginary parts are included into the potentials as 
\bea
V_{\text{in},\alpha\beta}^J(E) =-\frac{i}{2\pi E}\sum_{j=1}^2m_{J/\psi}m_p\mathcal{V}_{\alpha j}^J\mathcal{V}_{\beta j}^J k - \frac{i}{2\pi E}\sum_{j=1}^2m_{\eta_c}m_p\mathcal{V}_{\alpha j}^{\prime J}\mathcal{V}_{\beta j}^{\prime J} k.
\eea
As a result, the full effective potential for the elastic channels is the sum of the contact potential $V_C^J$, 
the effective potential from the inelastic channel $V_\text{in}^J$ and the OPE potential $V_\text{OPE}^J(E,p,p^\prime)$
\bea
V^J(E,p,p^\prime) = V_C^J + V_\text{in}^J(E)+ V_\text{OPE}^J(E,p,p^\prime) ,
\eea 
with $E$ the total energy of the system, and $p$ and $p^\prime$ the incoming and outgoing three-momenta.
The OPE potential $V_\text{OPE}^J(E,p,p^\prime)$ will be discussed in the
next subsection.

To describe the $\Lambda_b^0\to \jp K^-$ decay, one needs the bare weak production vertices as well.
In this work, we only consider the $S$-wave $\sigh$ bare production vertices for the $\Lambda_b^0\to \sigh K^-$
process, since the $\sigh$ thresholds are close to the energy region of interest. It is stated in Ref.~\cite{Burns:2019iih} that the decay of the $\Lambda_b$
into a spectator and $\sigh $ is suppressed compared to that of $\lamh$ due to isospin breaking or color suppression in the isospin conserving case.
However, the color suppression factor is hard to  quantify. In fact,
with the $b$ quark replaced by a $c$ quark, were the color suppression effective, one would expect that the branching ratio of $\Lambda_c^+\to \Sigma^0 K^+$ should be suppressed compared to that of $\Lambda_c^+\to \Lambda K^+$, which, however, is in conflict to data, since these branching ratios are almost equal~\cite{Zyla:2020zbs}. Reversely, this might suggest that the production strength of $\Lambda_b^0\to \sigh K^-$ could be comparable to that of $\Lambda_b^0\to \lamh K^-$. 
We take this as the motivation to
neglect the $\lamh K^-$ bare production vertex, since in the molecular picture advocated here the pentaquarks couple most strongly to the $\sigh$ channels.
Also, it was shown in Ref.~\cite{Dong:2020hxe} based on an EFT,  if the system contains only two channels, that a dip must appear near the elastic threshold in the inelastic line shape if the elastic-channel interaction is strongly attractive and the production process goes dominantly through inelastic channels. On the other hand, data in the production channels driven by elastic channels should have a near-threshold peak. The fact that there are no obvious hints of dips in the vicinity of the $\sigh$ thresholds may then be regarded as yet another indication for the dominance of the elastic production mechanisms in the pentaquark system.

The weak bare production matrix elements can also be parametrized in terms of
the  $|s_Q\otimes j_\ell\rangle$ basis, i.e., $\mathcal{F}_n^J = \langle \Lambda_b|\hat{\mathcal{H}}_W|(s_Q\otimes j_\ell)_n^J K^-\rangle$, 
where $(s_Q\otimes j_\ell)_n^J$ refers to the $n$th state in the $|s_Q\otimes j_\ell\rangle$ basis in Eqs.~\eqref{eq:HL1}-\eqref{eq:HL3}. 
With the seven parameters $\mathcal{F}_n^J$ in total~\cite{Du:2019pij}, 
the bare weak production amplitude for the $\alpha$th elastic channel for a given $J$ reads
\bea
P_\alpha^J =\sum_n R_{\alpha n}^J \mathcal{F}_n^J .
\eea
Therefore, the bare weak production amplitudes for $\Lambda_b^0\to \sigh K^-$ with the $S$ wave $\sigh$ in $J=\frac12$, $\frac32$, and $\frac52$ read
\bea\label{eq:bareproduction}
P^\frac12 & = & \bigg( \frac12 \mathcal{F}_1^\frac12 +\frac{1}{2\sqrt{3}}\mathcal{F}_2^\frac12 + \sqrt{\frac23}\mathcal{F}_3^\frac12, \frac{1}{2\sqrt{3}} \mathcal{F}_1^\frac12 + \frac56 \mathcal{F}_2^\frac12 -\frac{\sqrt{2}}{3}\mathcal{F}_3^\frac12 , \sqrt{\frac23}\mathcal{F}_1^\frac12 -\frac{\sqrt{2}}{3}\mathcal{F}_2^\frac12 -\frac13 \mathcal{F}_3^\frac12 \bigg)^T, \nonumber\\
P^\frac32 & = & \bigg( \frac{1}{\sqrt{3}} \mathcal{F}_1^\frac32 - \frac13 \mathcal{F}_2^\frac32 + \frac{\sqrt{5}}{3} \mathcal{F}_3^\frac32, -\frac12 \mathcal{F}_1^\frac32 +\frac{1}{\sqrt{3}} \mathcal{F}_2^\frac32 +\frac12 \sqrt{\frac53}\mathcal{F}_3^\frac32, \frac12 \sqrt{\frac53}\mathcal{F}_1^\frac32 +\sqrt{\frac53}\mathcal{F}_2^\frac32-\frac16\mathcal{F}_3^\frac32 \bigg)^T , \nonumber\\
P^\frac52 & = &  \mathcal{F}_1^\frac52 .
\eea

\subsection{One-pion-exchange potentials}\label{subsec:Lagrangian}

The LO OPE potential can be obtained using the effective Lagrangian for the axial coupling of the pions to the charmed mesons and baryons~\cite{Wise:1992hn,Yan:1992gz}, 
\bea\label{lag:ope}
\lag & = & \frac{g_1}{4} \langle \sigma\cdot u_{ab}\bar{H}_b\bar{H}_a^\dag \rangle + i g_2\epsilon_{ijk} S^{i\dag}_{ab}u^j_{bc}S^k_{ca} - \frac{1}{\sqrt{2}}g_3\big( S_{ab}^{i\dag}u_{bc}^i T_{ca} + T^\dag_{ab} u_{bc}^i S^{i}_{ca} \big),
\eea
where $\langle\ldots\rangle$ denotes the trace in the spinor space and $\sigma$ represents the Pauli matrices. The subindices $a$, $b$, $c$ denote the light-flavor content, and the superindices $i$, $j$, $k$ denote the polarization. $S^i$ and $\bar{H}$ are the heavy quark spin doublets for the ground states $(\Sigma_c,\Sigma_c^*)$ and $(\bar{D},\bar{D}^*)$~\cite{Manohar:2000dt},
\bea
S^i=\frac{1}{\sqrt{3}}\sigma^i\Sigma_c + \Sigma^{*i}_c,\qquad \bar{H}=-\bar{D}+\sigma\cdot \bar{D}^*,
\eea
with 
\bea
S_c^{(*)}=\begin{pmatrix}
\Sigma_c^{(*)++} & \frac{1}{\sqrt{2}}\Sigma_c^{(*)+} \\
  \frac{1}{\sqrt{2}}\Sigma_c^{(*)+} & \Sigma_c^{(*)0} \\
\end{pmatrix}, \quad
 \quad
\bar{D}^{(*)} = \begin{pmatrix}
\bar{D}^{(*)0} \\
D^{(*)-} \\
\end{pmatrix}.
\eea
The $\Lambda_c^+$ is described by the spinor field $T$, 
\bea
T=\begin{pmatrix}
0 & \frac{1}{\sqrt{2}}\Lambda_c^+ \\
- \frac{1}{\sqrt{2}}\Lambda_c^+& 0\\
\end{pmatrix}, 
\eea
and the pions are collected in 
$ \vec{u}=-\nabla \Phi /F_\pi + \mathcal{O}(\Phi^3)$, where $\Phi=\vec{\tau}\cdot\vec{\pi}$ with $\vec{\tau}$ and $\vec{\pi}$ the Pauli matrices in the isospin space and the pion fields, and $F_\pi = 92.1$ MeV is the pion decay constant. From the width of $D^{*+}\to D^0\pi^+$ one gets $g_1=0.57$, and the couplings $g_2=0.42$ and $g_3=0.71$ are taken from a lattice QCD calculation~\cite{Detmold:2012ge}. The signs of these couplings are fixed from the lattice results~\cite{Detmold:2012ge}.

Once the OPE is considered, the corresponding tensor force, whose importance is well-known for few-nucleon systems, 
generates a mixture between $S$ waves and $D$ waves, which
can have a sizable impact on the line shapes between thresholds~\cite{Baru:2017gwo,Wang:2018jlv,Baru:2019xnh}. 
The quantum numbers of the $\sigh$ and $\lamh$ systems taken into account in this work are listed in Table~\ref{tab:channels}.
\begin{table}[tb]
\caption{Coupled channels for both $S$- and $D$-wave $\Sigma_c^{(\ast)}\bar{D}^{(\ast)}$, $\lamh$ systems considered in the present work. 
Whenever the two particles can couple to different spins, we use a subindex to denote their total spin.
The states given in this table appear as basis vectors for the potentials discussed in the text. }
{\begin{tabular}{c|c|c}
\hline
$J^P$ & $S$ wave & $D$ wave  \\
\hline
$\bigg(\dfrac12\bigg)^-$ & $\Sigma_c\bar{D}$, $\Sigma_c\bar{D}^\ast$,$\Sigma_c^\ast\bar{D}^\ast$, $\Lambda_c\bar{D}$, $\Lambda_c\bar{D}^{*}$ & $\Sigma_c\bar{D}^\ast_{\frac32}$, $\Sigma_c^\ast\bar{D}$, $\Sigma_c^\ast\bar{D}^\ast_\frac32$, $\Sigma_c^\ast\bar{D}^\ast_\frac52$, $\Lambda_c\bar{D}^{*}_\frac32$ \\
$\bigg(\dfrac32\bigg)^-$ & $\Sigma_c\bar{D}^\ast$, $\Sigma_c^\ast\bar{D}$, $\Sigma_c^\ast\bar{D}^\ast$, $\Lambda_c\bar{D}^*$ & $\Sigma_c\bar{D}$, $\Sigma_c\bar{D}^\ast_\frac12$, $\Sigma_c\bar{D}^\ast_\frac32$, $\Sigma_c^\ast\bar{D}$, $\Sigma_c^\ast\bar{D}^\ast_\frac12$, $\Sigma_c^\ast\bar{D}^\ast_\frac32$, $\Sigma_c^\ast\bar{D}^\ast_\frac52$, $\Lambda_c\bar{D}$, $\Lambda_c\bar{D}^*_\frac12$, $\Lambda_c\bar{D}^*_\frac32$ \\
$\bigg(\dfrac52\bigg)^-$ & $\Sigma_c^\ast\bar{D}^\ast$ &  $\Sigma_c\bar{D}$, $\Sigma_c\bar{D}^\ast_\frac12$, $\Sigma_c\bar{D}^\ast_\frac32$, $\Sigma_c^\ast\bar{D}$, $\Sigma_c^\ast\bar{D}^\ast_\frac12$, $\Sigma_c^\ast\bar{D}^\ast_\frac32$, $\Sigma_c^\ast\bar{D}^\ast_\frac52$, $\Lambda_c\bar{D}$, $\Lambda_c\bar{D}^*_\frac12$, $\Lambda_c\bar{D}^*_\frac32$ \\
\hline
\end{tabular}\label{tab:channels}
}
\end{table}

All pentaquarks so far were observed in the $\jp$ final state and therefore have isospin 1/2. 
The  corresponding isospin wave functions can
therefore be related to the particle channels via
\bea
P_c^+=\sqrt{\frac23}\Sigma_c^{(*)++}D^{(*)-} -\sqrt{\frac13}\Sigma_c^{(*)+}\bar{D}^{(*)0}.
\eea
Starting from the potentials
\bea
&V&^{\text{\tiny{OPE}}}_{\Sigma_c^{(*)+}\bar{D}^{(*)0}\to\Sigma_c^{(*)+}\bar{D}^{(*)0}} =0, \nno
&V&^{\text{\tiny{OPE}}}_{\Sigma_c^{(*)++}D^{(*)-}\to\Sigma_c^{(*)++}D^{(*)-}} = -\frac{1}{\sqrt{2}}V^{\text{\tiny{OPE}}}_{\Sigma_c^{(*)+}\bar{D}^{(*)0}\to\Sigma_c^{(*)++}D^{(*)-}}=-\frac{1}{\sqrt{2}}V^{\text{\tiny{OPE}}}_{\Sigma_c^{(*)++}D^{(*)-}\to \Sigma_c^{(*)+}\bar{D}^{(*)0}},\nno
&V&^{\text{\tiny{OPE}}}_{\Sigma_c^{(*)++}D^{(*)-}\to \Lambda_c^{+}\bar{D}^{(*)0}} =  -\sqrt{2}V^{\text{\tiny{OPE}}}_{\Sigma_c^{(*)+}\bar{D}^{(*)0}\to \Lambda_c^{+}\bar{D}^{(*)0}} ,
\eea
one thus gets
\bea
V_{\sigh\to \sigh}^{\text{\tiny{OPE}}} & = & 2V^{\text{\tiny{OPE}}}_{\Sigma_c^{(*)++}D^{(*)-}\to\Sigma_c^{(*)++}D^{(*)-}},\nno
V^{\text{\tiny{OPE}}}_{\sigh\to \lamh}& = & \sqrt{\frac32}V^{\text{\tiny{{OPE}}}}_{\Sigma_c^{(*)++}D^{(*)-}\to  \Lambda_c^{+}\bar{D}^{(*)0}}.
\eea
In the framework of  time-ordered-perturbation theory (TOPT), the OPE potentials acquire two parts corresponding to the two contributions in Fig.~\ref{fig:topt}, where the vertical lines indicate the intermediate three-body states.
\begin{figure*}[!htb]
 \centering
  \includegraphics[width=0.8\textwidth]{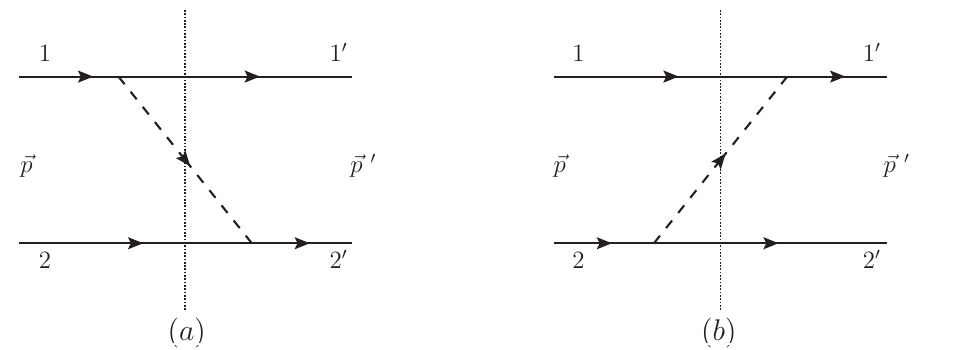}
  \caption{Diagrams in  TOPT responsible for the two contributions to the OPE potential. The solid line represents the $\Sigma_c^{(*)}$, $\Lambda_c$ and $\bar{D}^{(*)}$, the dashed line denotes  the pion field.}
  \label{fig:topt}
\end{figure*}
In some cases, for example, for the $\Sigma_c \bar{D}\to  \Lambda_c\bar{D}^{*}$
transition, the intermediate state $\Lambda_c\bar{D}\pi$ can be on shell in the energy region where all $P_c$ states reside. 
Thus, the effect of the three-body cut has to be taken into account in the potential and in the $\Sigma_c^{(*)}$ self-energy loops, see Eqs.~\eqref{eq:ope1}-\eqref{eq:ope3} in Appendix~\ref{app:OPE} for the relevant potentials and Sec.~\ref{sec:lse} for the self-energy loop.

\subsection{Next-to-leading-order contact terms}\label{subsec:counterterm}

In Subsection~\ref{subsec:heavylight}, we construct the LO contact potentials using the heavy-light spin structure decomposition $|s_Q\otimes j_\ell\rangle$. Alternatively, one can also obtain those potentials using an effective Lagrangian. For the $\sigh$ and $\lamh$ channels, the effective Lagrangian reads~\cite{Liu:2018zzu,Liu:2019tjn,Sakai:2019qph}
\bea\label{lag:contactelastic}
\lag_\text{LO} &=& - C_a \vec{S}^\dag_{ab}\cdot \vec{S}_{ba} \langle \bar{H}_c^\dag\bar{H}_c\rangle - C_b i\epsilon_{jik} S_{ab}^{j\dag}S_{ba}^k \langle \bar{H}_c^\dag \sigma^i\bar{H}_c\rangle \nno
&& + C_c \Big(S_{ab}^{i\dag} T_{ca}\langle \bar{H}_c^\dag\sigma^i \bar{H}_b\rangle - T^\dag_{ca}S_{ab}^i\langle \bar{H}^\dag_b\sigma^i\bar{H}_c\rangle\Big) - C_d T^\dag_{ab}T_{ba}\langle \bar{H}_c^\dag \bar{H}_c\rangle,
\eea
where the contact potentials $C_a$, $C_b$, $C_c$ and $C_d$  are related to those in Eqs.~\eqref{eq:contactpotential1}-\eqref{eq:contactpotential3} as   
\bea
C_a&=\frac13 C_{\frac12}+\frac23 C_{\frac32},\qquad C_b&=\frac13(C_{\frac32}-C_{\frac12}),\nno
C_c&= \frac23 C^\prime_{\frac12},\qquad \qquad C_d & = C_{\frac12}^{\prime\prime}.
\eea
The transitions between the elastic and inelastic channels in Eqs.~\eqref{eq:transitioneine} can also be obtained using the effective Lagrangian respecting the HQSS~\cite{Sakai:2019qph}:
\bea
\lag_\text{ine} = \frac{g_S}{\sqrt{3}}N^\dag \sigma^i \bar{H} J^\dag S^i - \sqrt{3}g_DN^\dag \sigma^i\bar{H}(\partial^i\partial^j-\frac13\delta^{ij}\partial^2)J^\dag S^j,
\eea
where $N$ denotes the nucleon doublet, and $J=-\eta_c+\vec\sigma\cdot \vec\psi$ contains the ground state charmonium fields. 

It was noticed in Refs.~\cite{Baru:2017gwo,Baru:2016iwj} for $B^{(*)}\bar{B}^*$ scattering that the NLO contact terms which contribute to the $S$-$D$ transitions need to be promoted to LO to tame the ultraviolet (UV) divergence associated with the $S$-$D$ OPE potentials in the course of iterations. The $S$-$S$ NLO contact terms, however, play a subleading role resulting only in a marginal improvement in the fits to the $Z_b$ line shapes. In this work, we proceed analogously, and we 
also find that once the full OPE potential is included dynamically the $S$-$D$ NLO contact terms are required to obtain regulator-independent results.
The $S$-$S$ NLO contact terms are expected  to play a marginal role also in the pentaquark system.
More details on this will be given in Sec.~\ref{subsec:ope}. To this end, we construct the NLO effective Lagrangian
\bea\label{lag:nlo}
\lag_\text{NLO} & = & -D_a^{SS} \Big( \partial^i\vec{S}^\dag_{ab}\cdot \vec{S}_{ba} \langle \partial^i \bar{H}_c^\dag\bar{H}_c\rangle + \vec{S}^\dag_{ab}\cdot \partial^i \vec{S}_{ba} \langle  \bar{H}_c^\dag\partial^i \bar{H}_c\rangle \Big) \nno
& &-D_b^{SS} i\epsilon_{jik} \Big( \partial^{\ell}S_{ab}^{j\dag}S_{ba}^k \langle \partial^\ell \bar{H}_c^\dag \sigma^i\bar{H}_c\rangle + S_{ab}^{j\dag} \partial^\ell S_{ba}^k \langle \bar{H}_c^\dag \sigma^i \partial^\ell \bar{H}_c\rangle \Big) \nno
& & - D_b^{SD} i\epsilon_{jik} \bigg[ \partial_i S^\dag_j S_k \langle \partial_\ell \bar{H}^\dag \sigma_\ell \bar{H}\rangle + \partial_\ell S^\dag_j  S_k \langle \partial_i\bar{H}^\dag \sigma_\ell  \bar{H}\rangle -\frac23 \partial_\ell S^\dag_j S_k \langle \partial_\ell\bar{H}^\dag \sigma_i \bar{H}\rangle  \nonumber \\
& & + S^\dag_j  \partial_i S_k \langle \bar{H}^\dag \sigma_\ell  \partial_\ell \bar{H}\rangle + S^\dag_j  \partial_\ell S_k \langle \bar{H}^\dag \sigma_\ell  \partial_i\bar{H}\rangle -\frac23 S^\dag_j \partial_\ell S_k \langle \bar{H}^\dag \partial_\ell\sigma_i \bar{H}\rangle \bigg]\nno
& & + \frac23 \sqrt{2}D_c^{SD} \bigg[ \partial^i S^{i\dag}_{ab} T_{ca} \langle \partial^j \bar{H}_c^\dag\sigma^j\bar{H}_b\rangle +\partial^j S^{i\dag}_{ab} T_{ca} \langle \partial^i \bar{H}_c^\dag\sigma^j\bar{H}_b\rangle  -\frac23 \partial^j S^{i\dag}_{ab} T_{ca} \langle \partial^j \bar{H}_c^\dag\sigma^i\bar{H}_b\rangle \nno
& & - \partial^i T^\dag_{ca}S_{ab}^i \langle \partial^j\bar{H}_b^\dag \sigma^j\bar{H}_c\rangle - \partial^j T^\dag_{ca}S_{ab}^i \langle \partial^i \bar{H}_b^\dag \sigma^j \bar{H}_c\rangle+\frac23  \partial^jT^\dag_{ca}S_{ab}^i \langle \partial^j\bar{H}_b^\dag \sigma^i\bar{H}_c\rangle \nno
& & +S^{i\dag}_{ab} \partial^i T_{ca} \langle  \bar{H}_c^\dag\sigma^j\partial^j\bar{H}_b\rangle +S^{i\dag}_{ab} \partial^j T_{ca} \langle \bar{H}_c^\dag\sigma^j \partial^i\bar{H}_b\rangle  -\frac23 S^{i\dag}_{ab} \partial^j T_{ca} \langle  \bar{H}_c^\dag\sigma^i\partial^j\bar{H}_b\rangle \nno
&& - T^\dag_{ca}\partial^i S_{ab}^i \langle \bar{H}_b^\dag \sigma^j\partial^j\bar{H}_c\rangle - T^\dag_{ca}\partial^j S_{ab}^i \langle \bar{H}_b^\dag \sigma^j \partial^i \bar{H}_c\rangle+\frac23 T^\dag_{ca} \partial^jS_{ab}^i \langle \bar{H}_b^\dag \sigma^i\partial^j\bar{H}_c\rangle\bigg] \nno
& & + D_d^{SS} \Big( \partial^i T_{ab}^\dag T_{ba} \langle \partial^i\bar{H}_c^\dag\bar{H}_c\rangle +  T_{ab}^\dag \partial^i T_{ba} \langle \bar{H}_c^\dag\partial^i\bar{H}_c\rangle\Big),
\eea
where the $D_{a(b,d)}^{SS}$ and the $D_{b(c)}^{SD}$ are responsible for the contact $S$-$S$ and $S$-$D$ transitions, respectively. Specifically,  $D_{b}^{SD}$ contributes to the
$O(Q^2)$ $S$-$D$ transitions 
$\sigh\to\sigh$, while   $D_{c}^{SD}$ to  $\sigh\to\lamh$.
The NLO contact potential matrix, on the basis of the $S$ and $D$ waves, for the scattering process $12\to 1^\prime 2^\prime$ with $p$ and $p^\prime$ for the incoming and outgoing three-momenta, respectively,   reads
\bea
V_\text{NLO}^J(p,p^\prime) = 
\begin{pmatrix}
(p^2+p^{\prime 2})V_{SS}^{J} & p^{\prime 2} V_{SD}^{J}\\
p^2 \big( V_{SD}^J \big)^T  & 0
\end{pmatrix}.
\eea
The potential $V_{SD}^J$ for the channels listed in Table~\ref{tab:channels} can be written in the matrix form (where the columns and rows are given for the channels in order they appear in the table) as
\bea\label{eq:nloSD12}
V_{SD}^\frac12 &=& -\frac{128}{3}
\left(
\begin{array}{ccccc}
 \frac{D_b^{SD}}{4 \sqrt{6}} & 0 & -\frac{D_b^{SD}}{8 \sqrt{30}} & -\frac{1}{8}
   \sqrt{\frac{3}{10}} D_b^{SD} & \frac{D_c^{SD}}{8 \sqrt{3}} \\
 -\frac{D_b^{SD}}{12 \sqrt{2}} & -\frac{D_b^{SD}}{8 \sqrt{6}} & \frac{D_b^{SD}}{6
   \sqrt{10}} & -\frac{D_b^{SD}}{8 \sqrt{10}} & -\frac{D_c^{SD}}{24} \\
 \frac{D_b^{SD}}{48} & -\frac{D_b^{SD}}{16 \sqrt{3}} & \frac{7 D_b^{SD}}{48 \sqrt{5}}
   & \frac{D_b^{SD}}{8 \sqrt{5}} & -\frac{D_c^{SD}}{24 \sqrt{2}} \\
 \frac{D_c^{SD}}{8 \sqrt{3}} & 0 & \frac{D_c^{SD}}{8 \sqrt{15}} & \frac{1}{8}
   \sqrt{\frac{3}{5}} D_c^{SD} & 0 \\
 -\frac{D_c^{SD}}{24} & \frac{D_c^{SD}}{8 \sqrt{3}} & -\frac{D_c^{SD}}{6 \sqrt{5}} &
   \frac{D_c^{SD}}{8 \sqrt{5}} & 0 \\
\end{array}
\right),  \\
V_{SD}^{\frac32}&=& -\frac{128}{3}
\left(
\begin{array}{cccccccccc}
 -\frac{D_b^{SD}}{8 \sqrt{3}} & \frac{D_b^{SD}}{24} & -\frac{D_b^{SD}}{12} &
   \frac{D_b^{SD}}{16 \sqrt{3}} & -\frac{D_b^{SD}}{48 \sqrt{2}} & -\frac{D_b^{SD}}{48
   \sqrt{5}} & \frac{1}{16} \sqrt{\frac{7}{10}} D_b^{SD} & -\frac{D_c^{SD}}{8 \sqrt{6}} &
   \frac{D_c^{SD}}{24 \sqrt{2}} & -\frac{D_c^{SD}}{12 \sqrt{2}} \\
 0 & \frac{D_b^{SD}}{16 \sqrt{3}} & \frac{D_b^{SD}}{16 \sqrt{3}} & 0 & \frac{D_b^{SD}}{16
   \sqrt{6}} & \frac{D_b^{SD}}{4 \sqrt{15}} & \frac{1}{16} \sqrt{\frac{21}{10}} D_b^{SD} & 0 &
   -\frac{D_c^{SD}}{8 \sqrt{6}} & -\frac{D_c^{SD}}{8 \sqrt{6}} \\
 \frac{D_b^{SD}}{16 \sqrt{15}} & -\frac{D_b^{SD}}{12 \sqrt{5}} & -\frac{D_b^{SD}}{48 \sqrt{5}}
   & \frac{D_b^{SD}}{4 \sqrt{15}} & -\frac{7 D_b^{SD}}{48 \sqrt{10}} & -\frac{D_b^{SD}}{15} &
   \frac{1}{80} \sqrt{\frac{7}{2}} D_b^{SD} & -\frac{D_c^{SD}}{8 \sqrt{30}} &
   \frac{D_c^{SD}}{6 \sqrt{10}} & \frac{D_c^{SD}}{24 \sqrt{10}} \\
 -\frac{D_c^{SD}}{8 \sqrt{6}} & \frac{D_c^{SD}}{24 \sqrt{2}} & -\frac{D_c^{SD}}{12 \sqrt{2}} &
   -\frac{D_c^{SD}}{8 \sqrt{6}} & \frac{D_c^{SD}}{48} & \frac{D_c^{SD}}{24 \sqrt{10}} &
   -\frac{1}{16} \sqrt{\frac{7}{5}} D_c^{SD} & 0 & 0 & 0 \\
\end{array}
\right) ,  \nonumber\\
V_{SD}^{\frac52}&=& -\frac{128}{3}
\left(
\begin{array}{cccccccccc}
 -\frac{D_b^{SD}}{8 \sqrt{10}} & -\frac{D_b^{SD}}{8 \sqrt{30}} & -\frac{D_b^{SD}}{16}
   \sqrt{\frac{7}{15}}  & -\frac{D_b^{SD} }{16} \sqrt{\frac{7}{5}} &
   \frac{D_b^{SD}}{8 \sqrt{15}} & -\frac{D_b^{SD}}{80} \sqrt{\frac{7}{3}}  & -\frac{D_b^{SD}}{20}
   \sqrt{\frac{7}{2}}  & \frac{D_c^{SD}}{8 \sqrt{5}} & \frac{D_c^{SD}}{8 \sqrt{15}} &
   \frac{D_c^{SD}}{8} \sqrt{\frac{7}{30}}  \\
\end{array}
\right) .\nonumber
\eea
By comparing to the $S$-$D$ OPE potentials in Eqs.~\eqref{eq:ope1}-\eqref{eq:ope3}, it is easy to see that the NLO $S$-$D$ contact potential has the same structure as the corresponding OPE potentials. Indeed, the coefficients for 
the $\Sigma_c^{(*)} \bar{D}^{(*)}\to \Sigma_c^{(*)} \bar{D}^{(*)}$ transitions are exactly the same, while those for  the transitions $\Sigma_c^{(*)} \bar{D}^{(*)}\to \Lambda_c \bar{D}^{(*)}$ differ by an overall coefficient, which can be absorbed into a redefined $D_c^{SD}$. 
The same holds for the structure of $V_{SS}^J$ compared to the LO contact potentials in Eqs.~\eqref{eq:contactpotential1}-\eqref{eq:contactpotential3}. Thus, $V_{SS}^J$ is not shown explicitly here.

\section{Lippmann-Schwinger equation with dynamic width of $\Sigma_c^{(*)}$}\label{sec:lse}

\begin{figure*}[tbh]
 \centering
  \includegraphics[width=0.5\textwidth]{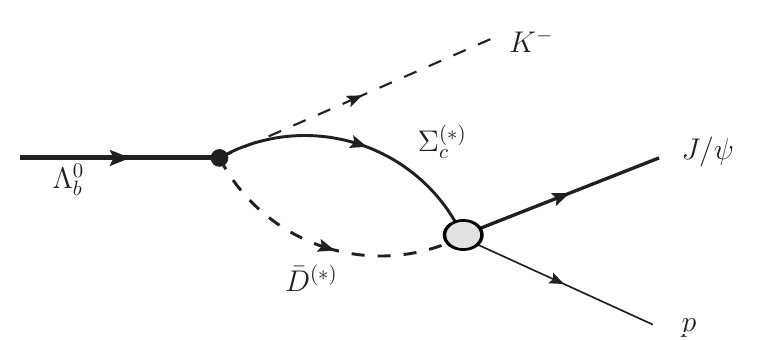}
  \caption{Diagram for the $\Lambda_b^0\to \jp K^-$ decay through $\Lambda_b^0\to\sigh K^-$ following the final state interactions.}
  \label{fig:decay}
\end{figure*}

With the above ingredients, one can obtain the production amplitude satisfying unitarity by solving the LSEs. 
A diagrammatic representation
thereof is shown in Fig.~\ref{fig:decay}. Here we treat the pion
relativistically and employ the relativistic mass-energy relations for $\Sigma_c^{(*)}$, 
$\bar{D}^{(*)}$ and $\Lambda_c$. To be consistent, we employ 
the LSEs with relativistic Green functions. The $T$-matrix for  $\sigh \to \sigh $ scattering is given by the LSE, 
\bea
T_{\alpha\gamma} ^J(E,p, p^\prime)= V_{\alpha\gamma}^J(E,p, p^\prime)-\sum_\beta \int\frac{d^3\vec{q}}{(2\pi)^3}V_{\alpha\beta}^J(E,p,q)G_\beta(E,q)T_{\beta\gamma}^J(E,q, p^\prime),
\eea
where the two-body propagator reads~\cite{Baru:2015tfa}
\bea\label{eq:twoprop}
G_\beta (E,\vec{q}) = \frac{m_{\Sigma_c^{(*)}}m_{D^{(*)}}}{E_{\Sigma_c^{(*)}}(\vec{q})E_{D^{(*)}}(\vec{q})}\frac{1}{E_{\Sigma_c^{(*)}}(\vec{q})+E_{D^{(*)}}(\vec{q})-E-  {\tilde{\Sigma}_R(s,m_{\Sigma_c^{(*)}})}/{(2E_{\Sigma_c^{(*)}}(\vec{q})})},
\eea
with 
\bea
s=\big (E-E_{D^{(*)}}(\vec{q}) \big)^2 - \vec{q}^2
\eea
for the off-shell $\Sigma_c^{(*)}$ resonance and $\tilde{\Sigma}_R(s,m_{\Sigma_c^{(*)}} )$ the renormalized self-energy loop diagram with $E_i=\sqrt{\vec{q}^2+m_i^2}$. 
The self-energy function $\tilde{\Sigma}_R(s,m_{\Sigma_c^{(*)}} )$ can be  evaluated from the diagram in Fig.~\ref{fig:selfenergy}, 
since the width of the $\Sigma_c^{(*)}$ is almost saturated by 
the two-body decay into the $\Lambda_c\pi$~\cite{Zyla:2020zbs}. 
For simplicity,  here we use a more pragmatic way and consider only the unitarity-driven part 
of the loop diagram, which generates a relevant three-body cut from the $\Lambda_c \bar D \pi$ threshold. 
We expect that the principal value part of the loop can be largely absorbed into  renormalization of the $\Sigma_c^{(*)}$ mass. Therefore, the self-energy function reduces to
\bea
\tilde{\Sigma}_R(s,m_{\Sigma_c^{(*)}})= i g^2 \frac{p^3}{\sqrt{s}},
\label{eq:self-energy}
\eea
where $p=\lambda^{1/2}(s,m_{\Lambda_c}^2,m_\pi^2)/{2\sqrt{s}}$ with K\"all\'en function
$\lambda(a,b,c)=a^2+b^2+c^2-2ab-2bc-2ca$ and $g$ is set to reproduce the physical
width of the $\Sigma_c^{(*)}$, i.e., $\tilde{\Sigma}_R(m_{\Sigma_c^{(*)}}^2,m_{\Sigma_c^{(*)}} )= i m_{\Sigma_c^{(*)}} \Gamma_{\Sigma_c^{(*)}}.$
The  effect of the $\bar{D}^{*}$ width can be safely neglected as it is only
several tens of keV, i.e., it is three orders of magnitude smaller than that of the $\Sigma_c^{(*)}$.
The widths of the $\bar{D}$ and $\Lambda_c$ are even  smaller than that of $\bar{D}^{*}$
since both states decay only via the weak interaction, 
and thus are also negligible.
Neglecting the energy-dependence of the self-energy, i.e., employing $\tilde{\Sigma}_R(s,m_{\Sigma_c^{(*)}} )= im_{\Sigma_c^{(*)}}\Gamma_{\Sigma_c^{(*)}}$ in the LSEs, 
the two-body propagator Eq.~\eqref{eq:twoprop} reduces to the form used in Ref.~\cite{Du:2019pij} in the nonrelativistic limit. 
\begin{figure*}[!htb]
 \centering
  \includegraphics[width=0.4\textwidth]{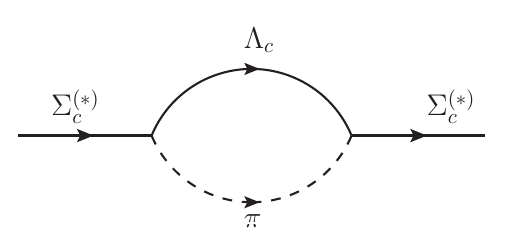}
  \caption{Loop diagrams of the self-energy for $\Sigma_c^{(*)}$.}
  \label{fig:selfenergy}
\end{figure*}

To describe the $\jp$ mass distribution, we work in terms of the production amplitudes rather than the scattering amplitudes. The LSE for the physical production amplitude of the $\alpha$th elastic channel $U_\alpha(E,p)^J$ reads
\begin{equation}
U^J_\alpha(E,p) {=} P^J_\alpha {-}\! \sum_\beta \!\! \int \!\! \frac{d^3\vec{q}}{(2\pi)^3}V^J_{\alpha\beta}(E,p,q)G_\beta(E,q) U^J_\beta(E,q),
\label{eq:lse}
\end{equation}
where $P_\alpha$ denotes the bare production vertices in Eq.~\eqref{eq:bareproduction}. 
Note that we only include the $S$-wave $\sigh$ production source in LSE, while the sources from the other channels are  neglected, as discussed above. The source corresponding to production of the $D$-wave $\sigh$ is also neglected due to the 
centrifugal barrier 
as discussed above. 
As a result of these simplifying assumptions,
the production amplitude, $U_i^{(\prime) J}$, for the $i$th $\jp$ ($\etacn$) inelastic channel is simply given by 
\bea
U^{(\prime) J}_i(E,k){=} {-}\! \sum_\beta \!\! \int \!\! \frac{d^3 \vec{q}}{(2\pi)^3}\mathcal{V}^{(\prime)J}_{i\beta}(k) G_\beta(E,q) U^J_\beta(E,q).
\label{eq:inelprod}
\end{eqnarray}
To render the integrals in the LSEs well defined one needs to introduce a regulator, with a cutoff larger than all typical three-momenta scales related to  coupled-channel dynamics. In what follows, we will discuss the results obtained for a hard cutoff $\Lambda$ set to 1~GeV for the pure contact potentials, namely scheme~I, and to 1.3~GeV for the case including OPE potentials, i.e., schemes II and III, as discussed below. However, we have checked that the final results barely depend on the cutoff value, when $\Lambda$ is varied from 0.7~GeV to 1.7~GeV for the contact scheme, and from 1.0 to 1.7 GeV for the schemes including OPE potentials, once the mentioned $S$-$D$ counter terms are
included (in the case of solution~$B$).

\section{Fit results}\label{sec:fit}

With the decay amplitude from Eq.~\eqref{eq:inelprod}, 
we are in a position to fit the
$\jp$ invariant mass distribution of the decay $\Lambda_b^0\to \jp K^-$ by LHCb~\cite{Aaij:2019vzc}.
The possible contributions from misidentified non-$\Lambda_b^0$ events, 
the $\Lambda^*$ resonances coupled to the $pK^-$, and possibly additional broad $P_c^+$ structures
are modelled by a smooth incoherent background ~\cite{Du:2019pij} 
\bea\label{eq:bkg}
f_\text{bgd}(E)= b_0+b_1E^2+b_2 E^4 +  \frac{g_r^2}{(m-E)^2+\Gamma^2/4},
\eea
where $b_0$, $b_1$, $b_2$, $g_r$, $m$ and $\Gamma$ are the parameters to be determined by fitting to the data. 
The other variations of the background used in the experimental 
analysis \cite{Aaij:2019vzc} are also considered, and the results are similar. In principle, the contributions 
from the $\Lambda^*$ resonances coupled to the $pK^-$ and the possible broad $P_c^+$ 
states living in the partial waves we are studying should be summed coherently at the 
amplitude level.\footnote{Note however that the preferred parity of the broad $P_c(4380)$ is opposite to that of the $P_c(4450)$ in Ref.~\cite{Aaij:2019vzc} and thus to all pentaquarks considered here.}
 However, the three $P_c$ states we are interested in are so narrow that, similar to Refs.~\cite{Aaij:2016phn,Aaij:2019vzc},
 our one-dimensional analysis is expected to be insensitive to  the possible broad structures. 
In addition, a direct sum of our signal amplitudes  and the coherent background would violate unitarity, unless the resonance structures in the background  emerge 
 in the channels included in the present framework. Since there is no evidence for such states so far,  we restrict ourselves to the incoherent background given in Eq.~\eqref{eq:bkg}.
 As for the $\Lambda^*$ contributions, it will be discussed below that the position of the poles in our analysis is  robust with respect to three different LHCb datasets used in the fits, including the fit with the kinematic constraint $m_{Kp}> 1. 9$~GeV which cuts more than 80\% 
of the $\Lambda^*$ hyperons.  The sensitivity of the poles to the  experimental input  will be included in the uncertainty estimate. The parameters of the background in various fits can be found 
in Appendix \ref{app:results}. 

 In this work, we consider three different fit schemes: 
\begin{itemize}
 \item scheme~I: the LO contact potentials for the $\sigh$ channels are implemented  including the couplings of the elastic channels to the $S$- and $D$-wave $\jp$ and $\etacn$ channels; 
 \item scheme~II: the OPE potentials supplemented by the NLO $S$-$D$ contact terms for the $\sigh$ channels are added to the potential of I;
 \item scheme~III: the explicit $\lamh$ channels are added to scheme~II, including
 the LO contact potentials, the OPE potentials and the NLO $S$-$D$ contact terms.
\end{itemize}
In this analysis, we do not consider isospin symmetry breaking effects, which can be important to the isospin-breaking decay modes~\cite{Burns:2015dwa,Guo:2019fdo}, but have little effect on the description of the line shapes in the isospin-conserving $\jp$ channel. The masses of particles used in the calculation are collected in Eq.~\eqref{eq:masses} in Appendix~\ref{app:results}. 
For all results shown a Gaussian convolution
of the theoretical distributions with the mass resolution of the experiment~\cite{Aaij:2019vzc} is employed.

Following the fit schemes introduced above, we improve our previous framework presented in Ref.~\cite{Du:2019pij},
where only the $\sigh$ and $\jp$ channels were considered explicitly and all other inelastic channels were parametrized as imaginary parts of the short ranged operators,
in several steps.

\subsection{Scheme I: Contact potentials}\label{subsec:contact}

\subsubsection{Including the $\etacn$ channel}  

The new $\etacn$ channel comes without additional unknown parameters,
since the transitions between the $\sigh$ and $\etacn$ channels are related to 
those between $\sigh$ and $\jp$ channels by  HQSS,
as discussed in Sec.~\ref{subsec:heavylight}. In addition, a relativistic form of 
the two-body propagator Eq.~\eqref{eq:twoprop}  and the energy-dependence 
of the width for the $\Sigma_c^{(*)}$, Eq.~\eqref{eq:self-energy},  are employed in this work. 
As the first step, we do not consider coupled-channel effects from the $\Lambda_c \bar{D}^{(*)}$ channels, i.e., for the fits discussed here we set 
both $C_\frac12^\prime$ and $C_\frac12^{\prime\prime}$ to zero.  From now on we further assume that the widths of the $\pc$ states are saturated by the
channels included --- in this subsection those are $\jp$, $\etacn$, and the elastic channels (including those via the decays of the $\Sigma_c^{(*)}$).
Therefore, no additional imaginary contact potentials as in Ref.~\cite{Du:2019pij} need to be introduced and unitarity is realized.

Three data sets of the $m_{\jp}$ distribution given in Ref.~\cite{Aaij:2019vzc} are used in fits. The $m_{\jp}$ distribution with $m_{Kp}>1.9$ GeV can effectively remove over $80\%$ of the $\Lambda^*$ contributions and the $\pc$ signals are enhanced in the $m_{\jp}$ distribution by applying $\cos\theta_{\pc}$-dependent weights to each event candidate, with $\theta_{\pc}$ the angle between the $K^-$ and the $J/\psi$ in the $\pc$ rest frame~\cite{Aaij:2019vzc}. In analogy to what was reported in Ref.~\cite{Du:2019pij}, also with our modified model,
two solutions, called $A$ and $B$,  
describing the data almost equally well
are found as shown in Fig.~\ref{fig:contactfit}. 
      
\begin{figure*}[tb]
 \centering
  \includegraphics[width=0.49\textwidth]{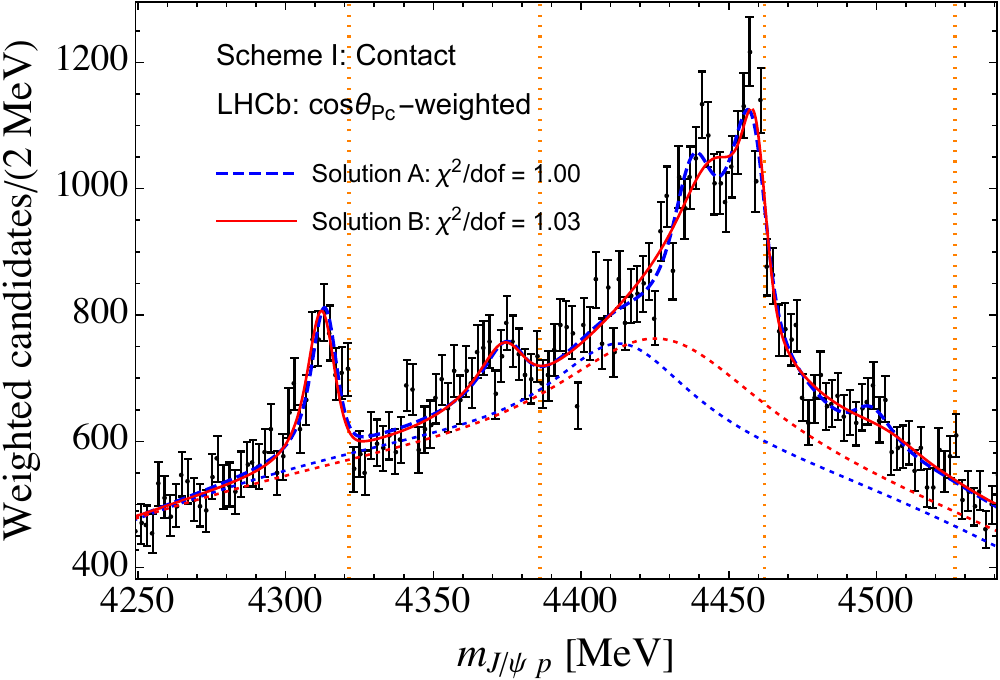}
  \includegraphics[width=0.49\textwidth]{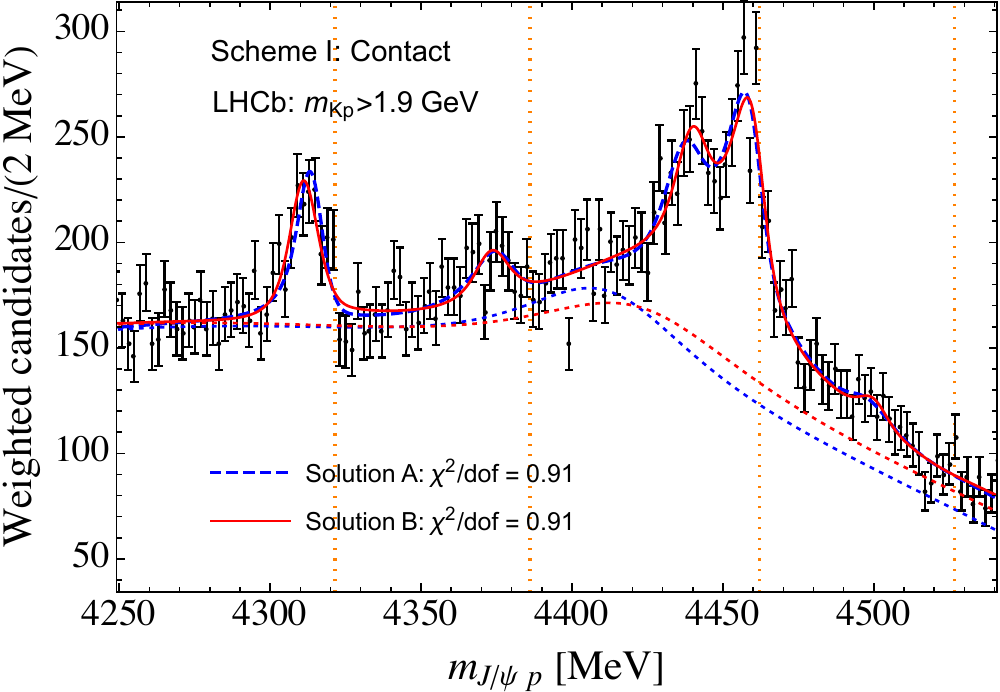}\\
  \includegraphics[width=0.49\textwidth]{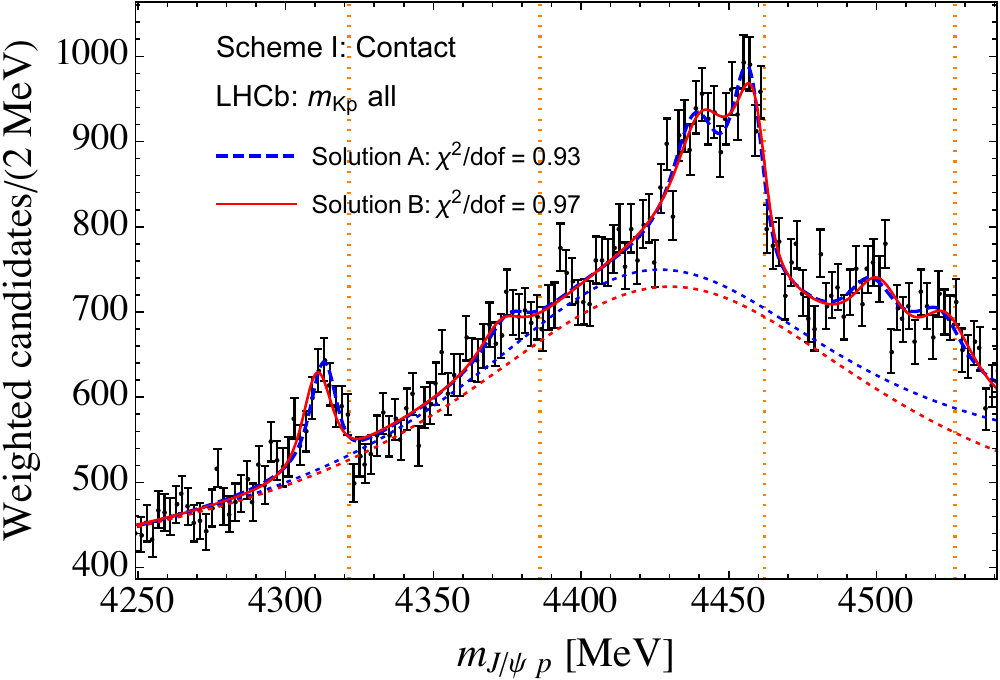}
  \caption{The best fits to the different experimental data~\cite{Aaij:2019vzc} for both solution $A$ (blue dashed curves) and solution $B$ (red solid curves) of scheme~I. The corresponding backgrounds are shown as red-dotted and blue-dotted curves, respectively. The vertical dashed lines from left to right are the $\Sigma_c\bar{D}$, $\Sigma_c^*\bar{D}$, $\Sigma_c\bar{D}^*$, and $\Sigma_c^*\bar{D}^*$ thresholds, respectively.}
  \label{fig:contactfit}
\end{figure*}

\begin{table*}[tb]
\caption{The names of the states, their dominant channels (DCs) and thresholds, the Riemann sheet (RS) where they lie, their quantum numbers found, the pole positions, and the dimensionless couplings to their DCs, $g_\text{DC}$, from the $T$-matrix residues for the solutions $A$ and $B$ of scheme~I are given. The uncertainties stated  result from taking the fit to different data and the different backgrounds used.}
{
\begin{tabular}{l |c|c| c c c | c c c}
\hline 
\hline
$\,$ & $\,$ & $\,$ & \, & $ \text{solution}$ & $A$  & \, & solution & $B$ \\
\hline
$\,$ & DC ([MeV]) & RS & $J^{P}$ & Pole {[}MeV{]} & $g_\text{DC}$ &  $J^{P}$& Pole {[}MeV{]}& $g_\text{DC}$ 
\tabularnewline
\hline 
$ \pc(4312) $ & $\Sigma_{c}\bar{D}\; (4321.6)$ & I & $\frac12^-$ & $4314(1)-4(1)i$ & $2.6(1)+0.4(2)i$ &  $\frac12^-$ & $4312(2)-4(2)i$ & $2.9(1)+0.4(2)i$
\tabularnewline
\hline
$\pc(4380)$ & $\Sigma_c^*\bar{D} \; (4386.2)$ & II & $\frac32^-$ & $4377(1)-7(1)i$ & $2.8(1)+0.1(1)i$ & $\frac32^-$ & $4375(2)-6(1)i$ & $3.0(1)+0.1(1)i$
\tabularnewline
\hline
$\pc(4440)$ & $\Sigma_c\bar{D}^* \; (4462.1)$ & III & $\frac12^-$ & $4440(1)-9(2)i$ & $3.7(2)+0.6(1)i$ & $\frac32^-$ & $4441(3)-5(2)i$ & $3.6(1)+0.3(1)i$
\tabularnewline
\hline
$\pc(4457)$ & $\Sigma_c\bar{D}^* \; (4462.1)$ & III & $\frac32^-$ & $4458(2)-3(1)i$ & $2.1(2)+0.3(1)i$ & $\frac12^-$ & $4462(4)-5(3)i$ & $2.0(2)+1.2(3)i$
\tabularnewline
\hline
$\pc$ & $\Sigma_c^*\bar{D}^* \; (4526.7)$ & IV & $\frac12^-$ & $4498(2)-9(3)i$ & $4.0(1)+0.4(2)i$ & $\frac12^-$ & $4526(3)-9(2)i$ & $1.5(2)+1.1(4)i$
\tabularnewline
\hline
$\pc$ & $\Sigma_c^*\bar{D}^* \; (4526.7)$ & IV & $\frac32^-$ & $4510(2)-14(3)i$ & $3.3(2)+0.6(2)i$ & $\frac32^-$ & $4521(2)-12(3)i$ & $2.5(2)+0.9(2)i$
\tabularnewline
\hline
$\pc$ & $\Sigma_c^*\bar{D}^* \; (4526.7)$ & IV & $\frac52^-$ & $4525(2)-9(3)i$ & $1.9(2)+0.6(7)i$ & $\frac52^-$ & $4501(3)-6(4)i$ & $3.9(2)+0.1(2)i$
\tabularnewline
\hline
\hline
\end{tabular}\label{tab:polecontact}
}
\end{table*}

To study the poles in an $n$-channel problem, a multi-sheet Riemann surface
in the complex energy plane is invoked. We only consider the RSs 
associated to the four elastic channels $\sigh$, since all inelastic thresholds 
are remote and their impact on the poles near the elastic thresholds is marginal.
Assuming that all four elastic channels can couple to each other, there are $2^4=16$ RSs in total labeled 
as ${ \rm RS_{\pm\pm\pm\pm}}$,
with the subscript composed of the signs of the imaginary parts of  the c.m. three-momentum in the $i$th channel, $q_{\text{cm},i}$, from the lowest in the energy to the 
highest.\footnote{ Even if  some elastic channels are uncoupled, we still use the generic four-channel notation 
 (as relevant in Schemes II and III) by assuming that the couplings to the absent channels vanish. For instance in scheme I, there are only 3, 3 and 1 elastic coupled-channels for $J=\frac12$, $J=\frac32$ and $J=\frac52$, respectively, see Table~\ref{tab:channels}.
}
However, in the special case when we are dealing only with quasi-bound states---unstable states that would  be bound states if  there were no low-lying thresholds which provide some possibility to decay---only 4 of 16 RSs are of interest because they can be reached from
the physical region by crossing the unitary branch cut from $E+i\epsilon$ to
$E-i\epsilon$ between the $(n-1)$th and the $n$th threshold, see Ref.~\cite{Badalian:1981xj} for a  review. 
 In order to simplify notation, we enumerate these RSs as follows:   RS I (the physical sheet) = ${ \rm RS_{++++}}$, 
RS II = ${ \rm RS_{-+++}}$, RS III = ${ \rm RS_{--++}}$ and RS IV = ${ \rm RS_{---+}}$.  
As a result, in the particular case of quasi-bound states that is of interest here, the poles located on these RSs have a much shorter path to the real energies on RS I and thus have more significant impact on
physical observables than those on the more remote sheets.
We note also that when we search for the poles of the $P_c$'s and calculate the corresponding residues, all inelastic channels  are assumed to be on the unphysical RS ($-$), since otherwise the poles would be remote,  see Ref.~\cite{Baru:2019xnh} for a  detailed discussion of this case and a more general classification of the poles in the conformally mapped $\omega$ plane.

The $N$th RS (RS-N) can be reached from the
physical amplitude $T$ by the analytical continuation in a matrix form,
\be
{(T_N)}^{-1}={(T)}^{-1}-2i\rho_N,
\ee
where $\rho_N$ is a diagonal matrix with the first $N-1$ elements equal to
the two-body phase space factors for the corresponding channels and the others zero,
i.e., $\rho_N = \text{diag}(\rho_1,\dots,\rho_{N-1},0,\dots)$ with $\rho_i=
m_{\Sigma_c^{(*)}} m_{D^{(*)}} q_{\text{cm},i}/(2\pi E)$.  
Moreover, one can define the product of the effective couplings $g_\alpha g_\beta$
 of a given state to channels $\alpha$ and $\beta$ 
from the residue of the scattering amplitude $T_{\alpha\beta}(E)$ 
at the pole of that state, i.e.,
\bea\label{eq:def:couplings}
g_\alpha g_\beta = \lim_{E\to E_\text{pole}} (E^2-E^2_\text{pole})T_{\alpha\beta}(E).
\eea

The two solutions produce different values of the parameters, which are given in Table~\ref{tab:paracontact} in Appendix~\ref{app:results},
in particular $C_\frac12$ and $C_\frac32$, and thus give different pole locations. 
Both solutions give seven poles in the $\sigh$ amplitudes, 
i.e., seven $\pc$ states: three with $J=1/2$, three with $J=3/2$ and one with $J=5/2$. 
The pole positions, the dominant channels (DCs) having the largest effective couplings 
and lying closer to the corresponding pole, and  the effective couplings of those for solutions $A$ and $B$ are listed in Table~\ref{tab:polecontact}. In both solutions, among the seven poles, the lowest one corresponds to the $\pc(4312)$ with $J^P=\frac12^-$. It appears as a $\Sigma_c\bar{D}$ bound state, located on the physical RS for the elastic channel and on the unphysical RS for the $\jp$ and $\etacn$ inelastic channels. It would become a true bound state,
if the $\jp$ and $\etacn$ channels were switched off. There are two $\Sigma_c\bar{D}^*$ bound states with quantum numbers $\frac12^-$ and $\frac32^-$, corresponding to the $\pc(4440)$ and the $\pc(4457)$, respectively, in solution $A$ and with the quantum numbers interchanged in solution $B$, as that in Ref.~\cite{Du:2019pij}. The interchange of the spin assignments for the $\pc(4440)$ and $\pc(4457)$ in the two solutions is a result of the proximity of the values of $C_\frac12$ ($C_\frac32$) in solution $A$ to those of $C_\frac32$ ($C_\frac12$) in solution $B$, see e.g. Table~\ref{tab:paracontact} in Appendix~\ref{app:results}. 
There are three $\pc$ states dominated by the $\Sigma_c^*\bar{D}^*$ channel with quantum numbers $\frac12^-$, $\frac32^-$, and $\frac52^-$. Analogous to that of the $\Sigma_c\bar{D}^*$ channel, the mass pattern of the three $\pc$ states is $m_{\frac12^-}<m_{\frac32^-}<m_{\frac52^-}$ for solution $A$ and the opposite, i.e., $m_{\frac12^-}>m_{\frac32^-}>m_{\frac52^-}$ for solution $B$. The narrow state $\pc(4380)$, required by HQSS, with $J^P=\frac32^-$ predicted in Refs.~\cite{Xiao:2019aya,Sakai:2019qph,Du:2019pij,Xiao:2020frg} is found around 4.38 GeV in both solutions.
As a consequence of HQSS,
it is as narrow as the other three $\pc$ states (see e.g. Table~\ref{tab:polecontact}) and  different from the broad $\pc(4380)$ reported by LHCb in 2015~\cite{Aaij:2015tga}. 
The effective couplings of the $P_c$ states to the nearby elastic channels are dominant though    the effects from the other elastic channels are not negligible and in some cases reach up to 30\% of the dominant one, 
see Tables~\ref{tab:coupling_I_A} and \ref{tab:coupling_I_B} in Appendix~\ref{app:couplings}.

\subsubsection{Predictions for the line shapes}
\label{sec:lineshapes_scheme_I}

With the parameters extracted from fits    to the $\jp$ channel and under the assumption that  production of the inelastic channels goes through the elastic ones, i.e., $\Lambda_b^0\to \sigh K^-$, one can predict line shapes for other channels. This is particularly interesting because the two solutions found here correspond to different heavy-light spin structure decompositions of the higher $P_c$ states. They are expected to have distinct consequences in their decays into channels other than the $\jp$ whose $s_Q$ is 1. 
Consequences of HQSS on the decays of $P_c$ states into the $\Lambda_c\bar D^{(*)}$ and $\eta_c p$ considering only $S$-waves have been reported in Refs.~\cite{Voloshin:2019aut,Sakai:2019qph}. The results below on these channels go beyond the previous results in giving line shapes, instead of ratios of partial widths, and including $D$-waves.

Among all possible decay channels other than the $\jp$ of the $P_c$ states, the $\sigh$ are of particular interest since their thresholds are close to the $\pc$ states and are sensitive to their nature~\cite{Guo:2014iya}. One can predict the line shapes 
of the $\sigh$ mass distributions from the decay $\Lambda_b \to \sigh K$. They are shown in Fig.~\ref{fig:pred:sigh_schemeI}. 
The existence of $S$-wave bound states below the thresholds leads to several  manifestations  in the line shapes.
First, one sees the threshold enhancements for the  line shapes of the $\sigh$ mass distributions. 
The threshold enhancement above the threshold of the $\Sigma_c\bar{D}$ is due to the $\pc(4312)$, and that of the $\Sigma_c^*\bar{D}$ is due to the narrow $\pc(4380)$. 
Second, the $P_c (4440)$  and  $P_c (4457)$  show prominent signals in the  $\Sigma_c \bar D$  and   $\Sigma_c^* \bar D$ invariant mass distributions only if they  couple to these channels in $S$ waves. 
Indeed, the $J=1/2$ $P_c (4440)$ ($P_c (4457)$) and  the $J=3/2$ $P_c (4457)$ ($P_c (4440)$ ) are clearly seen in  the $\Sigma_c \bar D$  and 
$\Sigma_c^* \bar D$ invariant mass distributions for solution $A$ ($B$), respectively.   
These observations can be used to single out the unique solution from the two once data in these channels become available.  Furthermore, since there are no   $J=3/2$ elastic channels below the $\Sigma_c^* \bar D$ threshold,  the $J=3/2$ $\Sigma_c^* \bar D$  molecular state $P_c (4380)$ can only manifest itself as a threshold enhancement in the $\Sigma_c^* \bar D$ channel but not as peaks in the other elastic channels.  The higher $P_c$ states show up as structures in all channels for both solutions though the strength of the peaks is quite uncertain and   depends on the data set used in fits.   This should not come as a surprise given that 
the  shape and the relative strength of the peaks  vary quite significantly in  three data sets by LHCb. Specifically, the data set ``$m_{Kp}$ all'' shows some hints for the existence of the higher resonances near 4.5 GeV while the other two data sets do not, see  Appendix \ref{app:couplings} for more details on the effective couplings to the source.

\begin{figure*}[tbh]
 \centering
  \includegraphics[width=1.\textwidth]{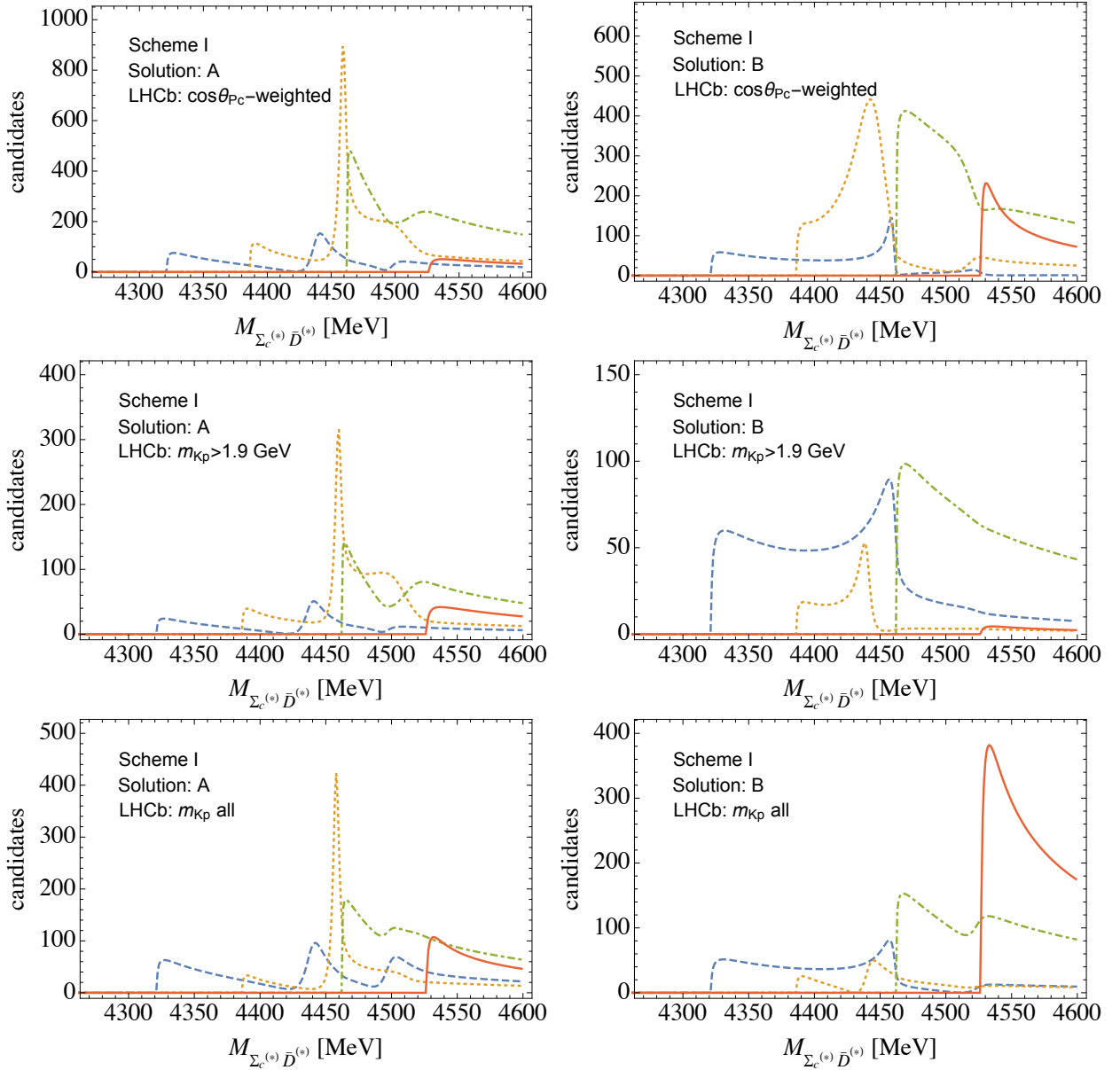}
\caption{Predictions for the line shapes of $\Sigma_c\bar{D}$ (blue dashed curves), $\Sigma_c^*\bar{D}$ (orange dotted curves), $\Sigma_c\bar{D}^*$ (green dot-dashed curves), and $\Sigma_c^*\bar{D}^*$ (red solid curves) mass distributions based on the fit results in  Fig.~\ref{fig:contactfit}   for  solutions $A$ and $B$ of scheme~I   without the $\Lambda_c \bar D^{(*)}$ channels. The qualitative differences for the higher states in various data sets are mainly caused by  the different shapes of the LHCb distributions in this energy range. 
Note also that  the background for this process  is not included since it is unknown.
}\label{fig:pred:sigh_schemeI}
\end{figure*}

With the parameters extracted from fits    to the $\jp$ channel and under the assumption that  production of the inelastic channels goes through the elastic ones, i.e., $\Lambda_b^0\to \sigh K^-\to \eta_c p K^-$, one can also predict line shapes for the $\eta_c p$  channels. 
In Fig.~\ref{fig:pred:etacp_schemeI},  we  show predictions for the $\eta_c p$ invariant mass distributions, which also allow one to make several important observations. 
First, this inelastic channel shows quite pronounced signals from all $P_c$ states  observed by LHCb and their spin partners predicted in our analysis.  Second,  
the centrifugal-barrier suppression of the $D$-wave coupling of $\jp$  to the $P_c$'s is no longer active because the $\jp$ threshold is much lower than the  relevant 
$\Sigma_c^{(*)}\bar D^{(*)}$ thresholds.  
Therefore, the coupling of the elastic channels to $\eta_c p$ in the $D$ wave, which is related to that for $\jp$ via HQSS,  is of a comparable strength or even larger than  the $S$-wave coupling, as shown in Table~\ref{tab:paracontact} in Appendix~\ref{app:results}. 
This observation can be used to account for the fact that, contrary to the elastic channels, the  $J=3/2$ $P_c (4457)$ for solution $A$ with its $D$-wave coupling to $\eta_c p$ is more pronounced in this line shape  than the $J=1/2$ $P_c (4440)$, though the latter can couple to $\eta_c p$ in  $S$-wave.
Similarly, the $J=3/2$ $P_c (4440)$ coupled to $\eta_c p$  in   $D$ wave dominates in the line shape for solution $B$. 
Thus, the shape of the $\Sigma_c^* \bar D$ and $\etacp$ spectra provides a straightforward criterion for distinguishing between the two solutions
of the simplified model including contact interactions
only, in the experiment. 

\begin{figure*}[tbh]
 \centering
  \includegraphics[width=1.\textwidth]{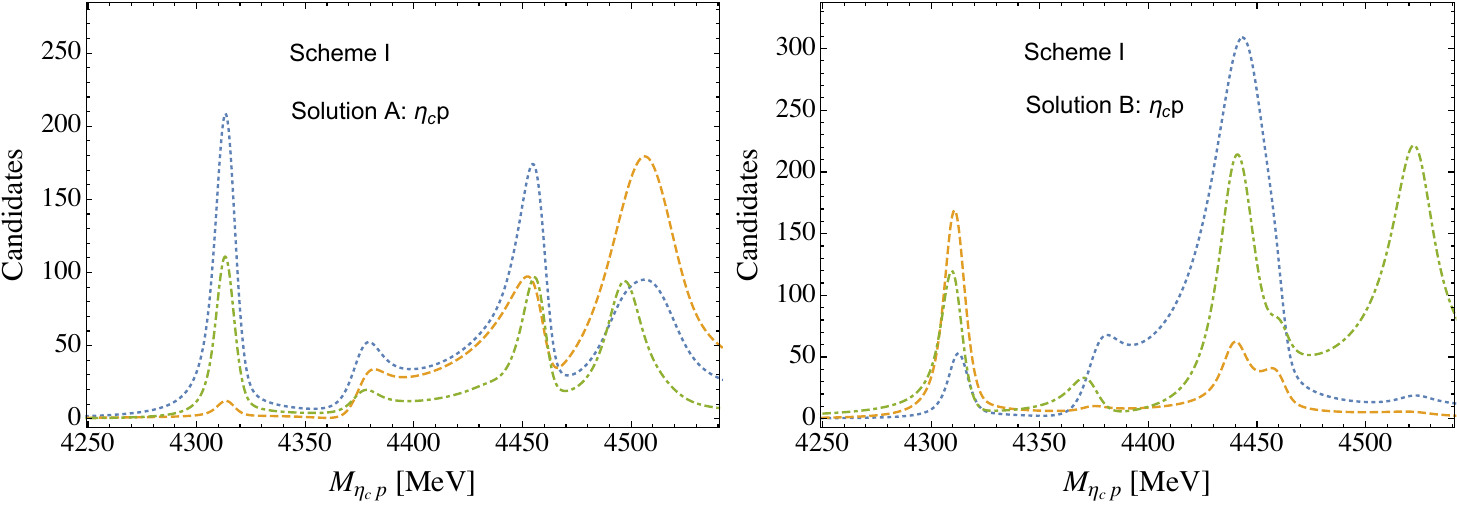}
\caption{ \label{fig:pred:etacp_schemeI}
Predictions for the line shapes of the $\etacn$ invariant mass distributions based on   the fit results in  Fig.~\ref{fig:contactfit} for  solutions $A$ and $B$ of scheme~I    without the $\Lambda_c \bar D^{(*)}$ channels. 
 The blue dotted, orange dashed and  green dot-dashed curves 
correspond to the predictions from the three fit results ($\cos\theta_{P_c}$-weighted, $m_{Kp}>1.9$ GeV and $m_{Kp}$-all) of Fig.~\ref{fig:contactfit}, in order. The contribution from background is not included.  }
\end{figure*}

\subsection{Scheme II: Including the OPE potential}\label{subsec:ope}

The importance of the OPE potential is well known for the nucleon interaction as its tensor force leads to 
the mixing between the $S$ and $D$ waves and can leave a significant impact on the line shapes. In this section, we investigate the role of the OPE  for the elastic 
channels without considering the $\lamh$ channels under the assumption that the widths of the $\pc$ states are saturated by the $\jp$, $\etacp$ and elastic 
channels.  We also note that the inclusion of the OPE potential does not involve any additional parameters, see Sec.~\ref{subsec:Lagrangian} for the discussion of the coupling constants used, however, it calls for 
the inclusion of the $S$-$D$ counter terms that
come with unknown strengths.

\subsubsection{How to renormalize the OPE and the role of $S$-$D$ transitions}\label{sec:SD}

\begin{figure*}[tbh]
 \centering
  \includegraphics[width=0.49\textwidth]{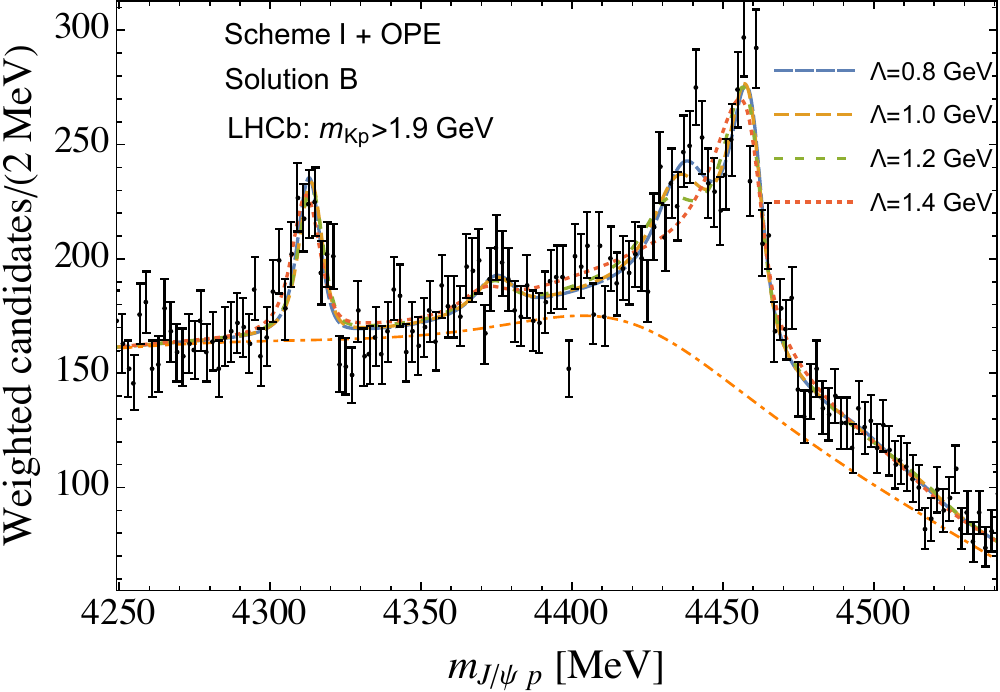}
  \includegraphics[width=0.49\textwidth]{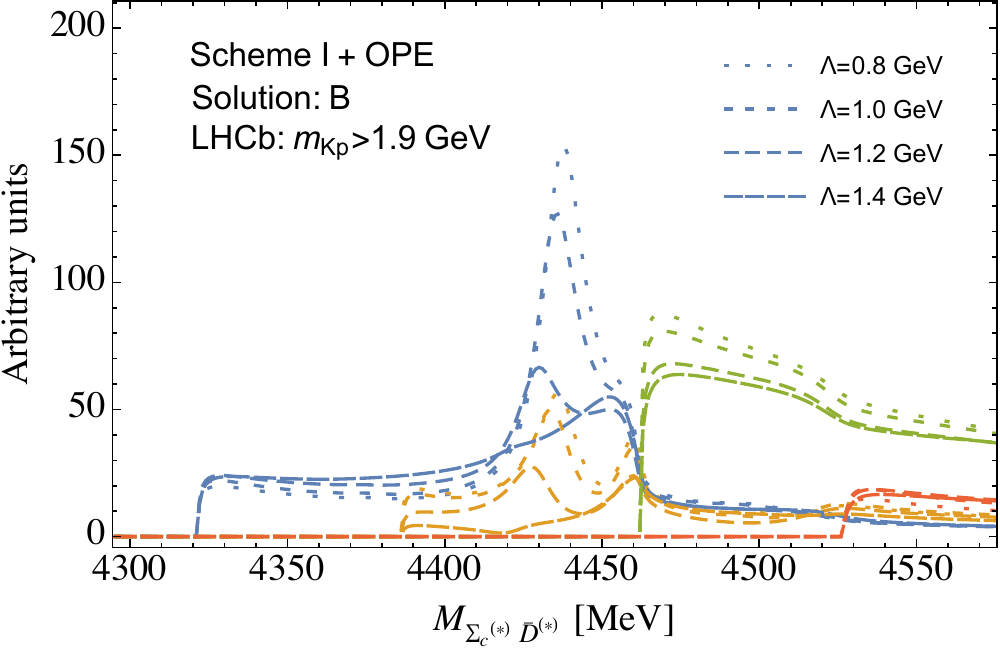}
  \caption{Left panel: Best fit results for different cutoffs $\Lambda$ using a potential of scheme~I plus OPE in the elastic channels.  Only solution $B$ exists in this case. 
   The background is shown as the orange dot-dashed curve. 
  Right panel: Predictions for the line shapes of the $\Sigma_c\bar D$ (blue), $\Sigma_c^*\bar D$ (orange), $\Sigma_c\bar{D}^*$ (green), and $\Sigma_c^*\bar{D}^*$ (red) mass distributions (background is not included). }
  \label{fig:cutoff-dep:OPE}
\end{figure*}

As noticed in Refs.~\cite{Baru:2017gwo,Baru:2016iwj},
because of the large mass of the heavy system and
the large splittings between the  thresholds treated dynamically,
the typical involved momenta are much larger than those
for  low-energy two--nucleon scattering,  e.g., in the  deuteron. In our case, the typical momenta $p_\text{typ}\sim\sqrt{2\mu\delta}$,
with $\delta$ and $\mu$ the largest threshold splitting and the reduced mass of the system, are about $670~\mathrm{MeV}$ when the energy range between the $\Sigma_c^{(*)} \bar D^{(*)}$ thresholds
is considered and can reach $900~\mathrm{MeV}$ when the energy is extended down
to the $\Lambda_c \bar D$ threshold.  These large typical momenta make the  contribution from 
the tensor force of the OPE
that leads to the 
$S$-$D$ transitions 
even more important than that in the $NN$ case, as it was already pointed out in similar studies of the $X(3872)$ and $Z_b$ systems~\cite{Baru:2017gwo,Baru:2016iwj}.
In previous studies, the iteration of such a potential  within   the integral equations was shown to yield 
a strong   regulator dependence for the observable quantities in the $b$-quark sector~\cite{Wang:2018jlv,Baru:2019xnh}. 
This is also what we observe in the current study, as can be seen from
the left panel in Fig.~\ref{fig:cutoff-dep:OPE}.  In line with Ref.~\cite{Du:2019pij}, 
once the OPE is included, only  solution  $B$ survives when the  cutoff $\Lambda$ in the LS equations is varied from 0.8 to 1.1 GeV.  Although the $\chi^2/\text{dof}$ changes barely with the cutoff variation, a closer look into the line shapes in Fig.~\ref{fig:cutoff-dep:OPE} shows visible deviations in the results: indeed, the peak corresponding to the $P_c(4440)$ disappears in the results with increasing the cutoff.  Moreover,  the difference in the predicted line shapes --  see the right panel in  Fig.~\ref{fig:cutoff-dep:OPE} -- becomes even more pronounced, in line with the observations of Ref.~\cite{Baru:2019xnh}.  In particular, the disappearance of the  $P_c(4440)$  in  the $\jp$ spectrum  for $\Lambda =1.3$ GeV  leads to the same effect in the $\Sigma_c \bar D$ and $\Sigma_c^* \bar D$ line shapes  in  Fig.~\ref{fig:cutoff-dep:OPE}. 
   
\begin{figure*}[tbh]
 \centering
  \includegraphics[width=0.9\textwidth]{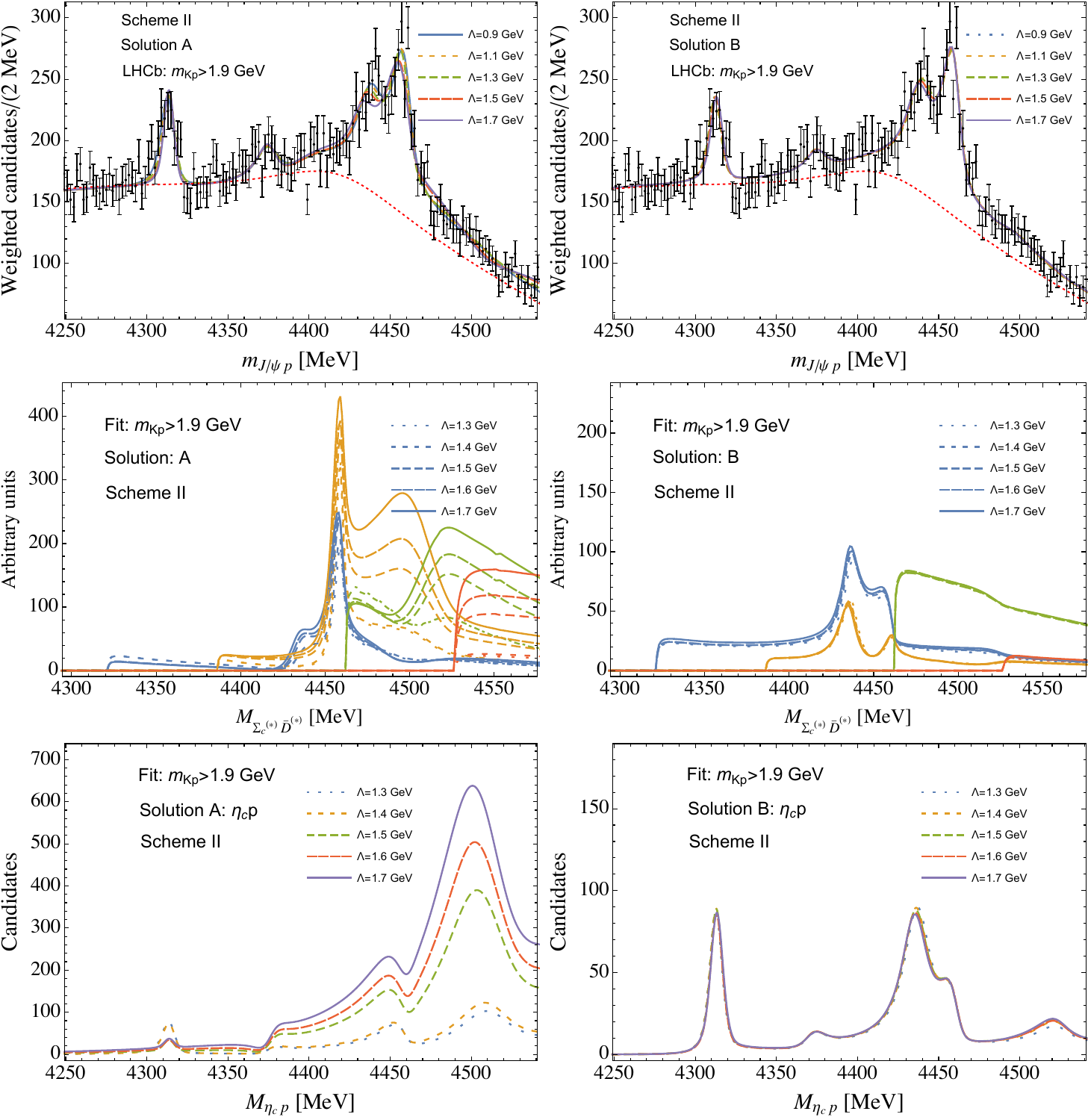}
  \caption{ First row:  Best fit results for solutions $A$ (left) and $B$ (right) for different cutoffs $\Lambda$ using a potential of scheme~II.  The background is shown as red-dotted curve. 
  Second row: Cutoff dependence of the predicted line shapes of $\Sigma_c\bar D$ (blue), $\Sigma_c^*\bar D$ (orange), $\Sigma_c\bar{D}^*$ (green), and $\Sigma_c^*\bar{D}^*$ (red) mass distributions.
  Third row: Cutoff dependence of the predicted line shapes  for the $\etacn$ mass distributions.  No background is included for predicted line shapes.}
  \label{fig:cutoff-dep:scheme_II}
\end{figure*}

In order to cure the cutoff dependence, caused by the short-range part of the $S$-$D$ OPE transitions, it was suggested to  promote the formally NLO $O(Q^2)$ $S$-$D$ contact terms\footnote{$Q$ denotes the soft scale for the given system.} to LO~\cite{Wang:2018jlv,Baru:2019xnh}. 
Here, we follow  the same procedure, namely we include  the $S$-$D$ contact terms in the elastic channels ($D_b^{SD}$ terms in Eq.~\eqref{eq:nloSD12}) together with the OPE. In what follows, this formulation will be referred to as scheme~II.   As expected, the line shapes obtained for  solution $B$  within scheme~II, as   shown in the right panel of Fig.~\ref{fig:cutoff-dep:scheme_II}, demonstrate only a milder regulator dependence, especially for  cutoffs $\Lambda > 1$ GeV, which provide a larger separation 
between the soft and hard scales.
  More importantly,  the line shapes  predicted in the elastic and inelastic channels also exhibit a very mild regulator dependence (see  the second and third rows in the right panel in Fig.~\ref{fig:cutoff-dep:scheme_II}). 
We expect that the residual cutoff dependence  can be further reduced if  the $S$-wave momentum dependent $O(Q^2)$ contact terms  are added,  in line with a related study of Ref.~\cite{Baru:2019xnh}. Since these contact terms obviously cannot be fixed on the basis of current data, further studies of the regulator dependence
within scheme~II will be postponed to future work when new data become available.

 In addition to solution $B$,  in scheme II   there is in principle also the best fit corresponding 
  to  solution $A$ with  the comparable  $\chi^2/{\rm dof}$  at least for smaller  cutoffs.  The corresponding results are shown in the left panel   in Fig.~\ref{fig:cutoff-dep:scheme_II}.
 However,  a closer look into the $J/\psi p$ line shape in Fig.~\ref{fig:cutoff-dep:scheme_II} shows a clear cutoff dependence,  making the peak from  $\pc(4440)$ hardly visible for  cutoffs greater than 1.5 GeV.  This observation finds  quantitative support in the values of the $\chi^2/{\rm dof}$, which grow with the cutoff and quickly become larger  than 
the those for solution $B$. For example, for the cutoff $\Lambda =1.3$ GeV the $\chi^2/\text{dof}$ values  are 1.01 and 0.90 for solutions $A$ and $B$, respectively.
This pattern is actually not surprising:  given that solution $A$ does not exist  
as long as  the OPE is included but  the $S$-$D$ contact terms are switched off,  the possibility to have a solution $A$ in scheme II 
could be  achieved only with the unnaturally large  $S$-$D$ transitions generated by the contact interactions in an attempt to improve the description of the data.  Accordingly,  in such an unnatural scenario, which would violate the power counting,  the cutoff dependence can not be properly absorbed into  redefinitions 
of the contact terms and reveals itself in the line shapes, especially for the predicted  mass distributions in the $\sigh$ and $\etacp$ channels,  as seen from Fig.~\ref{fig:cutoff-dep:scheme_II}.  Therefore, we discard this scenario and in what follows focus on the results for solution $B$.

Since the focus of the  discussion above in Sec.~\ref{subsec:ope} was put on the  renormalization of the OPE,  the results presented  in Figs.~\ref{fig:cutoff-dep:OPE} and 
\ref{fig:cutoff-dep:scheme_II} were obtained using some fixed background.  
It is also worth mentioning that while the separation of scales calls for  larger cutoffs, the  cutoff larger than the $c$-quark mass,  $m_c \simeq 1.5$ GeV, may introduce additional HQSS breaking effects.

\subsubsection{Description of the data in $\Lambda_b^0\to J/\psi p K^-$ }\label{sec:II:Jpsip}
 
\begin{figure*}[tbh]
 \centering
  \includegraphics[width=0.49\textwidth]{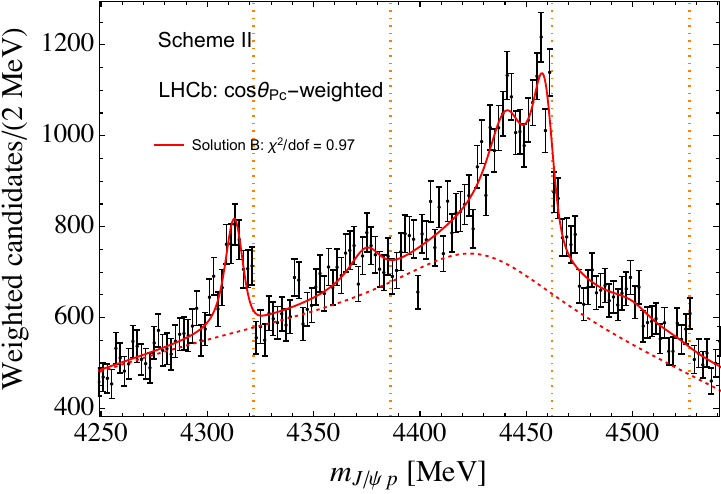}
  \includegraphics[width=0.49\textwidth]{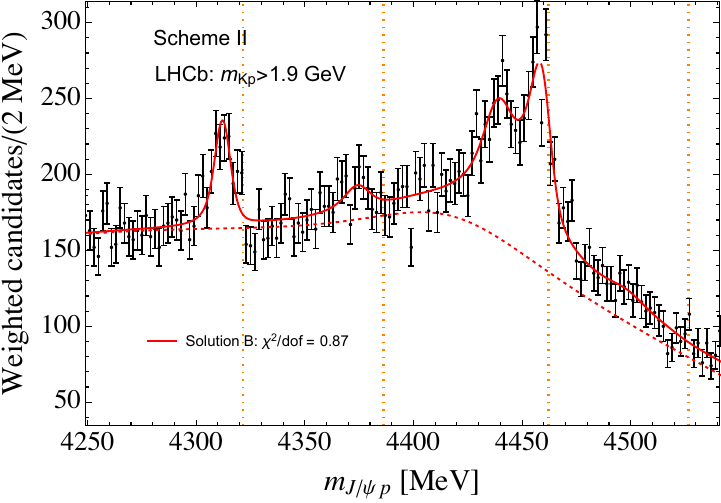}\\
  \includegraphics[width=0.49\textwidth]{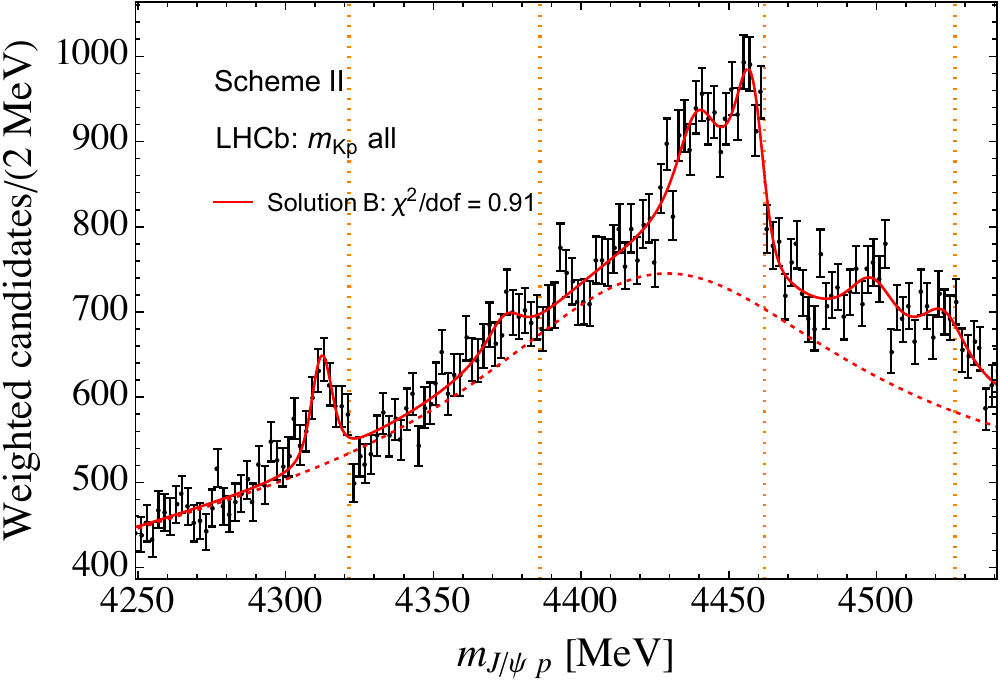}
  \caption{The best fits to the different experimental data~\cite{Aaij:2019vzc} for solution $B$ (red solid curves) of scheme~II. The corresponding backgrounds are shown as red-dotted  curves. The vertical dashed lines from left to right are the $\Sigma_c\bar{D}$, $\Sigma_c^*\bar{D}$, $\Sigma_c\bar{D}^*$, and $\Sigma_c^*\bar{D}^*$ thresholds, respectively.}
  \label{fig:fits:scheme_II}
\end{figure*}

In Ref.~\cite{Du:2019pij}, it was found  that the inclusion of the OPE singles out a unique solution (the solution $B$)
from the two almost equivalent ones present for the pure contact potentials.  However, no effects from the $S$-$D$ contact terms were taken into account which, as shown 
in the previous Section, does not meet the criteria of renormalizability. 
Here we improve on this  and discuss the renormalized results as well as their implications  
for the predicted line shapes, as will be shown in the next Section.  In Fig.~\ref{fig:fits:scheme_II},  we  present the best fit results for solution $B$ of scheme II for various data sets.

A general feature of both schemes I and II is that the source couplings $P_\alpha^J$ are badly determined, especially for those of the three $\Sigma_c^*\bar{D}^*$ bound states, which are almost invisible in the line shapes, as can be seen from Tables~\ref{tab:paracontact} and 
\ref{tab:II:paras} in Appendix~\ref{app:results}.  Also, the couplings to the inelastic channels  have sizeable uncertainties.   However, the elastic channel interactions, i.e., $C_\frac12$ and $C_\frac32$, are well determined and thus the pole positions and their effective couplings to the dominant channels are quite stable for   different fits, as shown in Table~\ref{table:II:pole}. The effective couplings of the $P_c$ resonances to all elastic channels are collected in Tables~\ref{tab:coupling_II_12}-\ref{tab:coupling_II_52} in Appendix~\ref{app:couplings}.
The two solutions lead to poles which are similar to those in scheme~I (see also Refs.~\cite{Valderrama:2019chc,Liu:2019zvb}). The results are insensitive to the form of the background. Similar to scheme~I, the $\pc(4312)$ couples dominantly to the $\Sigma_c\bar{D}$ with $J^P=\frac12^-$, and the $\pc(4440)$ and $\pc(4457)$ couple dominantly to the $\Sigma_c\bar{D}^*$ with quantum numbers $\frac32^-$ and $\frac12^-$, respectively, in solution $B$. 
They are all (quasi-)bound state poles and should be understood as hadronic molecules of the corresponding dominant channels~\cite{Guo:2017jvc}. In both schemes, there is a narrow $\pc(4380)$ with $J^P=\frac32^-$ located at around $4.38$~GeV as a $\Sigma_c^*\bar{D}$ bound state. Its width is smaller than that obtained using a constant width of $\Sigma_c^{(*)}$ in Ref.~\cite{Du:2019pij} by about a factor of 2 ( see, e.g., Table I in Ref.~\cite{Du:2019pij} for  solution $B$). The reduction of the width obtained within our dynamical framework including a three body cut is consistent with the findings of  Ref.~\cite{Baru:2011rs}, where the same effect was found for the partial width $X(3872)\to D\bar D\pi$.

\begin{table*}[tb]
\caption{\label{table:II:pole}
The names of the states, their dominant channels (DCs) and thresholds, the Riemann sheet (RS) where they lie, their quantum numbers found, the pole positions, and the dimensionless couplings to their DCs, $g_\text{DC}$, from the $T$-matrix residues for the solution $B$ of scheme~II are given. The uncertainties stated result from taking the fit to different data and from the different backgrounds used. The statistical uncertainties from the fit for a given background are negligible. }
\begin{ruledtabular}
{
\begin{tabular}{ l | c | c | c  c  c}
$\,$ & DC ([MeV]) & RS &  $J^{P}$& Pole {[}MeV{]}& $g_\text{DC}$ 
\tabularnewline
\hline 
$ \pc(4312) $ & $\Sigma_{c}\bar{D}\; (4321.6)$ & I & $\frac12^-$ & $4313(1)-3(1)i$ & $2.7(2)+0.2(1)i$
\tabularnewline
\hline
$\pc(4380)$ & $\Sigma_c^*\bar{D} \; (4386.2)$ & II &  $\frac32^-$ & $4376(1)-6(2)i$ & $2.8(2)+0.0(1)i$
\tabularnewline
\hline
$\pc(4440)$ & $\Sigma_c\bar{D}^* \; (4462.1)$ & III &  $\frac32^-$ & $4441(2)-6(2)i$ & $3.6(2)+0.5(2)i$
\tabularnewline
\hline
$\pc(4457)$ & $\Sigma_c\bar{D}^* \; (4462.1)$ & III &  $\frac12^-$ & $4461(2)-5(2)i$ & $1.9(4)+1.2(3)i$
\tabularnewline
\hline
$\pc$ & $\Sigma_c^*\bar{D}^* \; (4526.7)$ & IV & $\frac12^-$ & $4525(4)-9(1)i$ & $1.4(3)+0.8(5)i$
\tabularnewline
\hline
$\pc$ & $\Sigma_c^*\bar{D}^* \; (4526.7)$ & IV & $\frac32^-$ & $4520(3)-12(3)i$ & $2.4(2)+1.0(4)i$
\tabularnewline
\hline
$\pc$ & $\Sigma_c^*\bar{D}^* \; (4526.7)$ & IV &  $\frac52^-$ & $4500(2)-9(6)i$ & $4.0(5)+0.6(2)i$
\tabularnewline
%\hline
%\hline
\end{tabular}
}
\end{ruledtabular}
\end{table*}

\subsubsection{Predictions for the line shapes in the elastic and inelastic channels }\label{sec:II:predict}
 
Our predictions for the elastic line shapes within scheme~II are given  in Fig.~\ref{fig:II:pred:sigh}.  In order to see the effect from the renormalized OPE, these results 
should be compared to the analogous predictions for solution B of scheme~I in the contact potential framework  given in Fig.~\ref{fig:pred:sigh_schemeI}.

As a general pattern,  the shape of  the invariant mass distributions for most of the elastic channels is consistent   for   schemes I and II, though  quantitative differences  are visible.  
The exceptions are  the line shapes in the $\Sigma_c\bar{D}$  channel,  which  exhibit qualitatively different behavior near the state  $\pc(4440)$.  
The  $\pc(4440)$ with $J=3/2$ can couple to the $\Sigma_c\bar{D}$ channel only in $D$ wave and, given that only $S$-wave interactions are included in scheme~I, this state does not reveal  itself   
in the  $\Sigma_c\bar{D}$ line shape in  this scheme. However, the $\pc(4440)$ is clearly seen in the $\Sigma_c\bar{D}$ spectrum in scheme~II.  This can be accounted for by   the presence of the remaining $D$-wave contribution, which does not vanish after the OPE has been renormalized  and thus supports the coupled-channel transitions between the dominant ($S$-wave) channel $\Sigma_c\bar{D}^*$ and the $\Sigma_c\bar{D}$.\footnote{Note that the actual strength of the $D$-wave peaks depends on the 
coupling constants $g_1$ and $g_2$ in Eq.~\eqref{lag:ope}. While $g_1$ is extracted accurately from  the experimental width of the $D^{*+}\to D^0\pi^+$,   
$g_2$ was determined  in a lattice QCD calculation \cite{Detmold:2012ge}  and 
is  subject for an  about 25\% uncertainty  not included in Fig.~\ref{fig:II:pred:sigh}. }
Similarly, the  $D$-wave structure generated by the  $J=1/2$  $\pc(4457)$ state is seen in the $\Sigma_c^*\bar{D}$ line shape in scheme~II, though it  is not so pronounced as in the previous case.  Clearly, it is not present in scheme~I.  This shows that the $D$-wave interactions in the OPE may have nontrivial consequences for the elastic line shapes. 
Nevertheless,  all  enhancements for the line shapes generated by the  $S$-wave thresholds 
show up  consistently  in both Schemes. Moreover,  we would like to emphasize that,  as follows from Table~\ref{tab:coupling_II_32},
 by far the dominant contributions to the residues of the $\pc(4440)$ and  $\pc(4457)$ states originate from  the interactions in the nearby $S$-wave $\Sigma_c\bar{D}^*$ channel while the $D$-wave contributions to their residues are strongly suppressed.  This picture is very natural for the molecular scenario and also finds support in a  very strong  enhancement of the  $\Sigma_c\bar{D}^*$ line shape near its threshold in spite of the very limited phase space.

\begin{figure*}[tbh]
 \centering
  \includegraphics[width=0.49\textwidth]{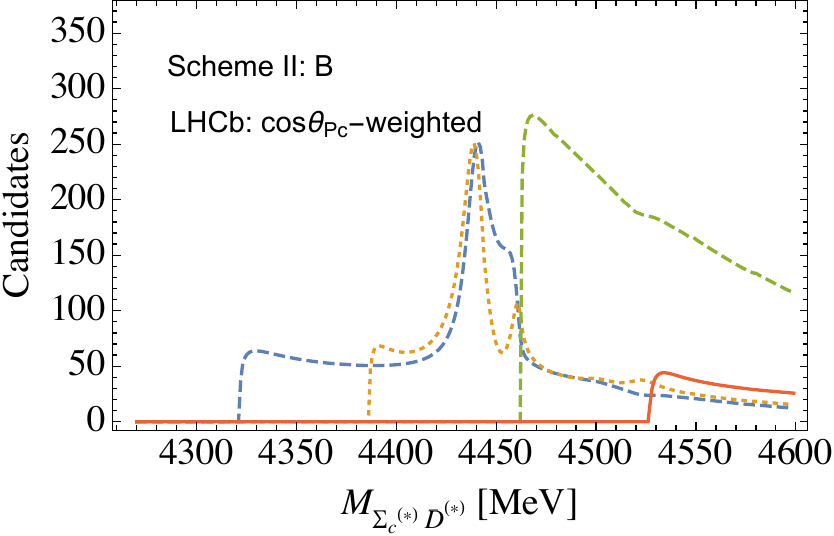}
  \includegraphics[width=0.49\textwidth]{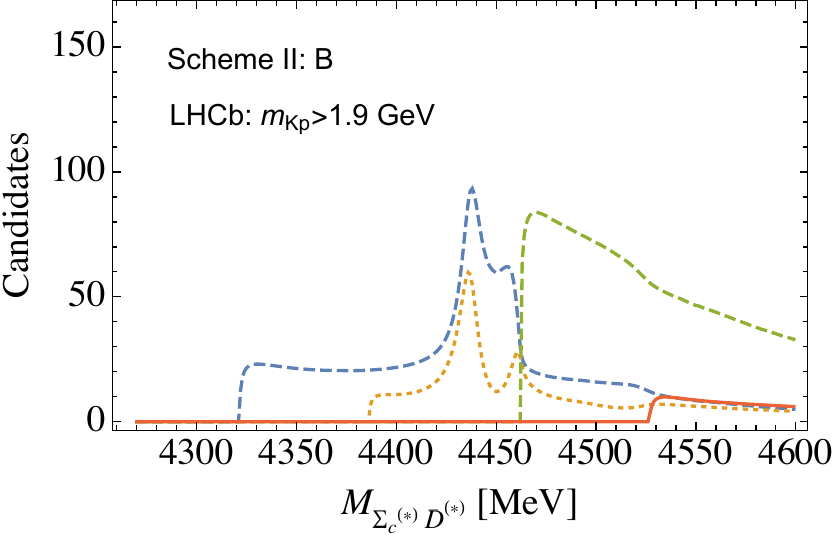}\\
  \includegraphics[width=0.49\textwidth]{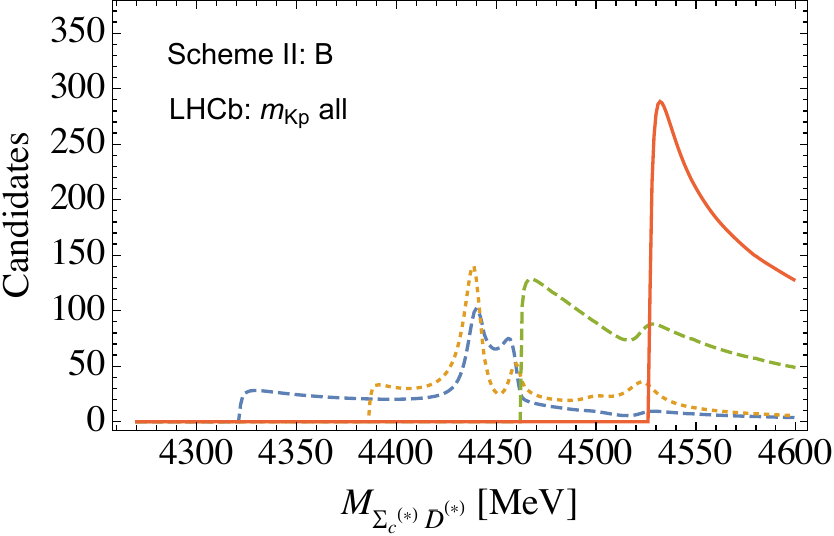}
  \caption{Predictions for the line shapes of $\Sigma_c\bar{D}$ (blue dashed curves), $\Sigma_c^*\bar{D}$ (orange dotted curves), $\Sigma_c\bar{D}^*$ (green dot-dashed curves), and $\Sigma_c^*\bar{D}^*$ (red solid curves) mass distributions based on the fit results in  Fig.~\ref{fig:fits:scheme_II} for  solution $B$ of scheme~II.
See caption to Fig.~\ref{fig:pred:sigh_schemeI}  for further details. 
}
  \label{fig:II:pred:sigh}
\end{figure*}

Predictions for the ${\etacn}$ invariant mass distribution within scheme~II are shown in Fig.~\ref{fig:II:pred:etacp}.  
Although the $D$-wave contributions from the OPE could bring about a visible impact on the elastic line shapes in the $\sigh$ channels, their contributions to the $\etacp$ are minor. Indeed, the $D$-wave  contributions to the residues are strongly suppressed and the phase-space suppression, which was important for the elastic channels, does not apply any more. 
Accordingly,  the $\jp$ and $\etacp$ spectra are related to each other by means of HQSS.
Thus,  the line shapes with and without the OPE for solution $B$ are qualitatively consistent with each other, though quantitative changes in the magnitudes of the peaks are visible.  
For this channel it appears promising, that the first data for the reaction $\Lambda_b\to \eta_c p K^-$
are already available~\cite{Aaij:2020mlx},
although the quality of the spectra is not yet sufficient for the kind of analysis indicated above.

\begin{figure*}[tb]
 \centering
  \includegraphics[width=0.5\textwidth]{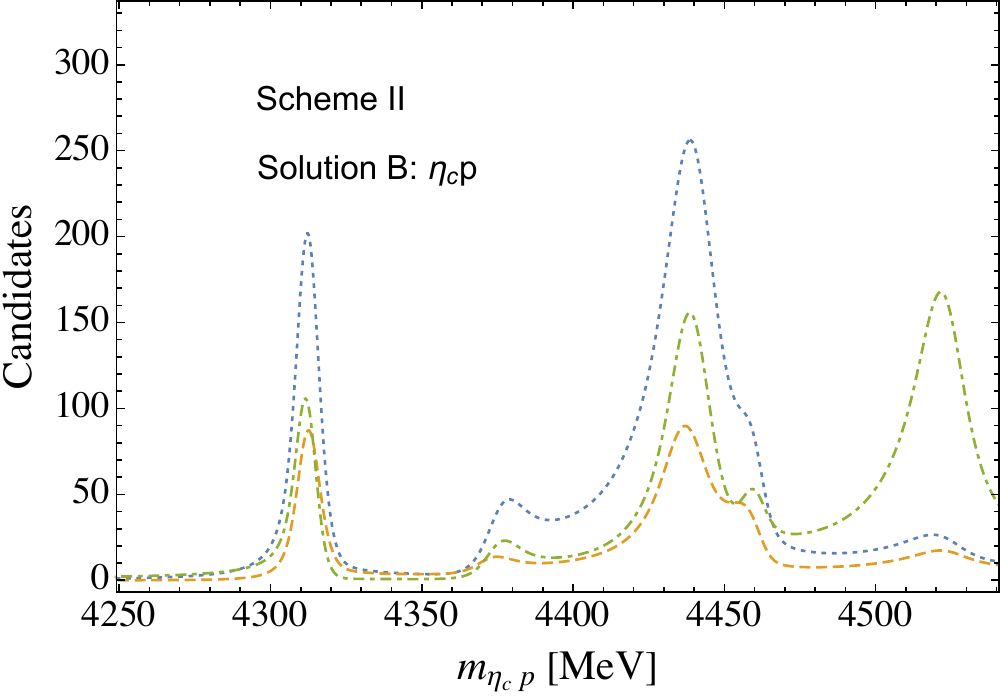}
  \caption{ \label{fig:II:pred:etacp} 
 Predictions for the line shapes of the $\etacn$ invariant mass distributions based on  various fit results for  solution $B$ of scheme~II. The blue dotted, orange dashed and  green dot-dashed curves 
correspond to the predictions from the three fit results ($\cos\theta_{P_c}$-weighted, $m_{Kp}>1.9$ GeV and $m_{Kp}$-all) of  Fig.~\ref{fig:fits:scheme_II}, in order. The contribution from background 
is not considered.}
\end{figure*}

\subsection{Scheme~III: Including the $\lamh$ channels}\label{subsec:lamh}

As discussed in schemes I and II, the $\jp$ invariant mass distribution can be well described without considering the 
coupled-channel effects from the $\lamh$ channels and thus the widths of the $\pc$ states can be accounted for 
by the channels $\jp$, $\etacn$ and the elastic channels.
However, the lack of experimental information at present does not mean that  the $\lamh$ channels can be neglected. 
Indeed, the proximity of the $\Lambda_c \bar{D}^{*}$ channel to the elastic thresholds, where the $P_c$ states reside, implies that 
the $\lamh$ channels potentially can be even more important than the distant  channels composed of a charmonium and a pion. 
Unlike other inelastic channels, the $\lamh$ channels can couple to the elastic channels via the OPE in addition to the short-range operators, which provides further support for their potentially 
important role.   
In addition, some model calculations without the unitarity constraint concluded large branching fractions of the $P_c$ states decaying into the $\Lambda_c\bar D^*$ channel~\cite{Lin:2017mtz,Lin:2019qiv}.
As discussed above, as the $\Sigma_c^{(*)}$ states dominantly decay
to $\Lambda_c\pi$, the simultaneous inclusion of the $\Lambda_c\bar{D}^{(*)}$ channels
and the OPE potential will lead to $\Lambda_c\bar{D}^{(*)}\pi$ three-body cuts\footnote{
Strictly speaking,  $\Lambda_c\bar{D}^{*}\pi$  is  a four-body state,  since $D^*$ can decay to $D \pi$.  However,  neglecting this width which is tiny, i.e., under the assumption that the $D^*$ is a stable particle, that we use here,  $\Lambda_c\bar{D}^{*}\pi$  corresponds effectively to a three-body channel.  }, 
when the energy goes above the corresponding three-body threshold. Therefore,  once the $\lamh$ channels are included one is also in a position to investigate the effects from the associated $\lamh \pi$ three-body cuts on the line shapes and the   $P_c$ states.  

Considering these points as a motivation, we developed  a coupled-channel framework that incorporates the  $\lamh$ channels explicitly including the OPE.  
While the currently available data do not allow us to adjust the parameters in the  $\Lambda_c \bar{D}^{(*)}$ potentials reliably, this framework, being the most advanced and 
well defined  
formulation of the relevant coupled-channel problem on the market,  should be very useful in the future in revealing  the role of   $\Lambda_c \bar{D}^{(*)}$  in the formation of the $P_c$ states and their decays.  
It should be mentioned that preliminary LHCb data on the $\Lambda_b^0\to \Lambda_c^+\bar D^0 K^-$ are already available, but the statistics is not high enough 
to see clear $P_c$ signals~\cite{Piucci:2019vsk}.

\subsubsection{Parameters in the fits}\label{sec:params}

As discussed in scheme II,
because of the large mass of the heavy system and
the large splittings between the dynamic thresholds,
the typical involved momenta $p_\text{typ}$ can reach
up to $900~\mathrm{MeV}$ when the energy is extended down
to the $\Lambda_c \bar D$ threshold.  These large typical momenta make the  contribution from 
the tensor force of the OPE
that leads to the $S$-$D$ transitions even more important than that in scheme~II. 
As a result, to obtain regulator-independent line shapes, 
promoting the formally NLO $O(Q^2)$ $S$-$D$ contact terms (see $D_b^{SD}$ and $D_c^{SD}$ in Eq.~\eqref{eq:nloSD12}) to LO is mandatory in line with the procedure discussed in 
Sec.~\ref{sec:SD}. 
 
From the  currently available data it is impossible to determine the strength of the $\Lambda_c \bar{D}^{(*)}\to \Lambda_c \bar{D}^{(*)}$ interactions, i.e., $C_\frac12^{\prime\prime}$.
Indeed, since the effect of  $\lamh$ on the elastic channels appears as a combination of $C_\frac12^\prime$ and $C_\frac12^{\prime\prime}$,   
a stronger (weaker) $\lamh$ interaction can be always compensated by a weaker (stronger) transition $\lamh\to \sigh$.
In addition,  because there is no isospin-conserving $\Lambda_c\Lambda_c\pi$ vertex,    the long-range OPE potentials for the $\lamh\to\lamh$ transitions vanish (see  Appendix~\ref{app:OPE}), 
and thus only short-ranged direct interactions  could be possible.  
Therefore, in what follows, we   use these arguments as a motivation to neglect direct interactions in  the $\lamh$ channels,  as it was also done for other inelastic channels. 
This approximation can be relaxed in the future easily, once new data, e.g. in the $\lamh$ channels become available. 
In the current study, the effect of $\lamh$ channels is therefore included through their coupling to the elastic channels.  The inclusion of the OPE potential will not introduce additional parameters
as the coupling constants are either fixed from experimental data or from  lattice QCD~\cite{Detmold:2012ge}.
In this case, under the assumption that the $\jp$, $\etacn$, $\lamh$ and $\sigh$ saturate the widths of the $\pc$ states, 
we have 7 parameters to describe the final state interactions:
\begin{itemize}
 \item $C_{\frac 12}$ and $C_{\frac 32}$ in Eq.\eqref{eq:contactpotential1} for the contact potentials among the elastic $\sigh$ channels;
 \item $C_{\frac 12}^\prime$ in Eq.~\eqref{eq:contactpotential3} for the transition between the elastic $\sigh$ channels and the inelastic $\lamh$ ones;
 \item $g_S$ and $g_D$ in Eq.~\eqref{eq:nloSD12} for the $S$- and $D$-wave inelastic $\jp$ and $\etacn$ channels;
 \item $D_b^{SD}$ and $D_c^{SD}$ in Eq.~\eqref{eq:nloSD12}, the NLO $S$-$D$ contact terms, for the transitions among the $\sigh$ and $\lamh$ channels. 
\end{itemize}
In addition, there are the  parameters that parametrize the background, Eq.~\eqref{eq:bkg}, 
and those of the bare production amplitudes,  Eq.~\eqref{eq:bareproduction}.

\subsubsection{Description of the data in $\Lambda_b^0\to J/\psi p K^-$ }\label{sec:Jpsip}
 
\begin{figure*}[tb]
 \centering
  \includegraphics[width=0.49\textwidth]{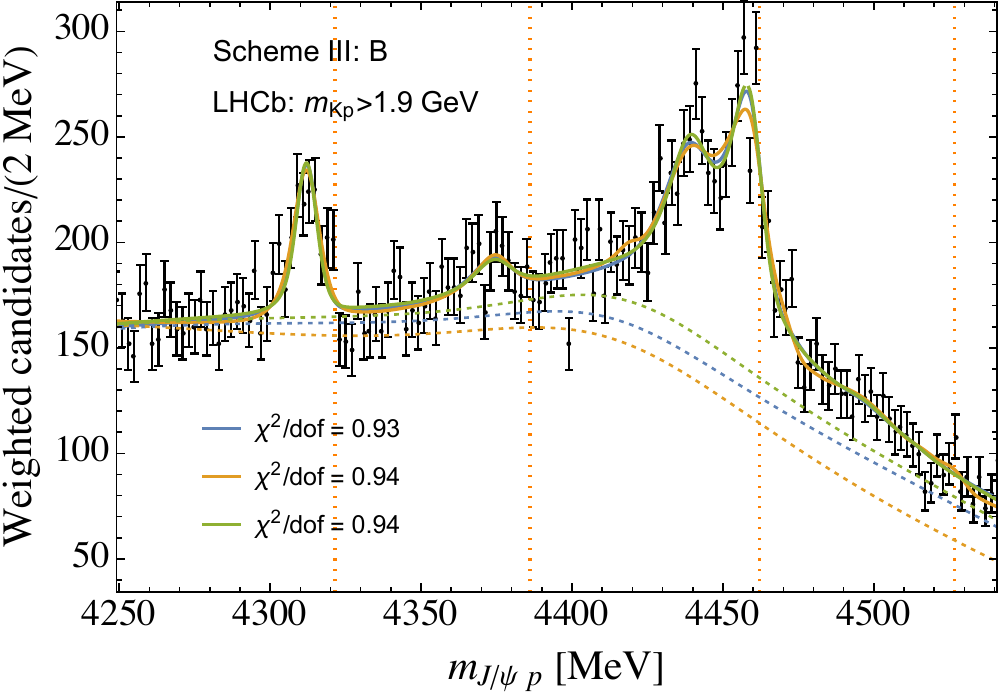}
  \caption{The best fits to the experimental data of $m_{Kp}>1.9$~GeV~\cite{Aaij:2019vzc} with three different fixed backgrounds for solution $B$ (solid curves) of scheme~III. The corresponding backgrounds are shown as dotted curves. The vertical dashed lines from left to right are the $\Sigma_c\bar{D}$, $\Sigma_c^*\bar{D}$, $\Sigma_c\bar{D}^*$, and $\Sigma_c^*\bar{D}^*$ thresholds, respectively.}
  \label{fig:III:fits}
\end{figure*}

There are  three prominent narrow $\pc$ structures in the $J/\psi p$ spectrum, which are used to 
extract the overall strength of the inelastic parameters and the production constants. 
As the fits without the $\lamh$ channels have already yielded   values for  $\chi^2/$dof  less than one, 
it is  not surprising that the explicit inclusion of  $\lamh$  does not improve the fit quality.
As a result, the fitted inelastic parameters, as well as the bare production parameters $P_\alpha^J$, are very sensitive 
to the  data sets used in the fits and to the background employed. 
Indeed, as the system is overdetermined,  it is hard to discriminate the contributions from the $\jp$ ($\etacp$) and $\lamh$ to the widths of the $\pc$ states from just the $\jp$ mass distributions.\footnote{We checked also that the  inclusion of just the $S$-wave contact interaction $C_\frac12^\prime$ between $\lamh$ and  $\sigh$  to  scheme~I  does not allow one to reliably extract $C_\frac12^\prime$ from data. 
}
To illustrate this more clearly, instead of fitting to the three data sets as used in schemes I and II, we make  fits to the data with $m_{Kp}>1.9$~GeV for solution $B$
only with three different, fixed
backgrounds, as shown in Fig.~\ref{fig:III:fits}. 
All the fits have values for   $\chi^2/$dof very close to the best fit results. However, their inelastic 
and bare production parameters are very different, as shown in Table~\ref{tab:III:paras} in Appendix~\ref{app:results}.
Accordingly they lead to   different predictions for the line shapes in the elastic and inelastic channels, 
as illustrated in  Fig.~\ref{fig:III:sigh}. 
It follows in particular from  Table~\ref{tab:III:paras} that  the product of the bare production vertices and the inelastic constants $\jp$ is much better constrained by the $\jp$ mass distributions than these quantities individually.  For example,  the fit with the background-2 (orange lines in Fig.~\ref{fig:III:fits})  yields values for the $\jp$ constants which are an order of magnitude smaller than for the other two fits 
(cf. the couplings $g_S$ and $g_D^\prime$ for the fit in the middle of Table~\ref{tab:III:paras} with the similar couplings in other fits),\footnote{In addition, we define a new parameter $g_D^\prime = g_D k_0^2$, with $k_0= \sqrt{\lambda(m^2_0,m_\psi^2,m_p^2)}/(2m_0) $ the c.m. momentum of the proton at the energy $m_0=(m_{\Sigma_c}+m_{\Sigma_c^*}+m_D+m_{D^*})/2 $, to allow the comparison of  $g_S$ and  $g_D^\prime$ in the same units.} 
which is balanced  by a corresponding increase in bare production vertices.  Consequently, the production rates to  $\sigh$ and $\lamh$ channels for this fit are an order of magnitude larger than for the other fits, as can be seen from Fig.~\ref{fig:III:sigh} (cf. the rates in the middle column with the others).  
Nevertheless, the results in figures allow for several constructive comments. First, we note that the line shapes in the elastic channels for background-1  are  completely consistent with the corresponding line shapes of scheme II shown in Fig.~\ref{fig:II:pred:sigh}.  The elastic spectra for  background-3 also look very similar except for the fact that the $\Sigma_c^* \bar D$ line shape in this case shows a dip  near the $\pc(4440)$ instead 
of the peaks in the other cases. As discussed in Sec.~\ref{subsec:heavylight} (see the next-to-the-last paragraph),  the dip in the $\Sigma_c^* \bar D$ line shape near the 
$\Sigma_c \bar D^*$ threshold  may appear because of the destructive interference between the  amplitudes in these channels. Indeed, as follows from Tables~\ref{fig:III:fits}  and \ref{fig:fits:scheme_II},
the  bare  production amplitudes $P^\frac32$ from Eq.~\eqref{eq:bareproduction} 
have  opposite signs  for the $\Sigma_c \bar D^*$ and  $\Sigma_c^* \bar D$  channels  for background-3, while they have the same signs for background-1 as well as in the case of  scheme II.  Second, since the differences in the $\Sigma_c^* \bar D$  channel play only a minor role for inelastic line shapes, 
 the predicted  line shapes in the $\Lambda_c \bar D^{(*)}$ channels in Fig.~\ref{fig:III:sigh} are quite similar  for backgrounds 1 and 3.  
 Therefore, one may conjecture that, in contrast to the results for background 2, the line shapes for backgrounds 1 and 3 represent natural extensions of the results of scheme II to the full multichannel case. Then, the results for the $\Lambda_c \bar D^{*}$  spectra should have  clear $S$-wave peaking structures from the $P_c(4312)$,  $P_c(4380)$, $P_c(4440)$
 and $P_c(4457)$.  In a complete analogy to the $\Sigma_c \bar D$ line shape  (see  discussion in Sec.~\ref{sec:II:predict}) , also the $\Lambda_c \bar D$ spectrum shows  the $D$-wave peak from the $P_c(4440)$.
 However, the $P_c(4312)$ can not be seen in the $\Lambda_c \bar D$ channel  because the LO $\Sigma_c \bar D\to \Lambda_c \bar D$ $S$-wave transition potential vanishes completely (see Eqs.~\eqref{eq:contactpotential1} and  \eqref{eq:ope1}).  Also,  vanishing of the $SD$ transition potential between 
 $\Sigma_c^* \bar D$ and $ \Lambda_c \bar D$, as follows from Eqs.~\eqref{eq:nloSD12} for $V_{SD}^{\frac32}$ and \eqref{eq:ope2}, explains why  the $\Lambda_c \bar D$ spectrum does not show  signals from the $P_c(4380)$. 
   These are therefore testable predictions in the molecular picture  which can be further updated and improved once data in these channels become available.

We stress also that the poles around the $\sigh$ thresholds are sensitive neither to the form of the background nor to the choice of  the experimental  data set in the fits. The poles 
in scheme~III are 
consistent with those in scheme~II and thus not explicitly presented here.  Also,  it is worth mentioning that the line shapes in the $\etacp$ channels,  which are related to  the $\jp$ distributions 
via HQSS, have  comparable 
rates for all fits considered here (see Fig.~\ref{fig:III:etacp}).  Furthermore, the shape of the $\etacp$ distributions is generally consistent with the predictions of schemes I and II, reported above. 
Specifically, the peak from the $P_c(4440)$, which is clearly seen in all line shapes for solution $B$, 
is an experimentally testable prediction. 

\begin{figure*}[tb]
 \centering
 \includegraphics[width=0.32\textwidth]{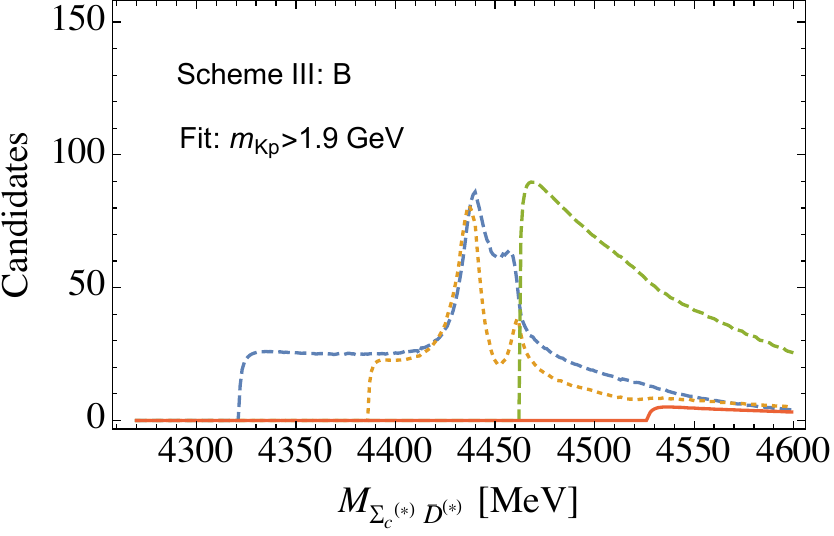}
   \includegraphics[width=0.34\textwidth]{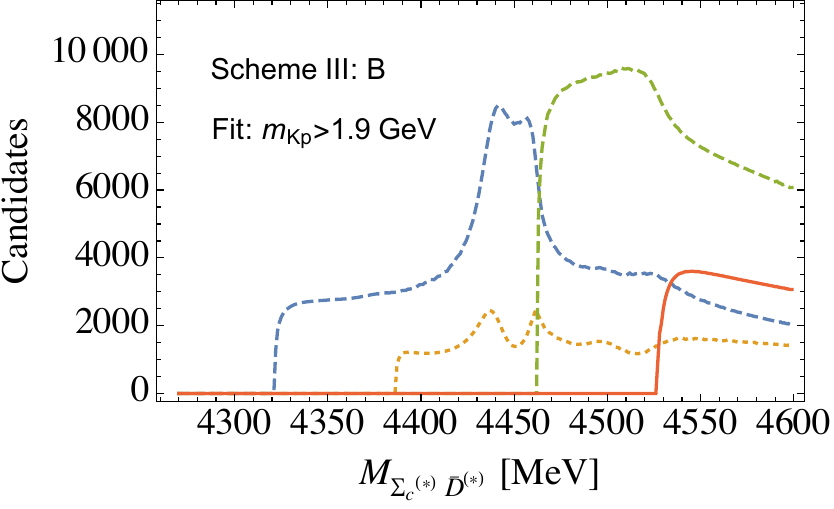} 
  \includegraphics[width=0.32\textwidth]{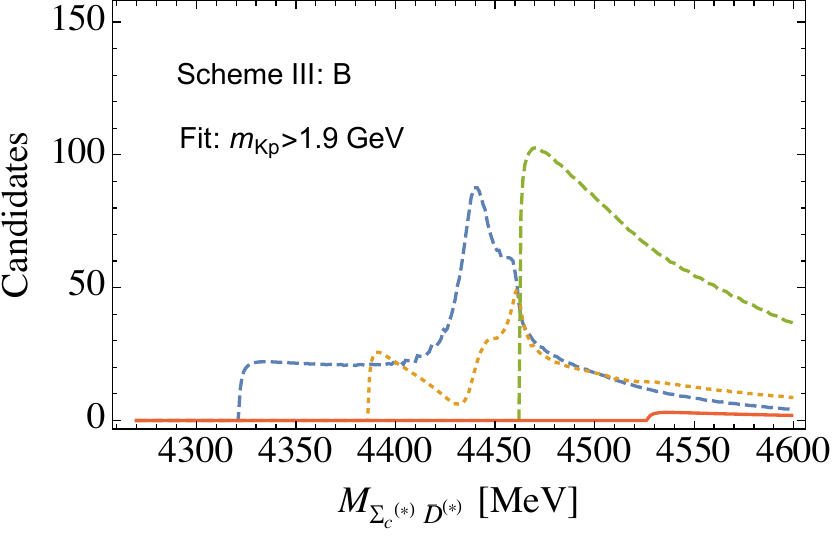}\\
  \includegraphics[width=0.32\textwidth]{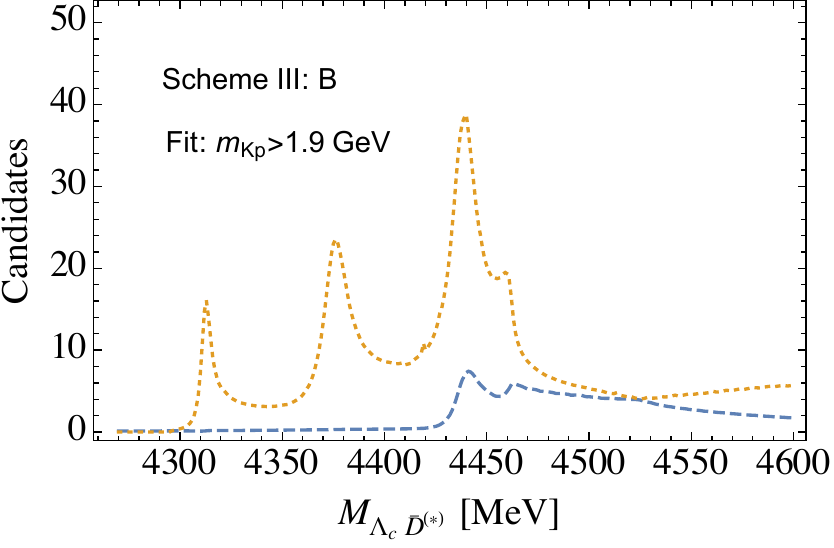}
  \includegraphics[width=0.34\textwidth]{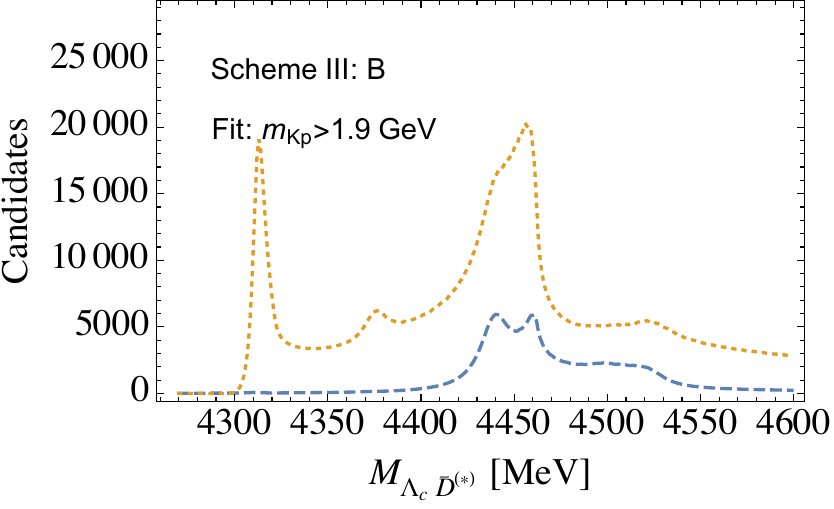}
  \includegraphics[width=0.32\textwidth]{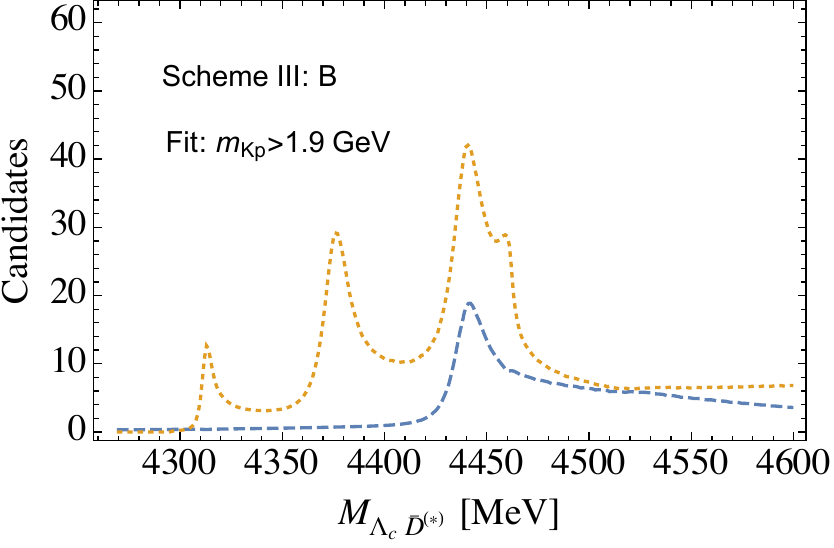}
  \caption{  \label{fig:III:sigh}
  Predictions for the line shapes in the $\sigh$ and $\lamh$ channels based on the fit results of scheme~III  (solution $B$) for three different backgrounds. The left, middle and right columns 
  show the results that are related to the blue, orange and green lines in Fig.~\ref{fig:III:fits}, respectively.  
  First row:   Mass distributions in the $\Sigma_c\bar{D}$ (blue dashed curves), $\Sigma_c^*\bar{D}$ (orange dotted curves), $\Sigma_c\bar{D}^*$ (green dot-dashed curves), and $\Sigma_c^*\bar{D}^*$ (red solid curves) channels. Second row:  Mass distributions in the $\Lambda_c\bar{D}$ (blue dashed curves) and $\Lambda_c\bar{D}^*$ (orange dotted curves) channels. Note the different scales on the $y$-axes
  of the different plots. No contribution from a possible background is included. }
\end{figure*}

\begin{figure*}[tb]
 \centering
  \includegraphics[width=0.49\textwidth]{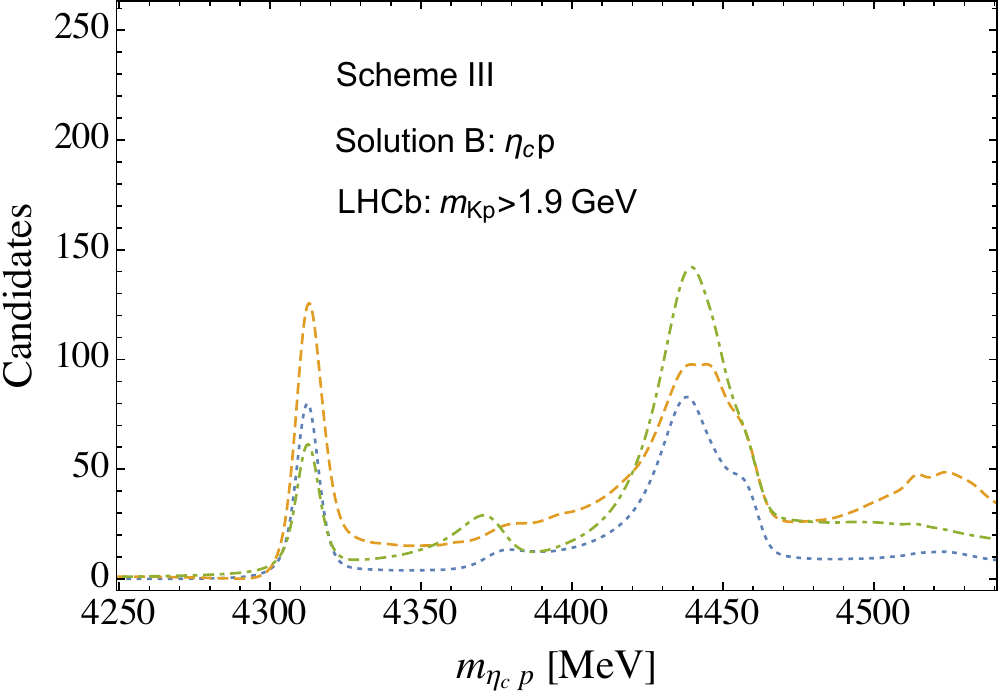}
  \caption{ \label{fig:III:etacp} 
   Predictions for the line shapes of the $\etacn$ invariant mass distributions   based on the fit results of Scheme~III (solution $B$) in  Fig.~\ref{fig:III:fits} for three different backgrounds.
    The blue dotted, orange dashed and  green dot-dashed curves 
correspond to the predictions from the three fit results in  Fig.~\ref{fig:III:fits} of   blue, orange and green colors, in order. No contribution from a possible background is included.}
\end{figure*}

\section{Summary}\label{sec:summary}

In this work we present an updated coupled-channel analysis of the process $\Lambda_b^0\to J/\psi p K^-$ for the hidden-charm pentaquark  states discovered by LHCb.
In Ref.~\cite{Du:2019pij}, an analysis of the  $J/\psi p$ invariant mass distributions in this process was performed within a coupled-channel approach including interactions between the elastic $\sigh$ channels and transitions from the elastic to the $S$- and $D$-wave $J/\psi p$ inelastic channels in a way consistent with HQSS.  To account for other inelastic channels not included explicitly, the LO
contact interactions in the elastic channels were augmented with imaginary parts in the spirit of  optical potentials.
The investigation presented here improves and extends the
study reported in Ref.~\cite{Du:2019pij} in several aspects. In particular, now we include the $\etacn$ and 
$\lamh$ channels dynamically in order to restore unitarity and investigate the effect from these inelastic channels.   The inclusion of  $\etacp$ is also required by HQSS.
To account for the effect of the width of $\Sigma_c^{(*)}$ in the Lippmann-Schwinger equations,  the self-energy of the $\Sigma_c^{(*)}$ is  considered dynamically, which generates  three-body cuts from the
$\Lambda_c \bar D^{(*)} \pi$ intermediate states. Additional  three-body cut contributions appear from the transition potentials $\sigh \to \lamh$ driven by OPE. Apart from  OPE, these transition potentials also involve one leading-order momentum-independent contact term as a consequence of HQSS.  

We emphasize that in the fits performed in this study we take into account all the data sets of the $\jp$ invariant mass distributions provided in Ref.~\cite{Aaij:2019vzc}.
To perform the fits, we consider  three strategies: employing the momentum-independent LO contact potentials without including the $\lamh$ channels (scheme~I), including the
OPE potential supplemented by the $S$-$D$ contact terms to scheme~I (scheme~II), and explicitly including in addition the $\lamh$ channels (scheme~III).  

For scheme~I,  we assume that the widths of the $\pc$ states 
are saturated by the $\jp$, $\etacn$ and $\sigh$ channels. Two solutions (denoted by $A$ and $B$) are 
found describing the data almost equally well, with $\chi^2$-values almost coincident to those of scheme~I in 
Ref.~\cite{Du:2019pij}.  Both solutions 
give seven poles corresponding to seven $\pc$ states. Among these states, the lowest one corresponds to the 
$\pc(4312)$ with $J^P=\frac12^-$. It appears as a $\Sigma_c\bar{D}$ bound state in both solutions. There are two 
$\Sigma_c\bar{D}^*$ bound states with quantum numbers $J^P=\frac12^-$ and $\frac32^-$, corresponding 
to the $\pc(4440)$ and the $\pc(4457)$, respectively, in solution $A$ and
with interchanged quantum numbers in solution $B$, as  
in Ref.~\cite{Du:2019pij}. Also the three predicted $\Sigma_c^*\bar{D}^*$
states show different mass patterns: 
$m_{\frac12^-}< m_{\frac32^-}< m_{\frac52^-}$ for solution $A$ and reversed for solution $B$. 
However, the three $\Sigma_c^*\bar{D}^*$ states are almost invisible in the spectra.
In addition, HQSS calls for a seventh state, located close to the $\Sigma_c^*\bar{D}$ threshold with $J^P=\frac32^-$ and with a width comparable to those of the other states. As in  Ref.~\cite{Du:2019pij}, in the data
we found clear evidence for this state, dubbed  $\pc(4380)$, in both solutions.  Testable predictions for the line shapes in the $\Sigma_c^{(*)}\bar{D}^{(*)}$ and $\etacp$ channels are proposed. 
For the $\sigh$ line shapes, the threshold enhancement at the $\sigh$ thresholds, which is seen in these line shapes,  signals the existence of the corresponding shallow bound states. 

Once the OPE potentials for the $\sigh$ channels are included in scheme~II, an enhanced $S$-wave-to-$D$-wave mixing is induced leading to
strongly regulator dependent line shapes. This calls for a promotion of formally NLO contact terms to LO to provide renormalization.  In this way regulator-independent results for $\jp$ spectra were found for solution $B$,  while those for solution $A$ still  show a visible cutoff dependence, which becomes even stronger when the line shapes in the elastic and inelastic channels are studied.   Meanwhile,  solution $B$ shows only a very mild regulator dependence for all predicted line shapes. This means that only  solution $B$, with the $P_c(4440)$ ($P_c(4457)$) identified as a $\frac32^-$ ($\frac12^-$) $\Sigma_c \bar D^*$ bound state, survives the strong requirements of renormalizability once the OPE is included.  While generally the predicted line shapes in  the $\Sigma_c^{(*)}\bar{D}^{(*)}$ and $\etacp$ channels are qualitatively similar in both schemes I and II for solution $B$,  the results of scheme II show also some signatures of the interactions in  $D$-waves,  which can be clearly seen, e.g.,  near the $\pc(4440)$ in the $\Sigma_c\bar{D}$ spectrum. 
Apart from this observation, the $\eta_c p$  invariant mass distributions are   very interesting since in most fits they show   significant signals from the $\pc(4312)$, the $\pc(4380)$ and also from higher states around 4500~MeV.  A very pronounced  peak from  the $\pc(4440)$ reveals itself in the $\eta_c p$ spectrum  in all schemes for solution $B$, and therefore  can be used to test the quantum numbers of the $\Sigma_c\bar{D}^*$ molecular states  experimentally and thus to unravel the  nature of all $P_c$ states.

In scheme III, we developed  the most advanced version of the coupled-channel approach for the pentaquark states
reported by LHCb, which also incorporates the  $\lamh$ channels as well as the OPE.  
It is demonstrated, however, that, because of the lack of structures around the $\lamh$ thresholds in the
$J/\psi p$ channel and the absence of data in the other channels,  the $\lamh$ interaction strength cannot be reliably extracted from the fits to the $\jp$ data only.  Once new data become available, this approach should not only allow one to determine  the pentaquarks quantum numbers but also to reveal their leading decay properties. 
Meanwhile, even based on the current data and a plausible assumption that the line shapes in the elastic channels should be qualitatively similar  in schemes II and III, we make predictions for the line shapes in the $\Lambda_c \bar D^{(*)}$ channels which can be further updated and improved once data in these channels become available.
In addition, we find that the effect of including a three-body cut from the $\Lambda_c \bar D \pi$ channel in the OPE potentials and especially 
in the  self-energy $\Sigma_c^{(*)}$ in the two-body Green function reduces the width of the narrow $\pc(4380)$ by about a factor 2 compared to the one extracted using a constant 
complex mass of $\Sigma_c^{(*)}$ in Ref.~\cite{Du:2019pij}.

It still remains to be understood  why the three $P_c$ states near 4.5~GeV are almost invisible in the data for the $\jp$ distributions,\footnote{One notices, however, that there are hints of two peaks around 4.50 and 4.52~GeV in the ``$m_{Kp}$ all" data set and that two of the three $\Sigma_c^*\bar D^*$ states have similar masses, see Table~\ref{table:II:pole}. } or why the $P_c(4440)$ and $P_c(4457)$ are more pronounced than the others. This must be related to the production mechanism, which in this work is assumed to proceed entirely through the $\sigd^{(*)}$ channels at the primary production vertex. There can also be effects from nearby triangle singularities discussed in Refs.~\cite{Guo:2015umn,Liu:2015fea,Bayar:2016ftu,Aaij:2019vzc} enhancing or interfering with the production of some of the $P_c$ states.

\begin{acknowledgements}

This work is supported in part by the National Natural Science Foundation of China (NSFC) and  the
Deutsche Forschungsgemeinschaft (DFG) through the funds provided to the Sino-German Collaborative Research Center TRR110 ``Symmetries and the Emergence of Structure in QCD'' (NSFC Grant No. 12070131001, DFG Project-ID 196253076),
by the NSFC under Grants No. 11835015, No. 12047503, No. 11961141012, and No.12035007, by the Chinese Academy of Sciences (CAS) under Grants
No. QYZDB-SSW-SYS013 and No. XDB34030000, and by the Munich Institute for Astro- and Particle Physics (MIAPP) of the DFG cluster of excellence ``Origin and Structure of the Universe.'' The work of M.L.D. is also supported by the Spanish Ministerio de Economía y Competitividad (MINECO) and the European Regional
Development Fund (ERDF) under contract FIS2017-84038-C2-1-P, by the EU Horizon
2020 research and innovation programme, STRONG-2020 project, under grant agreement No.~824093,  by Generalitat
Valenciana under contract PROMETEO/2020/023. 
Q.W. is also supported by the Guangdong Major Project of Basic and Applied Basic Research under Grant No. 2020B0301030008, Guangdong Provincial funding with Grant No.2019QN01X172, and by the Science and Technology Program of Guangzhou under Grant No. 2019050001.
The work of U.G.M. is further supported by the CAS President's International Fellowship
Initiative (PIFI) under Grant No. 2018DM0034 and by the VolkswagenStiftung under Grant No. 93562.
J.A.O. would like to acknowledge partial financial support  by the MICINN (Spain) under Grant No.  PID2019-106080GB-C22/AEI/10.13039/501100011033. 
\end{acknowledgements}

\appendix
\newpage

\section{One-pion-exchange potentials}\label{app:OPE}
Since the pion couples to the charmed mesons and baryons in the $P$ wave, see the Lagrangian in Eq.~\eqref{lag:ope}, the OPE potential mixes the $S$ wave and $D$ wave. To calculate the partial wave projection of the OPE potentials~\cite{Oller:2019rej}, we start from a two-body state $\left|JM,L\Sigma\right\rangle$ with $J$ ($M$) the total angular momentum (its third component), 
$L$ the orbital angular momentum and $\Sigma$ the total spin
\bea
\left|JM,L\Sigma\right\rangle=\frac{1}{\sqrt{4\pi}}\sum_{\sigma_{1},\sigma_{2},\Sigma_{3},L_{3}}\int d\Omega_{\vec{p}}Y_{L}^{L_{3}}\left(\Omega_{\vec{p}}\right)\langle s_{1}s_{2}\sigma_{1}\sigma_{2} | \Sigma \Sigma_3\rangle \langle L\Sigma  L_{3}\Sigma_{3}|JM\rangle\left|\vec{p}\sigma_{1}\sigma_{2}\right\rangle ,
\eea
where  $\left|\vec{p}\sigma_1\sigma_2\right\rangle$ is the direct product of the one-particle states
$\left|\vec{p},\sigma_1\right\rangle$ and $\left|-\vec{p},\sigma_2\right\rangle$ with $\sigma_i$ the third component of 
spin $s_i$ for the $i^{\rm th}$ particle in that channel; $\vec{p}$ is the momentum in the c.m. frame.
The partial wave transition between the states with the 
same $J$ and $M$ reads
\bea\label{eq:transition}
T_{L\Sigma,L^{\prime}\Sigma^{\prime}}^{J}& =& \langle JM,L\Sigma|\hat{T}|JM,L^{\prime}\Sigma^{\prime}\rangle\nonumber\\
&=&\frac{1}{4\pi}\sum_{\sigma_{1}\sigma_{2}\Sigma_{3}L_{3}}\sum_{\sigma_{1}^{\prime}\sigma_{2}^{\prime}\Sigma_{3}^{\prime}L_{3}^{\prime}}
\int d\Omega_{\vec{p}}Y_{L}^{L_{3}}\left(\Omega_{\vec{p}}\right)\langle s_{1}s_{2}\sigma_{1}\sigma_{2} | \Sigma \Sigma_3\rangle \langle L\Sigma  L_{3}\Sigma_{3}| JM\rangle \nonumber\\
&&\times\int d\Omega_{\vec{p}^{\prime}}Y_{L^{\prime}}^{L_{3}^{\prime}*}\left(\Omega_{\vec{p}^{\prime}}\right)\langle s_{1}^{\prime}s_{2}^{\prime}\sigma_{1}^{\prime}\sigma_{2}^{\prime} | \Sigma^{\prime} \Sigma_3^{\prime}\rangle \langle L^{\prime}\Sigma^{\prime} L_{3}^{\prime}\Sigma_{3}^{\prime}| JM\rangle\left\langle\vec{p}^{\prime}\sigma_{1}^{\prime}\sigma_{2}^{\prime}\right|\hat{T}\left|\vec{p}\sigma_1\sigma_2\right\rangle,
\eea
where $\hat{T}$ is the transition operator. In the framework of the TOPT, the OPE potential acquires two contributions as shown in Fig.~\ref{fig:topt}. For the scattering process $12\to 1^\prime 2^\prime$ with $E$ the total energy of the system, and $p$, $p^\prime$ the incoming and outgoing three-momenta, we define
\newcommand{\pr}{p^\prime}
\newcommand{\ppr}{p^{\prime 2}}
\bea\nonumber
V_{SS}(E,p, p^\prime )&\equiv &  -\int_{-1}^{1}d\cos\theta \frac{p^{\prime 2}+p^{2}-2p^\prime p\cos\theta}{2E_\pi(p_\pi)}\big[ D_a^\pi(E,p,\pr ,\theta)+D_b^\pi(E,p,\pr ,\theta) \big],\nonumber\\ \nonumber
V_{SD}(E,p,\pr )&\equiv & -\int_{-1}^{1}d\cos\theta \frac{4\ppr +p^2-8\pr p\cos\theta+3p^2\cos (2\theta) }{2E_\pi(p_\pi)}\big[ D_a^\pi(E,p,\pr ,\theta)+D_b^\pi(E,p,\pr ,\theta) \big],\\\nonumber
V_{DS}(E,p,\pr )&\equiv &  -\int_{-1}^{1}d\cos\theta \frac{\ppr +4p^2-8\pr p\cos\theta+3\ppr \cos (2\theta) }{2E_\pi(p_\pi)}\big[ D_a^\pi(E,p,\pr ,\theta)+D_b^\pi(E,p,\pr ,\theta) \big],\\\nonumber
V_{DD}^{(c_1,c_2,c_3,c_4)}(E,p,\pr )&\equiv & -\int_{-1}^{1}d\cos\theta \frac{c_1 (\ppr +p^2)-c_2\,\pr p\cos\theta+c_3(\ppr +p^2)\cos (2\theta)-c_4\,\pr p\cos (3\theta) }{2E_\pi(p_\pi)} \nonumber\\ %\nonumber
&& \times\big[ D_a^\pi(E,p,\pr ,\theta)+D_b^\pi(E,p,\pr ,\theta) \big],
\eea
where  $\theta$ denotes the angle between the three-momenta $\vec {p}^\prime $ and $\vec p$.
The values for the $c_i(i=1,\ldots,4)$ coefficients are specified below for each value of the total angular momentum $J$. Moreover, 
\bea
D_a^\pi(E,p, p^\prime,\theta)&=&\frac{1}{E_\pi(p_\pi)+E_{1^\prime}(\pr )+E_2(p)-E},\nonumber\\
D_b^\pi(E,p,\pr ,\theta)&=&\frac{1}{E_\pi(p_\pi)+E_{1}(p)+E_{2^\prime}(\pr )-E}\nonumber\\
\eea
are the contributions of the two TOPT orderings with $p_\pi=\sqrt{p^2+\ppr -2p\pr \cos\theta}$ and $E_i=\sqrt{p_i^2+m_i^2}$. 

The OPE potentials in the elastic channels listed in Table~\ref{tab:channels} for $J^P=\frac12^-$ can be written in a matrix form (where the columns and rows are given by the channels listed in order in Table~\ref{tab:channels}) as
\bea\label{eq:ope1}
V_{\frac{1}{2}^{-}}^\text{OPE}= \frac{g_1}{f_{\pi}^{2}}\left(\begin{array}{cc}
V_{\frac{1}{2}^{-}}^{SS} & V_{\frac{1}{2}^{-}}^{SD}  \\
V_{\frac{1}{2}^{-}}^{DS}  & V_{\frac{1}{2}^{-}}^{DD} 
\end{array}\right),
\eea
with 
\[
V_{\frac{1}{2}^{-}}^{SS}=\left(
\begin{array}{ccccc}
 0 & \frac{g_2}{2 \sqrt{3}}V_{SS} & \frac{g_2}{2 \sqrt{6}}V_{SS} & 0 & \frac{g_3}{8}V_{SS} \\
 \frac{g_2}{2 \sqrt{3}}V_{SS} & \frac{g_2}{3}V_{SS} & -\frac{g_2}{6 \sqrt{2}}V_{SS} & \frac{g_3}{8}V_{SS} & \frac{g_3}{4\sqrt{3}}V_{SS} \\
 \frac{g_2}{2 \sqrt{6}}V_{SS} & -\frac{g_2}{6 \sqrt{2}}V_{SS} & \frac{5 g_2}{12}V_{SS} & -\frac{g_3}{4 \sqrt{2}}V_{SS} &
   \frac{g_3}{4 \sqrt{6}}V_{SS} \\
 0 & \frac{g_3}{8}V_{SS} & -\frac{g_3}{4 \sqrt{2}}V_{SS} & 0 & 0 \\
 \frac{g_3}{8}V_{SS} & \frac{g_3}{4 \sqrt{3}}V_{SS} & \frac{g_3}{4 \sqrt{6}} V_{SS}& 0 & 0 \\
\end{array}
\right),
\]

\[
V_{\frac{1}{2}^{-}}^{SD}=\left(
\begin{array}{ccccc}
 \frac{g_2}{4 \sqrt{6}} V_{SD} & 0 & -\frac{g_2}{8 \sqrt{30}}  V_{SD}& -\frac{g_2}{8} \sqrt{\frac{3}{10}}
    V_{SD} & \frac{g_3}{16 \sqrt{2}} V_{SD} \\
 -\frac{g_2}{12 \sqrt{2}} V_{SD} & -\frac{g_2}{8 \sqrt{6}} V_{SD} & \frac{g_2}{6 \sqrt{10}}  V_{SD}&
   -\frac{g_2}{8 \sqrt{10}} V_{SD} & -\frac{g_3}{16 \sqrt{6}} V_{SD} \\
 \frac{g_2}{48} V_{SD} & -\frac{g_2}{16 \sqrt{3}} V_{SD} & \frac{7 g_2}{48 \sqrt{5}} V_{SD} & \frac{g_2}{8
   \sqrt{5}} V_{SD} & -\frac{g_3}{32 \sqrt{3}}  V_{SD}\\
 \frac{g_3}{16 \sqrt{2}} V_{SD} & 0 & \frac{g_3}{16 \sqrt{10}}  V_{SD}& \frac{3 g_3}{16 \sqrt{10}} V_{SD} & 0
   \\
 -\frac{g_3}{16 \sqrt{6}} V_{SD} & \frac{g_3}{16 \sqrt{2}} V_{SD} & -\frac{g_3}{4 \sqrt{30}} V_{SD} &
   \frac{g_3}{16} \sqrt{\frac{3}{10}}  V_{SD} & 0 \\
\end{array}
\right),
\]
\[
V_{\frac{1}{2}^{-}}^{DS}=\left(
\begin{array}{ccccc}
 \frac{g_2}{4 \sqrt{6}}V_{DS} & -\frac{g_2}{12 \sqrt{2}}V_{DS} & \frac{g_2}{48}V_{DS} & \frac{g_3}{16 \sqrt{2}} V_{DS}&
   -\frac{g_3}{16 \sqrt{6}}V_{DS} \\
 0 & -\frac{g_2}{8 \sqrt{6}}V_{DS} & -\frac{g_2}{16 \sqrt{3}} V_{DS}& 0 & \frac{g_3}{16 \sqrt{2}}V_{DS} \\
 -\frac{g_2}{8 \sqrt{30}}V_{DS} & \frac{g_2}{6 \sqrt{10}}V_{DS} & \frac{7 g_2}{48 \sqrt{5}} V_{DS}& \frac{g_3}{16
   \sqrt{10}}V_{DS} & -\frac{g_3}{4 \sqrt{30}}V_{DS} \\
 -\frac{g_2}{8} \sqrt{\frac{3}{10}} V_{DS} & -\frac{g_2}{8 \sqrt{10}}V_{DS} & \frac{g_2}{8 \sqrt{5}}V_{DS} &
   \frac{3 g_3}{16 \sqrt{10}} V_{DS}& \frac{g_3}{16} \sqrt{\frac{3}{10}} V_{DS} \\
 \frac{g_3}{16 \sqrt{2}}V_{DS} & -\frac{g_3}{16 \sqrt{6}}V_{DS} & -\frac{g_3}{32 \sqrt{3}}V_{DS} & 0 & 0 \\
\end{array}
\right) ,
\]
\[
V_{\frac{1}{2}^{-}}^{DD}=\left(
\begin{array}{ccccc}
 \frac{g_2}{24} V_{\text{DD}}^{\text{(1,8,3,0)}}  & -\frac{g_2V_{\text{DD}}^{\text{(4,23,12,9)}} }{32 \sqrt{3}} & -\frac{g_2V_{\text{DD}}^{\text{(8,37,24,27)}} }{96 \sqrt{5}} &
   \frac{g_2V_{\text{DD}}^{\text{(2,13,6,3)}} }{32 \sqrt{5}} & \frac{g_3V_{\text{DD}}^{\text{(1,8,3,0)}} }{32 \sqrt{3}} \\
 -\frac{g_2V_{\text{DD}}^{\text{(4,23,12,9)}} }{32 \sqrt{3}} & 0 &
   \frac{g_2V_{\text{DD}}^{\text{(1,-1,3,9)}} }{16 \sqrt{15}} & \frac{g_2}{32}
   \sqrt{\frac{3}{5}} V_{\text{DD}}^{\text{(2,13,6,3)}}  & \frac{g_3}{64}
   V_{\text{DD}}^{\text{(4,23,12,9)}}  \\
 -\frac{g_2V_{\text{DD}}^{\text{(8,37,24,27)}} }{96 \sqrt{5}} &
   \frac{g_2V_{\text{DD}}^{\text{(1,-1,3,9)}} }{16 \sqrt{15}} & \frac{ g_2}{120}
   V_{\text{DD}}^{\text{(13,77,39,27)}} & \frac{ g_2}{160} V_{\text{DD}}^{\text{(2,13,6,3)}} & \frac{g_3V_{\text{DD}}^{\text{(8,37,24,27)}} }{64 \sqrt{15}} \\
 \frac{ g_2V_{\text{DD}}^{\text{(2,13,6,3)}}}{32 \sqrt{5}} & \frac{g_2}{32} \sqrt{\frac{3}{5}}
   V_{\text{DD}}^{\text{(2,13,6,3)}}  & \frac{g_2 }{160} V_{\text{DD}}^{\text{(2,13,6,3)}}
   & \frac{3g_2}{80} V_{\text{DD}}^{\text{(1,9,3,-1)}}  & -\frac{g_3}{64}
   \sqrt{\frac{3}{5}} V_{\text{DD}}^{\text{(2,13,6,3)}}  \\
 \frac{g_3V_{\text{DD}}^{\text{(1,8,3,0)}} }{32 \sqrt{3}} & \frac{ g_3}{64}
   V_{\text{DD}}^{\text{(4,23,12,9)}} & \frac{ g_3V_{\text{DD}}^{\text{(8,37,24,27)}}
  }{64 \sqrt{15}} & -\frac{ g_3}{64} \sqrt{\frac{3}{5}} V_{\text{DD}}^{\text{(2,13,6,3)}}
   & 0 \\
\end{array}
\right).
\]

The OPE potentials for $J^P=\frac32^-$ read
\bea\label{eq:ope2}
V_{\frac{3}{2}^{-}}^\text{OPE}=\frac{g_{1}}{f_{\pi}^{2}}\left(\begin{array}{cc}
V_{\frac{3}{2}^{-}}^{SS} & V_{\frac{3}{2}^{-}}^{SD}\\
V_{\frac{3}{2}^{-}}^{DS} & V_{\frac{3}{2}^{-}}^{DD}
\end{array}\right),
\eea
where\footnote{As the $V_{\frac{3}{2}^{-}}^{DD}$ potential is symmetric, here only the elements on the upper triangle are presented. The case is similar for the $V_{\frac{5}{2}^{-}}^{DD}$ potential.}
\[
V_{\frac{3}{2}^{-}}^{SS}=
\left(
\begin{array}{cccc}
 -\frac{g_2}{12}V_{SS} & -\frac{g_2}{8 \sqrt{3}}V_{SS} & -\frac{\sqrt{5} g_2}{24} V_{SS} &
   -\frac{g_3}{12 \sqrt{2}}V_{SS} \\
 -\frac{g_2}{8 \sqrt{3}} V_{SS} & 0 & \frac{g_2}{8} \sqrt{\frac{5}{3}} V_{SS} &
   \frac{g_3}{4 \sqrt{6}}V_{SS} \\
 -\frac{\sqrt{5} g_2}{24} V_{SS}& \frac{g_2}{8} \sqrt{\frac{5}{3}}V_{SS}  &
   \frac{g_2}{12}V_{SS} & \frac{g_3}{12} \sqrt{\frac{5}{2}}V_{SS}  \\
 -\frac{g_3}{12 \sqrt{2}}V_{SS} & \frac{g_3}{4 \sqrt{6}}V_{SS} & \frac{g_3}{12} \sqrt{\frac{5}{2}}V_{SS}
    & 0 \\
\end{array}
\right) ,
\]
\[
V_{\frac{3}{2}^{-}}^{SD}=V_{SD} \left(
\begin{array}{cccccccccc}
 -\frac{g_2}{8 \sqrt{3}} & \frac{g_2}{24} & -\frac{g_2}{12} & \frac{g_2}{16 \sqrt{3}} &
   -\frac{g_2}{48 \sqrt{2}} & -\frac{g_2}{48 \sqrt{5}} & \frac{g_2}{16}
   \sqrt{\frac{7}{10}}  & -\frac{g_3}{32} & \frac{g_3}{32 \sqrt{3}} & -\frac{g_3}{16
   \sqrt{3}} \\
 0 & \frac{g_2}{16 \sqrt{3}} & \frac{g_2}{16 \sqrt{3}} & 0 & \frac{g_2}{16 \sqrt{6}} &
   \frac{g_2}{4 \sqrt{15}} & \frac{g_2}{16} \sqrt{\frac{21}{10}}  & 0 & -\frac{g_3}{32}
   & -\frac{g_3}{32} \\
 \frac{g_2}{16 \sqrt{15}} & -\frac{g_2}{12 \sqrt{5}} & -\frac{g_2}{48 \sqrt{5}} &
   \frac{g_2}{4 \sqrt{15}} & -\frac{7 g_2}{48 \sqrt{10}} & -\frac{g_2}{15} &
   \frac{g_2}{80} \sqrt{\frac{7}{2}}  & -\frac{g_3}{32 \sqrt{5}} & \frac{g_3}{8
   \sqrt{15}} & \frac{g_3}{32 \sqrt{15}} \\
 -\frac{g_3}{32} & \frac{g_3}{32 \sqrt{3}} & -\frac{g_3}{16 \sqrt{3}} & -\frac{g_3}{32}
   & \frac{g_3}{32 \sqrt{6}} & \frac{g_3}{32 \sqrt{15}} & -\frac{g_3}{32}
   \sqrt{\frac{21}{10}}  & 0 & 0 & 0 \\
\end{array}
\right) ,
\]
\[
V_{\frac{3}{2}^{-}}^{DS}=V_{DS}
\left(
\begin{array}{cccccccccc}
 -\frac{g_2}{8 \sqrt{3}} & \frac{g_2}{24} & -\frac{g_2}{12} & \frac{g_2}{16 \sqrt{3}} &
   -\frac{g_2}{48 \sqrt{2}} & -\frac{g_2}{48 \sqrt{5}} & \frac{g_2}{16}
   \sqrt{\frac{7}{10}}  & -\frac{g_3}{32} & \frac{g_3}{32 \sqrt{3}} & -\frac{g_3}{16
   \sqrt{3}} \\
 0 & \frac{g_2}{16 \sqrt{3}} & \frac{g_2}{16 \sqrt{3}} & 0 & \frac{g_2}{16 \sqrt{6}} &
   \frac{g_2}{4 \sqrt{15}} & \frac{g_2}{16} \sqrt{\frac{21}{10}}  & 0 & -\frac{g_3}{32}
   & -\frac{g_3}{32} \\
 \frac{g_2}{16 \sqrt{15}} & -\frac{g_2}{12 \sqrt{5}} & -\frac{g_2}{48 \sqrt{5}} &
   \frac{g_2}{4 \sqrt{15}} & -\frac{7 g_2}{48 \sqrt{10}} & -\frac{g_2}{15} &
   \frac{g_2}{80} \sqrt{\frac{7}{2}}  & -\frac{g_3}{32 \sqrt{5}} & \frac{g_3}{8
   \sqrt{15}} & \frac{g_3}{32 \sqrt{15}} \\
 -\frac{g_3}{32} & \frac{g_3}{32 \sqrt{3}} & -\frac{g_3}{16 \sqrt{3}} & -\frac{g_3}{32}
   & \frac{g_3}{32 \sqrt{6}} & \frac{g_3}{32 \sqrt{15}} & -\frac{g_3}{32}
   \sqrt{\frac{21}{10}}  & 0 & 0 & 0 \\
\end{array}
\right)^T ,
\]
\begin{sideways}
  \parbox{9.0in}{
\[
{\setstretch{1.35}
V_{\frac{3}{2}^{-}}^{DD}=
g_2\left(
\begin{array}{cccccccccc}
 0 & \frac{V_{\text{DD}}^{\text{(1,5,3,3)}} }{8 \sqrt{3}} & \frac{V_{\text{DD}}^{\text{(2,
   13,6,3)}} }{16 \sqrt{3}} & 0 & \frac{V_{\text{DD}}^{\text{(1,5,3,3)}} }{8
   \sqrt{6}} & \frac{-V_{\text{DD}}^{\text{(2,13,6,3)}} }{32 \sqrt{15}} & \frac{\sqrt{\frac{3}{70}} V_{\text{DD}}^{\text{(2,13,6,3)}}}{16}
     & 0 & \frac{ g_3V_{\text{DD}}^{\text{(1,5,3,3)}} }{32g_2}
   & \frac{g_3V_{\text{DD}}^{\text{(2,13,6,3)}}}{64g_2} 
    \\
  & \frac{V_{\text{DD}}^{\text{(1,5,3,3)}}}{12}
     & \frac{-V_{\text{DD}}^{\text{(2,13,6,3)}}}{48} 
    & \frac{-V_{\text{DD}}^{\text{(2,13,6,3)}} }{32 \sqrt{3}} &
   \frac{-V_{\text{DD}}^{\text{(1,5,3,3)}} }{24 \sqrt{2}} & \frac{V_{\text{DD}}^{\text{(2,
   13,6,3)}} }{24 \sqrt{5}} & \frac{V_{\text{DD}}^{\text{(2,13,6,3)}} }{16 \sqrt{70}}
   & \frac{g_3V_{\text{DD}}^{\text{(1,5,3,3)}}}{32g_2}   & \frac{g_3V_{\text{DD}}^{\text{(1,5,3,
   3)}} }{16 \sqrt{3}g_2} & \frac{-g_3V_{\text{DD}}^{\text{(2,13,6,3)}} }{64 \sqrt{3}g_2} \\
  &  & \frac{-V_{\text{DD}}^{\text{(1,5,3,3)}}}{24} 
    & \frac{-V_{\text{DD}}^{\text{(1,5,3,3)}} }{16 \sqrt{3}} &
   \frac{V_{\text{DD}}^{\text{(2,13,6,3)}} }{96 \sqrt{2}} & \frac{-\sqrt{5}V_{\text{DD}}^{\text{(1,5,3,3)}}}{48} 
     & \frac{\sqrt{\frac{5}{14}}V_{\text{DD}}^{\text{(2,13,6,3)}}}{32} 
     & \frac{ g_3V_{\text{DD}}^{\text{(2,13,6,3)}}}{64g_2} 
    & \frac{-g_3V_{\text{DD}}^{\text{(2,13,6,3)}}}{64 \sqrt{3}g_2} &
   \frac{-g_3V_{\text{DD}}^{\text{(1,5,3,3)}} }{32 \sqrt{3}g_2} \\
 &  & & 0 &
   \frac{-V_{\text{DD}}^{\text{(2,13,6,3)}} }{32 \sqrt{6}} & \frac{\sqrt{\frac{5}{3}}V_{\text{DD}}^{\text{(1,5,3,3)}} }{16}
     & \frac{ \sqrt{\frac{15}{14}}V_{\text{DD}}^{\text{(2,13,6,3)}}}{32}
     & 0 & \frac{g_3V_{\text{DD}}^{\text{(2,13,6,3)}}}{64g_2}
     & \frac{g_3V_{\text{DD}}^{\text{(1,5,3,3)}}}{32g_2}   \\
& &  & & \frac{5V_{\text{DD}}^{\text{(1,5,3,3)}} }{48}
    & \frac{7 V_{\text{DD}}^{\text{(2,13,6,3)}}
   }{96 \sqrt{10}} & \frac{-V_{\text{DD}}^{\text{(2,13,6,3)}} }{16 \sqrt{35}} &
   \frac{-g_3V_{\text{DD}}^{\text{(1,5,3,3)}} }{16 \sqrt{2}g_2} & \frac{g_3V_{\text{DD}}^{\text{(1,
   5,3,3)}} }{16 \sqrt{6}g_2} & \frac{-g_3V_{\text{DD}}^{\text{(2,13,6,3)}} }{64 \sqrt{6}g_2}
   \\
& & &  & & \frac{V_{\text{DD}}^{\text{(1,5,3,3)}}}{24}
     & \frac{V_{\text{DD}}^{\text{(2,13,6,3)}} }{32
   \sqrt{14}} & \frac{g_3V_{\text{DD}}^{\text{(2,13,6,3)}} }{64 \sqrt{5}g_2} &
   \frac{-g_3V_{\text{DD}}^{\text{(2,13,6,3)}} }{16 \sqrt{15}g_2} & \frac{\sqrt{\frac{5}{3}}g_3V_{\text{DD}}^{\text{(1,5,3,3)}} }{32g_2}
     \\
& &   &  & & & \frac{-3V_{\text{DD}}^{\text{(1,3,3,5)}} }{112}
    & \frac{-3 g_3V_{\text{DD}}^{\text{(2,13,6,3)}}
   }{32 \sqrt{70}g_2} & \frac{-\sqrt{\frac{3}{70}}g_3V_{\text{DD}}^{\text{(2,13,6,3)}}}{32g_2}  
    & \frac{\sqrt{\frac{15}{14}} g_3 V_{\text{DD}}^{\text{(2,13,6,3)}}}{64g_2}   \\
 &   &  &  &  &  &  & 0 & 0 & 0 \\
& & &  & & &  &  & 0&0 \\
& & &&  & &  &  &  & 0
   \\
\end{array}
\right).
}
\]}
\end{sideways}

The OPE potentials for the channels in order listed in Table~\ref{tab:channels} for $J^P=\frac52^-$ read
\bea\label{eq:ope3}
V_{\frac{5}{2}^{-}}^\text{OPE}=\frac{g_{1}}{f_{\pi}^{2}}\left(\begin{array}{cc}
V_{\frac{5}{2}^{-}}^{SS} & V_{\frac{5}{2}^{-}}^{SD}\\
V_{\frac{5}{2}^{-}}^{DS} & V_{\frac{5}{^{2}}^{-}}^{DD}
\end{array}\right),
\eea
where
\[
V_{\frac{5}{2}^{-}}^{SS}=-\frac{1}{4}g_2V_{SS},
\]
\[
V_{\frac{5}{2}^{-}}^{SD}=\left(
\begin{array}{cccccccccc}
 -\frac{g_2}{8 \sqrt{10}} & -\frac{g_2}{8 \sqrt{30}} & -\frac{g_2}{16} \sqrt{\frac{7}{15}}
    & -\frac{g_2}{16} \sqrt{\frac{7}{5}}  & \frac{g_2}{8 \sqrt{15}} & -\frac{ g_2}{80}
   \sqrt{\frac{7}{3}} & -\frac{g_2}{20} \sqrt{\frac{7}{2}}  & \frac{g_3}{16}
   \sqrt{\frac{3}{10}}  & \frac{g_3}{16 \sqrt{10}} & \frac{g_3 }{32} \sqrt{\frac{7}{5}}
   \\
\end{array}
\right)V_{SD},
\]
\[
V_{\frac{5}{2}^{-}}^{DS}=\left(
\begin{array}{cccccccccc}
 -\frac{g_2}{8 \sqrt{10}} & -\frac{g_2}{8 \sqrt{30}} & -\frac{g_2}{16} \sqrt{\frac{7}{15}}
    & -\frac{g_2}{16} \sqrt{\frac{7}{5}}  & \frac{g_2}{8 \sqrt{15}} & -\frac{ g_2}{80}
   \sqrt{\frac{7}{3}} & -\frac{g_2}{20} \sqrt{\frac{7}{2}}  & \frac{g_3}{16}
   \sqrt{\frac{3}{10}}  & \frac{g_3}{16 \sqrt{10}} & \frac{g_3 }{32} \sqrt{\frac{7}{5}}
   \\
\end{array}
\right)^TV_{DS},
\]

\begin{sideways}
  \parbox{9.0in}{
\[
{\setstretch{1.35}
V_{\frac{5}{2}^{-}}^{DD}=
g_2\left(
\begin{array}{cccccccccc}
 0 & \frac{V_{\text{DD}}^{\text{(1,5,3,3)}} }{8 \sqrt{3}} &
   \frac{- V_{\text{DD}}^{\text{(2,13,6,3)}}}{8 \sqrt{42}} & 0 &
   \frac{V_{\text{DD}}^{\text{(1,5,3,3)}} }{8 \sqrt{6}} & \frac{V_{\text{DD}}^{\text{(2,13,6,3)}} }{16 \sqrt{210}} & \frac{V_{\text{DD}}^{\text{(2,13,6,3)}} }{8\sqrt{35}} & 0 & \frac{g_3V_{\text{DD}}^{\text{(1,5,3,3)}}}{32g_2}   &
   \frac{-g_3V_{\text{DD}}^{\text{(2,13,6,3)}} }{32 \sqrt{14}g_2} \\
 & \frac{V_{\text{DD}}^{\text{(1,5,3,3)}}}{12}
   & \frac{V_{\text{DD}}^{\text{(2,13,6,3)}} }{24
   \sqrt{14}} & \frac{V_{\text{DD}}^{\text{(2,13,6,3)}} }{16 \sqrt{42}} &
   \frac{-V_{\text{DD}}^{\text{(1,5,3,3)}} }{24 \sqrt{2}} &
   \frac{-V_{\text{DD}}^{\text{(2,13,6,3)}} }{12 \sqrt{70}} &
   \frac{V_{\text{DD}}^{\text{(2,13,6,3)}} }{8 \sqrt{105}} & \frac{g_3V_{\text{DD}}^{\text{(1,5,3,3)}}}{32g_2}
     & \frac{g_3V_{\text{DD}}^{\text{(1,5,3,3)}} }{16
   \sqrt{3}g_2} & \frac{g_3V_{\text{DD}}^{\text{(2,13,6,3)}} }{32 \sqrt{42}g_2} \\
&  & \frac{-V_{\text{DD}}^{\text{(17,100,51,36)}}}{168}  
   & \frac{-V_{\text{DD}}^{\text{(4,5,12,27)}} }{224 \sqrt{3}} &
   \frac{-V_{\text{DD}}^{\text{(2,13,6,3)}} }{96 \sqrt{7}} & \frac{-\sqrt{5}V_{\text{DD}}^{\text{(16,83,48,45)}}}{672} 
    & \frac{\sqrt{\frac{5}{6}} V_{\text{DD}}^{\text{(2,13,6,3)}}}{112} 
     & \frac{-g_3V_{\text{DD}}^{\text{(2,13,6,3)}}
   }{32 \sqrt{14}g_2} & \frac{g_3V_{\text{DD}}^{\text{(2,13,6,3)}} }{32 \sqrt{42}g_2} &
   \frac{-g_3V_{\text{DD}}^{\text{(17,100,51,36)}} }{224 \sqrt{3}g_2} \\
  & & & 0 &
   \frac{V_{\text{DD}}^{\text{(2,13,6,3)}} }{32 \sqrt{21}} & \frac{\sqrt{\frac{5}{3}} V_{\text{DD}}^{\text{(11,61,33,27)}}}{112}
     & \frac{\sqrt{\frac{5}{2}}V_{\text{DD}}^{\text{(2,13,6,3)}}}{112}
     & 0 &
   \frac{-g_3V_{\text{DD}}^{\text{(2,13,6,3)}} }{32 \sqrt{14}g_2} & \frac{g_3V_{\text{DD}}^{\text{(4,5,12,27)}}}{448g_2}
    \\
 &  & & & \frac{5V_{\text{DD}}^{\text{(1,5,3,3)}}}{48}
     & \frac{-\sqrt{\frac{7}{5}}V_{\text{DD}}^{\text{(2,13,6,3)}}}{96} 
     & \frac{-V_{\text{DD}}^{\text{(2,13,6,3)}} }{4
   \sqrt{210}} & \frac{-g_3V_{\text{DD}}^{\text{(1,5,3,3)}} }{16 \sqrt{2}g_2} &
   \frac{ g_3V_{\text{DD}}^{\text{(1,5,3,3)}}}{16 \sqrt{6}g_2} & \frac{ g_3V_{\text{DD}}^{\text{(2,13,6,3)}}}{64 \sqrt{21}g_2} \\
 & &  & &  & \frac{-V_{\text{DD}}^{\text{(1,17,3,-9)}}}{168}   & \frac{V_{\text{DD}}^{\text{(2,13,6,3)}} }{112 \sqrt{6}} &
   \frac{-g_3V_{\text{DD}}^{\text{(2,13,6,3)}} }{32 \sqrt{70}g_2} &
   \frac{g_3V_{\text{DD}}^{\text{(2,13,6,3)}} }{8 \sqrt{210}g_2} & \frac{ g_3\sqrt{\frac{5}{3}} V_{\text{DD}}^{\text{(16,83,48,45)}}}{448g_2}
     \\
 &  &  &  & && \frac{-V_{\text{DD}}^{\text{(11,61,33,27)}}}{112}
     & \frac{-g_3\sqrt{\frac{3}{35}}
   V_{\text{DD}}^{\text{(2,13,6,3)}}}{16g_2}   & \frac{-g_3V_{\text{DD}}^{\text{(2,13,6,3)}}
   }{16 \sqrt{35}g_2} & \frac{-g_3\sqrt{\frac{5}{2}} V_{\text{DD}}^{\text{(2,13,6,3)}}}{224g_2} 
    \\
 & & &  & &  &  & 0 & 0 & 0 \\
  & &  & &&  & &  & 0 & 0 \\
 &  &  &  &
   &   &  & &  & 0 \\
\end{array}
\right).
}
\]
}
\end{sideways}

\section{Results of the fits}\label{app:results}

This study was performed neglecting 
isospin symmetry breaking effects. The masses of particles used in the calculation are~\cite{Zyla:2020zbs}
\bea\label{eq:masses}
M_{\Lambda_b^0}&= 5.6196~\text{GeV}, \quad M_{K^-} &= 0.4937~\text{GeV}, \nno 
M_{J/\psi}&=3.0969~\text{GeV}, \quad M_{p} & = 0.9383~\text{GeV}, \nno
M_{\eta_c}&=2.9839~\text{GeV}, \quad M_\pi & = 0.1380~\text{GeV},\nno
M_{\Sigma_c}& = 2.4535~\text{GeV}, \quad M_{\Sigma_c^*}& = 2.5181~\text{GeV},\nno
M_{D}& = 1.8680~\text{GeV}, \quad M_{D^*}&=2.0086~\text{GeV},\nno
M_{\Lambda_c}& = 2.2865~\text{GeV}.\qquad \quad &
\eea

The fitted parameters to the three data of schemes I and II shown in Figs.~\ref{fig:contactfit} and~\ref{fig:fits:scheme_II} are collected in Tables~\ref{tab:paracontact} and \ref{tab:II:paras}, respectively. We do not show the unit of $\mathcal{F}_n^J$ because an arbitrary overall normalization factor is inseparable.  The fitted parameters of the three fits of scheme III with fixed background in Fig.~\ref{fig:III:fits} are given in Table~\ref{tab:III:paras}. The parameters of the background in all fits 
are collected in Table~\ref{tab:bgds}.

\begin{table*}[!htb]
\caption{The best fit parameters of scheme I for various datasets shown in Fig.~\ref{fig:contactfit}.  Here only the statistical uncertainties are presented.  
}\label{tab:paracontact}
{
\begin{ruledtabular}
\begin{tabular}{c |c c| c c | c c}
%\hline 
%\hline
Parameter &  \multicolumn{2}{c|}{$\cos\theta_{\pc}$-weighted } &  \multicolumn{2}{c|}{$m_{Kp}>1.9$ GeV }&  \multicolumn{2}{c}{ $m_{Kp}$ all  }  \\
\hline
solution & $A$ & $B$ & $A$ & $B$ & $A$ & $B$ \\
\hline 
$C_\frac12$ $[\text{GeV}^{-2}]$ & $-14.72(26)$ & $-9.86(32)$ & $-14.89(18) $  & $-9.06(49)$ & $-14.71(17)$ & $-9.66(34)$
\tabularnewline
\hline
$C_\frac32$ $[\text{GeV}^{-2}]$ & $-10.33(25)$ & $-13.28(21)$ & $-10.16(20)$ & $-13.87(16)$ & $-10.54(18)$ & $-13.76(19)$
\tabularnewline
\hline
$g_S$ $[\text{GeV}^{-2}]$ & $1.17(201)$ & $2.23(132)$ & $1.39(90)$ & $4.51(94)$ & $2.42(89)$ & $3.63(80)$
\tabularnewline
\hline
$g_D^\prime$ $[\text{GeV}^{-2}]$ & $3.88(64)$ & $3.61(48)$ & $3.78(42)$ & $1.32(118)$ & $2.58(67)$ & $2.50(72)$
\tabularnewline
\hline
$\mathcal{F}^\frac12_1$ & $27(249)$ & $716(1424) $ & $-139(219)$ & $362(1109)$ & $606(327)$ & $-3388(875)$
\tabularnewline
\hline
$\mathcal{F}^\frac12_2$ & $883(257)$ & $1782(908)$ & $471(110)$ & $1631(743)$ & $208(247)$ & $4165(711)$
\tabularnewline
\hline
$\mathcal{F}^\frac12_3$ & $-3509(453)$ & $-3333(572)$ & $-1879(120)$ & $-2866(766)$ & $-3365(574)$ & $-1676(334)$
\tabularnewline
\hline
$\mathcal{F}^\frac32_1$ & $-3768(2140)$ & $-1948(1101)$ & $-2785(755)$ & $-497(335)$ & $1255(1087)$ & $2385(985)$
\tabularnewline
\hline
$\mathcal{F}^\frac32_2$ & $-1743(1670)$ & $32(1255)$ & $-1459(541)$ & $32(-463)$ & $-305(735)$ & $-1495(1139)$
\tabularnewline
\hline
$\mathcal{F}^\frac32_3$ & $-4662(648)$ & $-3400(351)$ & $-2396(186)$ & $-1530(242)$ & $-2901(306)$ & $934(458)$
\tabularnewline
\hline
$\mathcal{F}^\frac52_1$ & $721(1593)$ & $-992(205)$ & $0(825)$ &  $421(130)$ & $1953(292)$ & $-1284(226)$b 
\tabularnewline
%\hline
%\hline
\end{tabular}
\end{ruledtabular}
}
\end{table*}

\begin{table*}[!htb]
\caption{\label{tab:II:paras}The best fit parameters of scheme II (solution $B$) for various datasets shown in Fig.~\ref{fig:fits:scheme_II}.  Here only the statistical uncertainties are presented. }
{
\begin{ruledtabular}
\begin{tabular}{c |c  c  c}
Parameter & $\cos\theta_{\pc}$-weighted &  $m_{Kp}>1.9$ GeV & $m_{Kp}$ all
\tabularnewline
\hline
$C_\frac12~[\text{GeV}^{-2}]$ &  $7.35(27)$ &  $7.45(21)$ &  $7.06(19)$
\tabularnewline
\hline
$C_\frac32~[\text{GeV}^{-2}]$ &  $-16.55(99)$ &  $-16.41(71)$ &  $-16.01(45)$
\tabularnewline
\hline
$D_b^{SD}~[\text{GeV}^{-4}] $ &  $-1.63(77)$ &  $-1.75(36)$ &  $-1.90(17)$
\tabularnewline
\hline
$g_S~[\text{GeV}^{-2}]$ &  $3.33(63)$ &  $3.42(58)$ &  $2.79(55)$
\tabularnewline
\hline
$g_D^\prime~[\text{GeV}^{-2}]$ & $1.96(55)$ &  $1.67(30)$ &  $1.67(37)$
\tabularnewline
\hline
$\mathcal{F}_1^\frac12$ & $ -396(636)$ & $890(636)$ &  $-2192(454)$
\tabularnewline
\hline
$\mathcal{F}_2^\frac12$ &  $2996(456)$ &  $1089(391)$ &  $3085(435)$
\tabularnewline
\hline
$\mathcal{F}_3^\frac12$ &  $-1992(387)$ &   $-1368(258)$ &   $-788(144)$
\tabularnewline
\hline
$\mathcal{F}_1^\frac32$ &  $-1086(806)$ &  $-489(209)$ &  $-627(601)$
\tabularnewline
\hline
$\mathcal{F}_2^\frac32$ &  $-381(899)$ &  $-21(324)$ &  $ -113(1101)$
\tabularnewline
\hline
$\mathcal{F}_3^\frac32$ & $-2046(594)$ &  $-983(146)$ & $-1399(295)$
\tabularnewline
\hline
$\mathcal{F}_1^\frac52$ &  $692(335)$ & $243(114)$ &  $ 759(146)$
\tabularnewline
%\hline
%\hline
\end{tabular}
\end{ruledtabular}
}
\end{table*}

\begin{table*}[!htb]
\caption{\label{tab:III:paras}The best fit parameters of scheme III (solution $B$) for various backgrounds shown in Fig.~\ref{fig:III:fits}. 
The fit is made to the LHCb dataset with $m_{Kp}>1.9$ GeV. Here only the statistical uncertainties are presented. }
\begin{ruledtabular}
\begin{tabular}{c |c  c  c}
Parameter & background-1 &  background-2 & background-3  \\
%\hline
\hline 
$C_\frac12~[\text{GeV}^{-2}]$ & $7.58(42)$ & $10.38(46)$ & $7.76(110)$
\tabularnewline
\hline
$C_\frac32~[\text{GeV}^{-2}]$ & $-16.04(62)$ & $-15.37(57)$ & $-16.26(64)$
\tabularnewline
\hline
$C_\frac12^\prime~[\text{GeV}^{-2}]$ & $14.41(197)$ &  $19.32(44)$ & $14.93(387)$
\tabularnewline
\hline
$D_b^{SD}~[\text{GeV}^{-4}] $ & $-1.62(45)$ & $-0.71(19)$ & $-1.46(80)$
\tabularnewline
\hline
${D}_c^{SD}~[\text{GeV}^{-4}]$ & $-2.33(15)$ & $-1.65(30)$ & $-2.24(31)$
\tabularnewline
\hline
$g_S~[\text{GeV}^{-2}]$ & $3.42(71)$ & $0.52(19)$ & $3.38(77)$
\tabularnewline
\hline
$g_D^\prime~[\text{GeV}^{-2}]$ & $1.67(46)$ & $0.24(10)$ & $2.00(65)$
\tabularnewline
\hline
$\mathcal{F}_1^\frac12$ & $800(596)$ & $16970(3632)$ & $622(1126)$
\tabularnewline
\hline
$\mathcal{F}_2^\frac12$ & $1214(411)$ & $9004(2535)$ & $1376(692)$
\tabularnewline
\hline
$\mathcal{F}_3^\frac12$ & $-1462(177)$ & $-22751(2744)$ & $-1406(547)$
\tabularnewline
\hline
$\mathcal{F}_1^\frac32$ & $-432(244)$ & $-5603(3320)$ & $-1538(793)$
\tabularnewline
\hline
$\mathcal{F}_2^\frac32$ & $-51(338)$ & $-1640(2687)$ & $874(432)$
\tabularnewline
\hline
$\mathcal{F}_3^\frac32$ & $-1132(206)$ & $-11493(3221)$ & $-451(278)$
\tabularnewline
\hline
$\mathcal{F}_1^\frac52$ & $197(150)$ &$3878(2093)$ & $247(229)$
\tabularnewline
\end{tabular}
\end{ruledtabular}
\end{table*}

\newcommand{\tn}{$\times10^9$}
\newcommand{\te}{$\times10^8$}
\newcommand{\ts}{$\times10^7$}
\newcommand{\tsx}{$\times10^6$}
\begin{table*}[!htb]
\caption{\label{tab:bgds}The parameters of the backgrounds in all fits.}
%\begin{ruledtabular}
\begin{tabular}{ c | c | c | c | c | c | c | c | c}
\hline 
\hline
Scheme & Solution & Fit & $b_0$ &  $b_1~[\text{GeV}^{-2}]$  & $b_2~[\text{GeV}^{-4}]$ & $g_r$ [GeV] & $m$ [GeV] & $\Gamma$ [GeV]  \\
\hline
\multirow{6}{*}{I} & \multirow{3}{*}{A} & $\cos\theta_{\pc}$-weighted  & -4.05$\times10^9$ & 4.39$\times 10^8$ & -1.16$\times 10^7$ & 150.7 & 4.41 & 0.058 \\
\cline{3-9}
& & $m_{Kp}>$1.9 GeV &  -3.14\te & 4.30\ts & -1.33\tsx & 104.9 & 4.41 & 0.074 \\
\cline{3-9}
& & $m_{Kp}$ all &  1.18\tn & -1.13\te & 2.88\tsx & 595.9 & 4.42 & 0.166 \\
\cline{2-9}
&  \multirow{3}{*}{B} & $\cos\theta_{\pc}$-weighted  & -2.87$\times10^9$ & 3.16\te & -8.40\tsx & 254.0 & 4.43 & 0.088 \\
\cline{3-9}
& & $m_{Kp}>$1.9 GeV & -1.44\te & 2.52\ts & -8.56$\times10^5$ & 126.2 & 4.42 & 0.093 \\
\cline{3-9}
& & $m_{Kp}$ all &  1.30\tn & -1.23\te & 3.03\tsx & 706.1 & 4.43 &  0.188  \\
\hline
\multirow{3}{*}{II} &  \multirow{3}{*}{B} & $\cos\theta_{\pc}$-weighted  & -3.19\tn & 3.51\te &  -9.33\tsx & 280.4 & 4.43 & 0.102 \\
\cline{3-9}
& & $m_{Kp}>$1.9 GeV &  -3.14\te & 4.30\ts & -1.33\tsx & 160.43 & 4.42 & 0.117 \\
\cline{3-9}
& & $m_{Kp}$ all &  3.95\ts & 5.55\tsx & -1.83$\times10^5$ & 497.8 & 4.43 & 0.152   \\
\hline
\multirow{3}{*}{III} & \multirow{3}{*}{B} & background-1 & -2.93\te & 4.07\ts & -1.27$\times10^6$  &  145.0 & 4.41 & 0.117  \\
\cline{3-9}
& & background-2 &  2.41\te  & -1.36\ts & 1.05$\times10^5$ & 228.0 & 4.41 & 0.150 \\
\cline{3-9}
& & background-3 &  -5.91\te & 7.17\ts & -2.07\tsx & 145.0 & 4.41 & 0.117 \\
\hline
\hline
\end{tabular}
%\end{ruledtabular}
\end{table*}

\section{The effective couplings to the elastic channels and the source}\label{app:couplings}

The effective couplings of the $P_c$ states to the elastic channels defined in Eq.~\eqref{eq:def:couplings} for solutions $A$ and $B$ in scheme~I are collected in Tables~\ref{tab:coupling_I_A} and \ref{tab:coupling_I_B}, respectively, where the RSs on which those poles are found and the corresponding quantum numbers are given explicitly. For scheme II, the effective couplings for different quantum numbers, i.e. $J^P=\frac12^-$, $\frac32^-$ and $\frac52^-$, are presented in Table~\ref{tab:coupling_II_12}, \ref{tab:coupling_II_32} and \ref{tab:coupling_II_52}, respectively.

\begin{table*}[!htb]
\caption{The effective couplings to the elastic channels for solution $A$ of scheme I, as defined in Eq.~\eqref{eq:def:couplings}.}\label{tab:coupling_I_A}
{
%\begin{ruledtabular}
\begin{tabular}{| c | c | c | c | c | c | c |}
%\hline 
\hline
State & Pole [MeV] & ~$J^P$~ & ~RS~ &  $\Sigma_c\bar{D}$ & $\Sigma_c\bar{D}^*$ & $\Sigma_c^*\bar{D}^*$ \tabularnewline
%\hline 
\hline
$P_c(4312)$ & $4314(1)-4(1)i$ & $\frac12^-$ & I & $2.6(1)+0.4(2)i$ & $0.7(1)+0.2(1)i$ & $0.4(1)+0.1(1)i$
\tabularnewline
\hline
$P_c(4440)$ & $4440(1)-9(2)i$ & $\frac12^-$ & III & $0.1(1)+0.3(1)i$ & $3.7(2)+0.6(1)i$ & $-0.7(1)+0.2(2)i$
\tabularnewline
\hline
$P_c $ & $4498(2)-9(3)i$ & $\frac12^-$ & IV & $0.1(1)+0.2(1)i$ & $0.0(1)-0.3(1)i$ & $4.0(1)+0.4(2)i$
\tabularnewline
\hline
\hline
State & Pole [MeV] & $J^P$ & RS &  $\Sigma_c\bar{D}^*$ & $\Sigma_c^*\bar{D}$ & $\Sigma_c^*\bar{D}^*$ \tabularnewline
%\hline 
\hline
$P_c(4380)$ & $4377(1)-7(1)i$ & $\frac32^-$ & II & $0.5(1)+0.2(1)i$ & $2.8(1)+0.1(1)i$ & $-0.9(1)+0.1(2)i$
\tabularnewline
\hline
$P_c(4457)$ & $4458(2)-3(1)i$ & $\frac32^-$ & III & $2.1(2)+0.3(1)$i & $0.1(1)-0.1(0)i$ & $-0.7(1)+0.2(2)i$
\tabularnewline
\hline
$P_c$ & $4510(2)-14(3)i$ & $\frac32^-$ & IV & $-0.4(2)-0.3(1)i$ & $0.2(1)+0.3(1)i$ & $3.3(2)+0.6(2)i$
\tabularnewline
\hline
\hline
State & Pole [MeV] & $J^P$ & RS &  $\Sigma_c^*\bar{D}^*$ & \, & \, \tabularnewline
%\hline 
\hline
$P_c$ & $4525(2)-9(3)i$ & $\frac52^-$ & IV & $1.9(2)+0.6(7)i$ & \, & \,  \tabularnewline
\hline
%\hline
%\hline
\end{tabular}
%\end{ruledtabular}
}
\end{table*}

\begin{table*}[!htb]
\caption{The effective couplings to the elastic channels for solution $B$ of scheme I, as defined in Eq.~\eqref{eq:def:couplings}.}\label{tab:coupling_I_B}
{
%\begin{ruledtabular}
\begin{tabular}{| c | c | c | c | c | c | c |}
%\hline 
\hline
State & Pole [MeV] & ~$J^P$~ & ~RS~ &  $\Sigma_c\bar{D}$ & $\Sigma_c\bar{D}^*$ & $\Sigma_c^*\bar{D}^*$ \tabularnewline
%\hline 
\hline
$P_c(4312)$ & $4312(2)-4(2)i$ & $\frac12^-$ & I & $2.9(1)+0.4(2)i$ & $-0.6(2)+0.0(1)i$ & $-0.5(1)+0.1(2)i$
\tabularnewline
\hline
$P_c(4457)$ & $4462(4)-5(3)i$ & $\frac12^-$ & III & $0.1(1)-0.2(2)i$ & $2.0(2)+1.2(3)i$ & $0.2(1)+0.2(1)i$
\tabularnewline
\hline
$P_c $ & $4526(3)-9(2)i$ & $\frac12^-$ & IV & $0.0(0)-0.1(1)i$ & $0.0(1)+0.1(1)i$  & $1.5(2)+1.1(4)i$
\tabularnewline
\hline
\hline
State & Pole [MeV] & $J^P$ & RS &  $\Sigma_c\bar{D}^*$ & $\Sigma_c^*\bar{D}$ & $\Sigma_c^*\bar{D}^*$ \tabularnewline
%\hline 
\hline
$P_c(4380)$ & $4375(2)-6(1)i$ & $\frac32^-$ & II & $0.5(1)-0.2(2)i$ & $3.0(1)+0.1(1)I$ & $-0.8(2)+0.2(2)i$
\tabularnewline
\hline
$P_c(4440)$ & $4441(3)-5(2)i$ & $\frac32^-$ & III & $3.6(1)+0.3(1)$i & $0.0(1)+0.1(1)i$ & $0.8(1)-0.1(2)i$
\tabularnewline
\hline
$P_c$ & $4521(2)-12(3)i$ & $\frac32^-$ & IV & $-0.1(2)-0.2(1)$i & $0.1(2)+0.4(1)i$ & $2.5(2)+0.9(2)i$
\tabularnewline
\hline
\hline
State & Pole [MeV] & $J^P$ & RS &  $\Sigma_c^*\bar{D}^*$ & \, & \, \tabularnewline
%\hline 
\hline
$P_c$ & $4501(3)-6(4)i$ & $\frac52^-$  & IV & $3.9(2)+0.1(2)i$ & \, & \,  \tabularnewline
\hline
%\hline
%\hline
\end{tabular}
%\end{ruledtabular}
}
\end{table*}

\begin{table*}[!htb]
\caption{The effective couplings to the elastic channels for $J^P=\frac12^-$ resonances of scheme II, as defined in Eq.~\eqref{eq:def:couplings}.}\label{tab:coupling_II_12}
{
\begin{ruledtabular}
\resizebox{\textwidth}{!}{
\begin{tabular}{ l | c  c  c  }
%\hline 
% \multirow{2}{*}{$J=\frac12^-$} & \multicolumn{3}{c|}{Solution A} & \multicolumn{3}{c}{Solution B}  \\
%\cline{2-7}
$J=\frac12^-$ &  $P_c(4312)$  & $P_c(4457)$ &  $P_c$ 
\tabularnewline
\hline
Pole [MeV] &  $4313(1)-3(1)i$  & $4461(2)-5(2)i$ & $4525(4)-9(1)i$
\tabularnewline
\hline
$\Sigma_c\bar{D}(S)$ &  $2.7(2)+0.2(1)i$ & $0.0(1)-0.2(1)i$ & $0.0(1)-0.1(1)i$ 
\tabularnewline
\hline
$\Sigma_c\bar{D}^*(S)$ & $-0.6(2)+0.1(1)i$ & $1.9(4)+1.2(3)i$  & $-0.0(1)+0.1(1)i$ 
\tabularnewline
\hline
$\Sigma_c^*\bar{D}^*(S)$ &  $-0.4(3)+0.1(1)i$ & $0.3(1)+0.7(2)i$ & $1.4(3)+0.8(5)i$ 
\tabularnewline
\hline
$\Sigma_c\bar{D}^*_{\frac32}(D)$  & $-0.7(1)+0.1(0)i$ &  $0.0(0)+0.0(1)i$ & $0.1(0)-0.0(0)i$ 
\tabularnewline
\hline
$\Sigma_c^*\bar{D}(D)$ &  $-0.0(0)+0.0(0)i$ & $ -0.3(1)-0.0(0)i$ & $-0.2(1)-0.0(0)i$ 
\tabularnewline
\hline
$\Sigma_c^*\bar{D}^*_{\frac32}(D)$ &  $0.2(1)-0.0(1)i$ & $-0.3(1)+0.3(1)i$ &  $0.0(0)-0.0(0)i$
\tabularnewline
\hline
$\Sigma_c^*\bar{D}^*_{\frac52}(D)$ &  $0.6(2)-0.1(0)i$ & $0.2(1)-0.2(1)i$ & $0.0(0)-0.0(0)i$
\tabularnewline
%\hline
%\hline
\end{tabular}
}
\end{ruledtabular}
}
\end{table*}

\begin{table*}[!htb]
\caption{The effective couplings to the elastic channels for $J^P=\frac32^-$ resonances of scheme II, as defined in Eq.~\eqref{eq:def:couplings}. }\label{tab:coupling_II_32}
{
\begin{ruledtabular}
\resizebox{\textwidth}{!}{
\begin{tabular}{ c | c  c  c }
$J=\frac32^-$ &  $P_c(4380)$  & $P_c(4440)$ &  $P_c$ 
\tabularnewline
\hline
Pole [MeV] &  $4376(1)-6(2)i$ & $4441(2)-6(2)i$ & $4520(3)-12(3)i$
\tabularnewline
\hline
$\Sigma_c\bar{D}^*(S)$ &  $0.4(2)-0.3(1)i$ & $3.6(2)+0.5(2)i$ &  $-0.2(1)+0.1(1)i$ 
\tabularnewline
\hline
$\Sigma_c^*\bar{D}(S)$ &  $2.8(2)+0.0(1)$ & $0.1(2)+0.1(1)i$ & $-0.0(1)-0.3(1)i$ 
\tabularnewline
\hline
$\Sigma_c^*\bar{D}^*(S)$ &   $-0.7(2)-0.0(1)i$ & $0.8(2)+0.5(2)i$ & $ 2.4(2)+1.0(4)i $ 
\tabularnewline
\hline
$\Sigma_c\bar{D}(D)$ &  $0.0(0)+0.0(0)i$ & $-0.3(1)+0.0(0)i$ & $0.1(1)+0.1(1)i$ 
\tabularnewline
\hline
$\Sigma_c\bar{D}^*_{\frac12}(D)$  & $-0.2(0)+0.0(0)i$ & $-0.0(1)-0.0(0)i$ & $-0.2(1)+0.1(1)i$ 
\tabularnewline
\hline
$\Sigma_c\bar{D}^*_{\frac32}(D)$ &  $-0.2(1)-0.0(0)i$ & $0.2(1)-0.0(0)i$ & $-0.0(0)+0.1(1)i$ 
\tabularnewline
\hline
$\Sigma_c^*\bar{D}(D)$ &  $0.0(0)+0.0(0)i$ & $0.1(1)+0.0(0)i$ & $0.3(1)-0.1(1)i$
\tabularnewline
\hline
$\Sigma_c^*\bar{D}^*_{\frac12}(D)$ &  $-0.2(1)+0.0(0)i$ & $0.1(1)-0.1(1)i$ & $0.0(1)+0.0(0)i$
\tabularnewline
\hline
$\Sigma_c^*\bar{D}^*_{\frac32}(D)$ &  $-0.4(1)+0.0(0)i$ & $0.1(1)+0.0(0)i$ & $0.1(0)+0.0(1)i$ 
\tabularnewline
\hline
$\Sigma_c^*\bar{D}^*_{\frac52}(D)$ & $-0.6(1)+0.0(1)i$ & $-0.2(1)+0.2(1)i$ & $0.0(0)+0.0(0)i$
\tabularnewline
%\hline
%\hline
\end{tabular}
}
\end{ruledtabular}
}
\end{table*}

\begin{table*}[!htb]
\caption{The effective couplings to the elastic channels for $J^P=\frac52^-$ resonances of scheme II, as defined in Eq.~\eqref{eq:def:couplings}.}\label{tab:coupling_II_52}
{
%\begin{ruledtabular}
%\resizebox{\textwidth}{!}{
\begin{tabular}{ c |  c }
%\hline 
\hline
\hline
% \multirow{2}{*}{$J=\frac52^-$} & Solution A & Solution B  \\
%\cline{2-3}
 $J=\frac52^-$ &   $P_c$ 
\tabularnewline
\hline
Pole [MeV] &  $4500(2)-9(6)i$
\tabularnewline
\hline
$\Sigma_c^*\bar{D}^*(S)$ & $4.0(5)+0.6(2)i$
\tabularnewline
\hline
$\Sigma_c\bar{D}(D)$ & $-0.2(1)+0.0(1)i$
\tabularnewline
\hline
$\Sigma_c\bar{D}^*_{\frac12}(D)$ &  $-0.1(1)+0.0(0)i$
\tabularnewline
\hline
$\Sigma_c\bar{D}^*_{\frac32}(D)$ & $-0.1(0)+0.0(1)i$
\tabularnewline
\hline
$\Sigma_c^*\bar{D}(D)$ & $-0.3(1)+0.0(1)i$
\tabularnewline
\hline
$\Sigma_c^*\bar{D}^*_{\frac12}(D)$ & $-0.0(1)+0.0(0)i$
\tabularnewline
\hline
$\Sigma_c^*\bar{D}^*_{\frac32}(D)$ & $0.1(1)-0.0(0)i$
\tabularnewline
\hline
$\Sigma_c^*\bar{D}^*_{\frac52}(D)$ & $0.2(1)-0.1(0)i$
\tabularnewline
\hline
\hline
%\hline
%\hline
\end{tabular}
%}
%\end{ruledtabular}
}
\end{table*}

Three experimental data sets  obtained by LHCb using different kinematic selections for the candidates are employed in fits made in this analysis. 
It is worth mentioning, however, that the  shape and the relative strength of the peaks vary quite significantly in these three data sets.
Moreover, there are some hints for the existence of the higher resonances near 4.5 GeV 
in the data set ``$m_{Kp}$ all''. However, they are almost invisible for the other two data sets. 
Therefore, it should not come as a surprise that  the  effective couplings to the source also strongly depend on the input used. 
Their particular values are also quite sensitive to the variation of the background. 
For illustration, in Table~\ref{tab:productions} we collect the effective couplings to the source $g_\text{source}$ for scheme I defined as 
\bea\label{eq:def:production}
g_\alpha g_\text{source} = \lim_{E\to E_\text{pole}} (E^2-E^2_\text{pole})U_{\alpha}(E).
\eea
The uncertainties given in parentheses  correspond to the different backgrounds.

\begin{table*}[!htb]
\caption{\label{tab:productions} The couplings of the $P_c$ states to the source for Scheme~I. They are normalized by the 
event numbers and thus only the relative values are meaningful. The uncertainties given are from taking different backgrounds.  
}
%\begin{ruledtabular}
\resizebox{\textwidth}{!}{
\begin{tabular}{ c | c | c | c | c | c | c }
\hline  
\hline
 \multirow{2}{*}{Pole} & \multicolumn{3}{c|}{Solution $A$} & \multicolumn{3}{c}{Solution $B$}  \\
\cline{2-7}
 & $\cos\theta_{\pc}$-weighted  & $m_{Kp}>1.9$ GeV&  $m_{Kp}$ all & $\cos\theta_{\pc}$-weighted  & $m_{Kp}>1.9$ GeV&  $m_{Kp}$ all \\
 \hline
 $P_c(4312)$ & 68(19)+4(2)$i$ &   39(6)+3(1)$i$ & 64(19)+2(3)$i$ &  69(13)+9(5)$i$ & 68(16)+6(4)$i$ & 65(9)+6(4)$i$  \\
 \hline
 $P_c(4380)$ & -66(17)+13(6)$i$ & -35(6)+7(2) $i$ & -39(9)+5(2)$i$ & 60(11)-11(20)$i$ & 30(7)-2(2)$i$ & 41(11)+4(2)$i$\\
 \hline
 $P_c(4440)$ & -96(9)+4(3)$i$ & -51(10)+3(1)$i$ & -75(14)+3(3)$i$ & 160(20)+23(16)$i$ & 55(20)+5(3)$i$ & 97(23)+4(7)$i$\\
 \hline
 $P_c(4457)$ &126(31)+30(20)$i$ & 65(5)+18(4)$i$ & 81(12)+8(4)$i$ & -76(15)-24(12)$i$ & 75(13)+48(8)$i$  & 77(6)+32(13)$i$\\
 \hline
 $P_c(1/2)$ & -37(35)+21(8)$i$ & -14(14)+12(4)$i$  & -72(40)+12(6)$i$ &  13(26)+32(7)$i$ & 1(14)+14(16)$i$ & 76(52)+11(24)$i$\\
 \hline
 $P_c(3/2)$ & $-98(77)+16(9)i$ & -88(39)+8(2)$i$ & -19(12)+7(4)$i$  &  -5(6)+2(2)$i$ & 1(5)-2(7)$i$ & -6(20)+11(8)$i$\\
 \hline
 $P_c(5/2)$ &  -23(23)-7(7)$i$  & 0(12) +0(11)$i$ & 52(31)+5(4)$i$ & 39(60)+2(4)$i$ & -17(20)+0(1)$i$ & 55(32)+1(2)$i$\\
\hline
\hline
\end{tabular}}
%\end{ruledtabular}
\end{table*}

\bibliography{references}
%\bibliography{main}
\end{document}

\begin{table*}[!htb]
  \caption{\label{tab:III:paras}The best fit parameters of scheme III shown in Fig.~\ref{fig:III:fits}. Here we define $g_D^\prime = g_D k_0^2$, where $k_0= \sqrt{\lambda(m^2_0,m_\psi^2,m_p^2)}/2m_0 $ with $m_0=(m_{\Sigma_c}+m_{\Sigma_c^*}+m_D+m_{D^*})/2 $, to make the $g_D^\prime$ have the same unit as $g_S$. Here only the statistical uncertainties are presented. \blue{Remove the results for solution $A$}}
  {
  \begin{tabular}{c |c c| c c | c c}
  \hline 
  \hline
  Parameter & background    & $1$ &  background  &  $2$ & background &  $3$  \\
  \hline
  solution & $A$ & $B$ & $A$ & $B$ & $A$ & $B$ \\
  \hline 
  $C_\frac12~[\text{GeV}^{-2}]$ & $3.64(13)$ & $7.41(47)$ & $3.70(10)$ & $9.34(82)$ & $3.95(49)$ & $8.23(39)$
  \tabularnewline
  \hline
  $C_\frac32~[\text{GeV}^{-2}]$ & $-15.12(34) $ & $-17.65(48)$ & $-14.87(13)$ & $-15.47(80)$ & $-16.74(30)$ & $-16.69(40)$
  \tabularnewline
  \hline
  $C_\frac12^\prime~[\text{GeV}^{-2}]$ & $13.26(37)$ & $15.39(300)$ & $12.68(20)$ &  $19.94(144)$ & $14.43(57)$ & $18.68(57)$
  \tabularnewline
  \hline
  $D_b^{SD}~[\text{GeV}^{-4}] $ & $-2.65(5) $ & $-1.45(43)$ & $-2.62(3)$ & $-0.71(27)$ & $-2.15(11)$ & $-0.90(16)$
  \tabularnewline
  \hline
  ${D}_c^{SD}~[\text{GeV}^{-4}]$ & $-1.50(16)$ & $-2.23(90)$ &  $-1.28(4)$ & $-1.99(33)$ & $-1.94(29)$ & $-2.83(11)$
  \tabularnewline
  \hline
  $g_S~[\text{GeV}^{-2}]$ & $1.36(115)$ & $3.98(78)$ & $1.17(18)$ & $0.53(19)$ & $4.28(59)$ & $3.61(49)$
  \tabularnewline
  \hline
  $g_D^\prime~[\text{GeV}^{-2}]$ & $1.61(41)$ & $1.51(37)$ & $0.03(242)$ & $0.24(10)$ & $0.66(231)$ & $2.50(29)$
  \tabularnewline
  \hline
  $\mathcal{F}_1^\frac12$ & $192(241)$ & $1118(882)$ & $406(365)$  & $12931(7254)$ & $-318(694)$ & $1199(1096)$
  \tabularnewline
  \hline
  $\mathcal{F}_2^\frac12$ & $-14(220)$ & $913(346)$ & $-1075(301)$ & $6493(6612)$ & $282(518)$ & $1160(480)$
  \tabularnewline
  \hline
  $\mathcal{F}_3^\frac12$ & $-1384(410)$ & $-1404(250)$ & $-3469(412)$ & $-23057(7843)$ & $-1386(322)$ & $-1404(261)$
  \tabularnewline
  \hline
  $\mathcal{F}_1^\frac32$ & $-1444(321)$ & $-390(195)$ & $-3930(554)$ & $-3599(2659)$ & $-686(333)$ & $-1938(381)$
  \tabularnewline
  \hline
  $\mathcal{F}_2^\frac32$ & $-1942(697)$ & $60(255)$ & $1668(577)$ & $-1422(1943)$ & $377(457)$ & $638(304)$
  \tabularnewline
  \hline
  $\mathcal{F}_3^\frac32$ & $-764(320)$& $-720(207)$ & $-2543(667)$ & $-7984(3574)$ & $-202(195)$ & $-675(103)$
  \tabularnewline
  \hline
  $\mathcal{F}_1^\frac52$ & $224(648)$ & $120(135)$ & $0(35892)$ &$2894(1812)$ & $-79(469)$ & $235(199)$
  \tabularnewline
  \hline
  \hline
  \end{tabular}
  }
  \end{table*}